\documentclass[pdflatex,sn-mathphys-num]{sn-jnl}

\usepackage{graphicx}
\usepackage{amsmath,amssymb,amsfonts}
\usepackage{amsthm}
\usepackage[title]{appendix}
\usepackage{xcolor}
\usepackage{textcomp}
\usepackage{booktabs}
\usepackage{tabularx}
\usepackage{caption}
\usepackage{placeins}
\usepackage{enumitem}
\usepackage[T1]{fontenc}
\usepackage[most]{tcolorbox}
\usepackage{algorithm}
\usepackage[noend]{algpseudocode}
\algrenewcommand\algorithmicrequire{\textbf{Input:}}
\algrenewcommand\algorithmicensure{\textbf{Output:}}
\algrenewcommand\algorithmiccomment[1]{\hfill{\footnotesize\itshape$\triangleright$ #1}}

\newcommand{\partitle}[1]{\medskip\noindent\textbf{#1}\quad}

\definecolor{checkbg}{RGB}{255,235,0}
\definecolor{checkfg}{RGB}{200,0,0}

\newif\ifshowchecklist
\showchecklisttrue             % set to \showchecklistfalse once every item is resolved

\setlist{itemsep=1pt,topsep=3pt}

\theoremstyle{thmstyleone}
\newtheorem{lemma}{Lemma}

\newcommand{\ICM}[1]{\mathrm{ICM}_{#1}}
\newcommand{\pin}{p_{\mathrm{in}}}
\newcommand{\pout}{p_{\mathrm{out}}}
\newcommand{\IBR}{\mathrm{IBR}}

\begin{document}

\title[Disparity in Influence]{Disparity in Influence: A Fairness Evaluation of Centrality Measures in Complex Networks}

\author[1]{\fnm{Kirtidev} \sur{Mohapatra}}\email{kirtidevm@iitbhilai.ac.in}
\author[1]{\fnm{Harsh Vardhan} \sur{Singh}}\email{harshvardhans@iitbhilai.ac.in}
\author[1]{\fnm{Rishi Ranjan} \sur{Singh}}\email{rishi@iitbhilai.ac.in}
\author[2]{\fnm{Akrati} \sur{Saxena}}\email{a.saxena@liacs.leidenuniv.nl}

\affil[1]{\orgdiv{Department of Computer Science and Engineering}, \orgname{Indian Institute of Technology Bhilai}, \orgaddress{\country{India}}}
\affil[2]{\orgdiv{Leiden Institute of Advanced Computer Science}, \orgname{Leiden University}, \orgaddress{\country{The Netherlands}}}

\abstract{Social networks are shaped by homophily-driven connections, which naturally lead to the emergence of communities with uneven sizes and influence. Such structural imbalances impact the fairness of network analysis tools, particularly centrality measures that are widely used to identify influential nodes for tasks like information diffusion and network robustness. In this work, we conduct a comprehensive assessment of fairness in centrality measures, using influence spreading across networks with heterogeneous community structures. We first analytically show that disparity is bounded by outreach; cross-community outreach is bounded by the inter-community cut; and when all seeds lie in one community, disparity approaches the normalized outreach within that community as cross-community transmission vanishes. This explains why attenuation concentrates influence in the community where diffusion starts. We further evaluate a broad spectrum of centrality measures, including local, semi-local, global, and community-aware measures, on both synthetic network generative models (HBA and HICH-BA) and real-world datasets, using the Independent Cascade and Linear Threshold Models. The results consistently show that influence tends to concentrate within larger communities, with disparities becoming more pronounced under constrained or deterministic diffusion settings. We further observe that fairness is highly sensitive to the choice of centrality measure rather than its general category. In particular, fairness-aware adaptations of PageRank demonstrate a strong balance between maximizing influence reach and maintaining equitable distribution across communities. On real-world networks, these conclusions are largely stable under random removal or addition of up to $3\%$ of the edges. Our findings highlight the need to explicitly account for fairness when designing and selecting centrality-based strategies in networked systems.}

\keywords{Algorithmic fairness, Centrality measures, Influence maximization, Influence Bias, Homophily, Independent Cascade Model, Linear Threshold Model}

\maketitle

%% =====================================================================
\section{Introduction}\label{sec:intro}
%% =====================================================================

Social networks model social systems in which individuals (nodes) are connected through interactions (edges) such as friendship, collaboration, or communication. Studying them helps to understand how relationships form, how communities are structured, and how influence is distributed among individuals. Real social networks are shaped by social, economic, and political forces that introduce structural inequalities~\cite{saxena2024fairsna}. Two mechanisms are particularly important. \emph{Homophily}, the tendency of individuals to connect with others who share similar attributes, produces tightly knit communities~\cite{mcpherson2001birds,saxena2025homophily}, and \emph{preferential attachment} lets already well-connected nodes accumulate further links~\cite{barabasi1999emergence}. Together, they reinforce systemic disparities, such as the glass ceiling effect, in which a minority group is underrepresented among the best-connected nodes despite comparable attributes~\cite{avin2015homophily,stoica:2018}. In addition, community sizes in real social networks are typically heavily skewed, with a few large communities and many small ones~\cite{newman2004fast}. If these imbalances are ignored, network-based analyses and decisions inherit them, thereby motivating fairness-aware algorithms for tasks such as ranking, link prediction, community detection, misinformation control, and influence maximization~\cite{saxena2024fairsna,dong2023fairness}.

Centrality measures are basic tools of network analysis: they rank nodes by their importance based on their positions in the network~\cite{singh2022centrality}, and they are widely used in influence maximization, viral marketing, and recommendation. They are commonly grouped into \emph{local} measures, which use the immediate neighborhood of a node, \emph{semi-local} measures, which use a few hops, and \emph{global} measures, which use the whole network~\cite{saxena2020centrality}; \emph{community-aware} (modular) measures additionally take the community structure into account~\cite{ghalmane2019centrality,rajeh2021investigating}. Influence maximization, the selection of a small seed set that maximizes the expected spread of a diffusion process, is NP-hard, and even evaluating a candidate seed set requires Monte Carlo simulation~\cite{kempe2003maximizing,chen2010scalable}. Algorithms with fairness guarantees exist~\cite{tsang2019group,fish2019gaps,rahmattalabi2021fair,becker2022fairness}, but they inherit this cost, and in practice seeds are therefore often selected by ranking nodes with a centrality measure~\cite{ganguly2024influence,dey2021survey}. Most centrality measures, however, ignore group membership. By rewarding connectivity, they tend to favor nodes in large and dense communities and may thereby amplify the structural imbalances created by homophily~\cite{espin2022inequality}. Fairness-aware adaptations of PageRank have been proposed to counter this bias~\cite{tsioutsiouliklis2021fairness}: Fairness-Sensitive PageRank modifies the teleportation component in favor of a protected group, and Locally Fair PageRank balances the distribution of scores in every neighborhood. It is not known how these measures, or any other centrality measure, behave when they are used to seed a diffusion process in a network with imbalanced communities, and whether a measure that appears fair distributes influence well or reaches few nodes.

This paper addresses these questions. We evaluate $25$ local, semi-local, global, community-aware, and fairness-aware centrality measures with respect to both influence outreach and the fair distribution of that outreach across communities. We examine how the outreach--fairness trade-off depends on the diffusion model, how structural properties of the network, in particular homophily, community-size imbalance, and clustering, amplify disparity, and which measures provide favorable trade-offs that remain stable when the network is perturbed. The study substantially extends our CSoNet 2025 paper~\cite{PSSS:2025}, which compared centrality measures on a two-community homophilic network under the Independent Cascade Model. We give formal definitions of all measures and metrics and derive analytical results that explain the main empirical patterns: disparity is bounded by outreach, cross-community outreach is bounded by the number of inter-community edges and the inter-community activation probability, and, at each attachment step (HBA) and each ordinary edge-creation step (HICH-BA), higher homophily makes the creation of an inter-community edge less likely. We add the High-Clustering Homophily Barab\'asi--Albert model~\cite{saxena2023fairness}, which produces several highly clustered communities of different sizes, and the Linear Threshold Model, which captures diffusion through collective reinforcement. We further analyze the group composition of the seed sets and test the stability of the results through a sensitivity analysis of the activation probability and random perturbations to the structure of the real networks. The experiments show that the fairness of centrality-based seeding is a property of the individual measure and of the network and diffusion setting rather than of the class of the measure, and that several measures appear fair only because they spread little. The two fairness-aware PageRank variants also behave differently. In homophilic networks at the default activation probability, Locally Fair PageRank yields near-balanced reach under attenuated cascades, whereas Fairness-Sensitive PageRank over-serves the minority under cascade diffusion; at higher activation probabilities, Fairness-Sensitive PageRank becomes the fairer of the two in the two-community model, and under threshold diffusion it is more effective on the synthetic networks, though not on the real ones.

The paper is organized as follows. Section~\ref{sec:related} reviews related work. Section~\ref{sec:prelim} introduces the notation, diffusion models, and evaluation metrics and formulates the fair seeding problem, and Section~\ref{sec:theory} presents the analytical results. Section~\ref{sec:method} describes the centrality measures, networks, and experimental settings. Section~\ref{sec:results} reports the results, Section~\ref{sec:discussion} discusses them, and Section~\ref{sec:conclusion} concludes.

%% =====================================================================
\section{Related Work}\label{sec:related}
%% =====================================================================

\subsection{Centrality measures and influence maximization}

Kempe et al.~\cite{kempe2003maximizing} formulated influence maximization as a discrete optimization problem: select $k$ seed nodes that maximize the expected number of nodes activated under the Independent Cascade Model (ICM) or the Linear Threshold Model (LTM). They showed that the problem is NP-hard and that the expected spread is monotone and submodular under both models, so that greedy selection achieves a $(1-1/e)$ approximation~\cite{nemhauser1978analysis}, or $(1-1/e-\varepsilon)$ when the spread is estimated by simulation. Because each greedy step requires Monte Carlo estimation of the spread, a large body of work has developed faster heuristic and sampling-based alternatives~\cite{chen2010scalable,li2018influence,peng2018influence}. The simplest and most widely deployed alternative is to rank nodes by a centrality measure and to select the top $k$~\cite{kundu2011new,hussain2013empirical,dey2019influence,dey2021survey}. Dey et al.~\cite{dey2019influence} compared different centrality measures as seed nodes for information propagation in online social networks, and later surveyed the role of centrality in seed selection for large-scale networks~\cite{dey2021survey}. 

The spreading ability of classical centralities has been compared extensively~\cite{de2014role,saxena2020centrality,singh2022centrality}, and dedicated measures have been proposed to identify influential spreaders. Chen et al.~\cite{chen2012identifying} and Gao et al.~\cite{gao2014ranking} introduced semi-local measures that approximate global information at local cost. Yang and Xiao~\cite{yang2021improved} combined the $k$-shell decomposition with a gravity model, and Xiao et al.~\cite{xiao2024new} used local average shortest paths over an extended neighborhood. In networks with community structure, measures that distinguish intra- from inter-community connections often identify better spreaders. Tulu et al.~\cite{tulu2018identifying} proposed the community-based mediator, Ghalmane et al.~\cite{ghalmane2019centrality} introduced modular centralities that combine the local and global roles of a node, and Rajeh et al.~\cite{rajeh2021investigating} compared community-aware centralities systematically. Ganguly et al.~\cite{ganguly2024influence} seeded within detected communities using classical centralities. All of these studies evaluate centrality measures by their outreach only; how the resulting influence is distributed across communities, which is the focus of our work, has not been examined systematically.

\subsection{Fairness-aware social network analysis}

Saxena et al.~\cite{saxena2024fairsna} survey algorithmic fairness in social network analysis, highlighting how homophily and structural inequalities contribute to disparate outcomes across network analysis tasks. Dong et al.~\cite{dong2023fairness} survey fairness notions and mitigation techniques across graph-mining tasks. The structural origins of disadvantage are well documented. Avin et al.~\cite{avin2015homophily} proved that homophily combined with preferential attachment produces a glass ceiling for minorities, and Stoica et al.~\cite{stoica:2018} showed that recommendation algorithms can reinforce it. Karimi et al.~\cite{karimi2018homophily} showed that homophily and relative group size jointly determine the degree ranking of minorities. Lee et al.~\cite{lee2019homophily} showed that they also distort how groups perceive each other's prevalence. Esp\'in-Noboa et al.~\cite{espin2022inequality} documented the inequality and inequity produced by network-based ranking and recommendation algorithms, including PageRank, on generative models with controlled homophily. In recent years, fairness-aware methods have been developed for various network analysis tasks, including node ranking \cite{tsioutsiouliklis2021fairness}, link prediction \cite{saxena2022hm,saxena2022nodesim}, influence maximization \cite{fish2019gaps,tsang2019group}, influence blocking \cite{saxena2023fairness}, opinion dynamics \cite{stkepien2026fairness}, and community detection \cite{de2024group,corriera2026individual}. Here, we focus on fairness in influence maximization and review its literature in detail.

Research on fairness in influence maximization has evolved from defining equitable information access to developing optimization methods that balance fairness and outreach. Early formulations introduced information-access gaps between groups \cite{fish2019gaps} and group-rationality and maximin constraints to ensure equitable influence propagation \cite{tsang2019group}. Beyond defining fairness objectives, subsequent studies examined how network structure and seed selection contribute to unequal influence, demonstrating that diversity-aware seeding can improve both equity and overall outreach \cite{stoica2019fairness,stoica:2020}. To accommodate different fairness objectives, Farnadi et al.~\cite{farnad2020unifying} unified equality, equity, maximin, and diversity criteria in an integer-programming framework. Rahmattalabi et al. \cite{rahmattalabi2021fair} adopted a welfare-based formulation that interpolates between utilitarian and egalitarian objectives. Further research has considered specific challenges arising from homophily \cite{anwar2021balanced}, time-critical diffusion \cite{ali2021fairness}, and unequal information access, including interventions based on edge addition \cite{jalali2023fairness} and analyses using generative network models \cite{wang2022information}. Despite these advances, incorporating fairness makes influence maximization computationally more challenging than its classical counterpart. In particular, deterministic maximin fairness can be inapproximable, whereas randomized formulations admit constant-factor approximation guarantees \cite{becker2022fairness}. Building on randomized strategies, a stochastic Frank--Wolfe framework has also been developed for fairness-aware influence maximization \cite{zhang2026fairness}.

%Several notions of fairness have been proposed for influence maximization. Fish et al.~\cite{fish2019gaps} formalized the information-access gap between groups; Tsang et al.~\cite{tsang2019group} imposed group-rationality and maximin constraints; Stoica and Chaintreau~\cite{stoica2019fairness} and Stoica et al.~\cite{stoica:2020} analyzed how seeding interacts with structural bias and showed that diversity-aware seeding can improve both equity and total reach; Farnadi et al.~\cite{farnad2020unifying} unified equality, equity, maximin and diversity criteria in an integer-programming framework; and Rahmattalabi et al.~\cite{rahmattalabi2021fair} adopted a welfare-based formulation that interpolates between utilitarian and egalitarian objectives. Anwar et al.~\cite{anwar2021balanced} studied balanced influence maximization in the presence of homophily, Ali et al.~\cite{ali2021fairness} studied fairness under time-critical diffusion, Jalali et al.~\cite{jalali2023fairness} defined an information-unfairness measure and reduced it by adding edges, and Wang et al.~\cite{wang2022information} studied information-access equality on generative network models. Fair influence maximization is harder than its classical counterpart: Becker et al.~\cite{becker2022fairness} showed that the deterministic maximin formulation is inapproximable, whereas randomized relaxations admit constant-factor guarantees, a direction that Zhang et al.~\cite{zhang2026fairness} extended with a stochastic Frank--Wolfe framework.

Recent algorithmic advances in fairness-aware influence maximization can be broadly grouped into sampling-based, optimization-based, and learning-based approaches. Sampling-based methods extend reverse influence sampling to incorporate fairness objectives, including attribute-aware sampling \cite{lin2023fair} and unbiased estimators that transform welfare-fair influence maximization into a weighted maximum coverage problem \cite{rui2025scalable}. A second research direction addresses fairness and outreach as competing optimization objectives. Multi-objective grey wolf optimizers \cite{razaghi2022group,zhao2024fair}, fairness-constrained multi-objective optimization \cite{zhao2026fairness}, and prior-knowledge-guided optimization \cite{gong2023influence} offer different strategies for balancing these objectives. Further developments accommodate overlapping communities \cite{jiang2025overlapping} and employ evolutionary optimization to construct group-specific seed sets before combining them \cite{osmani2026balancing}. More recently, learning-based approaches have sought to reduce reliance on predefined heuristics by learning influence patterns or seeding policies. These include learning from observed cascades rather than simulated diffusion \cite{feng2023influence}, reinforcement learning for fairness-aware and adaptive seed selection \cite{wang2025faim,saxena2026dq4fairim,guo2026higher}, and graph contrastive learning for fairness-aware seed selection \cite{dam2026fairgcl}. Beyond these approaches, robustness to uncertainty in community partitions has emerged as an additional consideration \cite{gao2026robust}. %Despite their methodological differences, these approaches involve simulation, sampling, or training, which can impose substantial computational costs. Our study complements this literature by systematically characterizing the fairness of computationally inexpensive centrality-based seeding heuristics, providing both practical alternatives when more sophisticated methods are computationally prohibitive and reference behavior for interpreting fairness-aware seeding algorithms.

Fairness has also been built into node ranking. Tsioutsiouliklis et al.~\cite{tsioutsiouliklis2021fairness} introduced Fairness-Sensitive PageRank (FSPR), which modifies the jump vector so that a prescribed fraction $\varphi$ of the PageRank mass is assigned to a protected group, and Locally Fair PageRank (LFPR), which lets every node distribute its mass between the groups in the proportion $\varphi$. They further proposed a link recommendation to achieve fair PageRank~\cite{tsioutsiouliklis2022link}. Kumar et al. proposed mean-field approximation methods for FSPR and LFPR PageRanks \cite{kumar2026locally, kumar2026fairness}. Beyond PageRank, group-fair formulations of classical centralities are scarce, and, to our knowledge, apart from our conference paper~\cite{PSSS:2025}, no work has systematically evaluated how local, semi-local, global, and community-aware centralities behave as seeding rules with respect to the fairness of the resulting diffusion. Finally, centrality-based decisions rely on an observed snapshot of the network, which may contain errors or be manipulated. Sybil attacks introduce adversary-controlled identities into a social graph~\cite{yu2006sybilguard}; Waniek et al.~\cite{waniek2018hiding} showed that a small number of strategically rewired edges suffices to evade centrality-based detection, and adversarial perturbations of graph structure can change the outputs of graph learning models substantially~\cite{zugner2018adversarial}. Fairness-aware measures introduce additional exposure, since they depend on a declared group attribute as well as on the structure; we evaluate robustness to random perturbations and discuss targeted manipulation in Section~\ref{sec:discussion}.

%% =====================================================================
\section{Preliminaries}\label{sec:prelim}
%% =====================================================================

\subsection{Notation}\label{sec:notation}

Let $G=(V,E)$ be a simple undirected graph with $n=|V|$ nodes and $|E|$ edges. For a node $v\in V$, $N(v)=\{u\in V:\{u,v\}\in E\}$ denotes its neighborhood and $k_v=|N(v)|$ its degree; $A$ is the adjacency matrix and $\lambda_1$ its largest eigenvalue. We write $d(u,v)$ for the shortest-path distance between $u$ and $v$ and $\Gamma_r(v)=\{u\in V:1\le d(u,v)\le r\}$ for the set of nodes within distance $r$ of $v$. The node set is partitioned into $q$ disjoint communities (groups) $\mathcal{C}=\{C_1,\dots,C_q\}$, and $C(v)$ is the community of $v$. For a node $v$ and a community $C_i$, $k_{v,C_i}=|N(v)\cap C_i|$ is the number of neighbours of $v$ in $C_i$; $k_v^{\mathrm{in}}=k_{v,C(v)}$ and $k_v^{\mathrm{out}}=k_v-k_v^{\mathrm{in}}$ are the intra- and inter-community degrees of $v$. For a community $C_j$, $\partial C_j$ is the set of edges with exactly one endpoint in $C_j$, and $e_j=|\partial C_j|$ is the size of its cut. In the two-community setting, we write $R$ for the minority and $B$ for the majority group. A seed set is a subset $S\subseteq V$ of size $|S|=k$, where $k$ is the seed budget. A centrality measure is a function $c: V\to\mathbb{R}$, and the \emph{top-$k$ rule} returns the set $S_c^{k}$ of the $k$ nodes with the largest values of $c$. Table~\ref{tab:notation} summarises the notation used throughout the paper.

\begin{table}[!t]
\centering
\caption{Main notation.}
\label{tab:notation}
\footnotesize
\begin{tabularx}{\linewidth}{@{}l>{\raggedright\arraybackslash}X@{\hspace{1em}}l>{\raggedright\arraybackslash}X@{}}
\toprule
Symbol & Meaning & Symbol & Meaning \\
\midrule
$G=(V,E)$ & network, $n=|V|$ nodes & $S$, $k$ & seed set and seed budget, $k=|S|$ \\
$N(v)$, $k_v$ & neighborhood and degree of $v$ & $S_c^{k}$ & top-$k$ nodes of centrality $c$ \\
$d(u,v)$, $\hat{d}$ & shortest-path distance, average shortest-path length & $\pin$, $\pout$ & intra-/inter-community activation probability \\
$A$, $\lambda_1$ & adjacency matrix, largest eigenvalue & $\theta_v$ & LTM threshold of $v$ \\
$\Gamma_r(v)$ & nodes within distance $r$ of $v$ & $A_\infty(S)$ & final active set \\
$\mathcal{C}$, $q$ & partition into $q$ communities & $\sigma(S)$ & expected outreach \\
$C_i$, $C(v)$ & community $i$, community of $v$ & $\sigma_i(S)$, $\rho_i(S)$ & outreach in $C_i$, normalised by $|C_i|$ \\
$k_v^{\mathrm{in}}$, $k_v^{\mathrm{out}}$ & intra-/inter-community degree & $\Delta(S)$ & disparity \\
$e_j$ & cut size of community $C_j$ & $\IBR(S)$ & influence balance ratio \\
$R$, $B$ & minority and majority group & $\mathrm{SD}(S)$ & seed difference $|S\cap B|-|S\cap R|$ \\
$f$ & minority fraction (HBA) & $T$ & number of Monte Carlo runs \\
$h$ & homophily parameter & $\varphi$ & target share of the protected group \\
$\eta$ & perturbation level (robustness) & $M$ & number of perturbed networks \\
\bottomrule
\end{tabularx}
\end{table}

\subsection{Diffusion models}\label{sec:diffusion}

Both diffusion models describe the spread of influence as a process in discrete time steps. At every time $t$, a node $v$ is either active or inactive, which we encode by the state $S_v(t)\in\{0,1\}$. Given a seed set $S\subseteq V$ with $|S|=k$, the process starts from
\begin{equation*}
S_v(0)=\begin{cases}1, & v\in S,\\ 0, & v\notin S,\end{cases}
\end{equation*}
and active nodes remain active. The process stops when no further node becomes active, and we denote the final set of active nodes by $A_\infty(S)$.

\paragraph{A. Independent Cascade Model (ICM).} When a node $u$ becomes active at step $t$, it receives a single opportunity to activate each of its inactive neighbors $v$ at step $t+1$, and the attempt succeeds with probability $p_{uv}$, independently of all other attempts~\cite{kempe2003maximizing}. If $\mathcal{A}_t$ denotes the set of nodes activated at step $t$ (with $\mathcal{A}_0=S$), an inactive node $v$ therefore becomes active at step $t+1$ with probability
\begin{equation*}
\Pr\bigl(S_v(t+1)=1 \mid S_v(t)=0,\ \mathcal{A}_t\bigr)=1-\prod_{u\in N(v)\cap\mathcal{A}_t}\bigl(1-p_{uv}\bigr).
\end{equation*}
Each edge is used at most once, and newly activated nodes participate in subsequent rounds. In realistic settings, the activation probabilities are not uniform but differ between intra- and inter-community edges~\cite{gupta2016modeling}. We therefore use community-dependent probabilities,
\begin{equation}\label{eq:puv}
p_{uv}=\begin{cases}\pin, & C(u)=C(v),\\ \pout, & C(u)\neq C(v),\end{cases}
\end{equation}
with $\pout=\pin$ in $\ICM{1}$ (uniform transmission), $\pout=\pin/2$ in $\ICM{2}$ and $\pout=\pin/4$ in $\ICM{3}$, i.e.\ progressively weaker transmission across community boundaries; note that $p_{uv}=p_{vu}$. An equivalent description, which we use in Section~\ref{sec:theory}, replaces every edge $\{u,v\}$ by the two arcs $(u,v)$ and $(v,u)$ and declares each arc $(u,v)$ \emph{live} independently with probability $p_{uv}$; the final active set then has the same distribution as the set of nodes reachable from $S$ along live arcs~\cite{kempe2003maximizing}.

\paragraph{B. Linear Threshold Model (LTM).} Every node $v$ holds a threshold $\theta_v\in[0,1]$, the minimum fraction of its neighbors that must be active for $v$ to become active. Thresholds are randomly drawn from a uniform distribution for each simulation run. With the fraction of active neighbors
\begin{equation*}
\phi_v(t)=\frac{1}{k_v}\sum_{u\in N(v)}S_u(t),
\end{equation*}
An inactive node with $k_v\ge1$ becomes active at step $t+1$ if $\phi_v(t)\ge\theta_v$~\cite{kempe2003maximizing,zhong2017linear}; a node without neighbors is active only if it is a seed. Given the thresholds, the dynamics are deterministic, and the randomness of the outcome comes from the thresholds. In contrast to the ICM, a node $v$ needs at least $\lceil\theta_vk_v\rceil$ active neighbors, so activation typically requires \emph{social reinforcement} from several of them.

\subsection{Evaluation metrics}\label{sec:metrics}

\paragraph{Influence outreach.} For a seed set $S$, the influence outreach, the group outreach, and the normalized group outreach are
\begin{equation}\label{eq:sigma}
\sigma(S)=\mathbb{E}\bigl[|A_\infty(S)|\bigr],\qquad
\sigma_i(S)=\mathbb{E}\bigl[|A_\infty(S)\cap C_i|\bigr],\qquad
\rho_i(S)=\sigma_i(S)/|C_i| .
\end{equation}

\paragraph{Disparity.} The disparity is the largest gap in normalized outreach between two communities~\cite{saxena2024fairsna},
\begin{equation}\label{eq:disparity}
\Delta(S)=\max_{i}\rho_i(S)-\min_{j}\rho_j(S)\in[0,1].
\end{equation}
$\Delta=0$ means that every community is reached in proportion to its size, and values closer to $1$ indicate increasingly unequal reach.

\paragraph{Influence Balance Ratio.} Disparity is symmetric and does not show which group is favored. For two groups we therefore also report the directed ratio $\IBR(S)=\rho_B(S)/\rho_R(S)$, for which values above $1$ favour the majority and values below $1$ the minority; for $q>2$ communities we use $\IBR(S)=\max_i\rho_i(S)/\min_i\rho_i(S)\ge1$. The ratio is only defined when the denominator is positive and becomes unstable when it is close to zero. For two groups and $\rho_R(S)>0$, $\Delta=\rho_R(S)\,|\IBR-1|$; the identity holds for the expected quantities and the realized quantities of every run, but not for the run-averaged estimates defined below.

\paragraph{Seed difference.} For two groups, $\mathrm{SD}(S)=|S\cap B|-|S\cap R|$ describes the group composition of the seed set. Proportional representation corresponds to $\mathrm{SD}^\star=k(|B|-|R|)/n$, i.e.\ $0.6k$ for a minority of $20\%$.

%\paragraph{Estimation.} 
Since both diffusion models are stochastic, all quantities are estimated from $T=1{,}000$ independent Monte Carlo runs. For run $t$ with final active set $A^{(t)}$, we compute the normalized outreach for each community, from which we obtain the realized disparity $\Delta^{(t)}$ and IBR. We report the mean outreach $\widehat\sigma=\frac1T\sum_t|A^{(t)}|$ and the run-averaged disparity $\widehat\Delta=\frac1T\sum_t\Delta^{(t)}$ (and analogously the IBR), with error bars of one standard deviation across runs. Because $\max_i-\min_j$ is a convex function of the vector of normalized outreaches, Jensen's inequality implies that $\widehat\Delta$ is never smaller than the disparity computed from the averaged outreaches; accordingly, $\widehat\Delta$ estimates $\mathbb{E}[\Delta^{(t)}]\ge\Delta(S)$ rather than $\Delta(S)$ itself. All comparisons use the same estimator, and Lemma~\ref{lem:bounds} states which of its bounds also hold for $\mathbb{E}[\Delta^{(t)}]$.

\subsection{Problem formulation}\label{sec:formulation}

A good seeding rule combines high outreach with low disparity, and the two objectives can conflict: the seed set that reaches most nodes need not distribute its reach fairly, and the seed set with the most equal reach may reach few nodes. Given a network $G$, a partition $\mathcal{C}$, a diffusion model and a budget $k$, fair seeding is therefore the bi-objective problem
\begin{equation}\label{eq:biobjective}
\max_{S\subseteq V,\;|S|=k}\ \sigma(S)\qquad\text{and}\qquad\min_{S\subseteq V,\;|S|=k}\ \Delta(S),
\end{equation}
or, in constrained form for a tolerance $\delta\in[0,1]$, $\max\{\sigma(S):|S|=k,\ \Delta(S)\le\delta\}$. A seed set $S$ \emph{dominates} a seed set $S'$ if $\sigma(S)\ge\sigma(S')$ and $\Delta(S)\le\Delta(S')$, with at least one of the two inequalities strict. The centrality-based heuristic studied in this paper restricts the feasible set to the $25$ candidates $S_c^{k}$, one for each centrality measure $c$ of Section~\ref{sec:centralities}, and our evaluation asks which of these candidates are non-dominated and why. In the figures, favorable measures lie towards high outreach and low disparity.

Two properties explain why we compare the measures in the $(\sigma,\Delta)$ plane rather than optimize a single objective. First, the problem is computationally hard: for $\delta=1$ the constraint $\Delta(S)\le\delta$ is non-binding (since $\Delta(S)\in[0,1]$), so the constrained problem contains classical influence maximization, which is NP-hard for general (directed) ICM instances~\cite{kempe2003maximizing}; for the LTM with weights $1/k_v$ and thresholds uniform on $[0,1]$ this holds already on undirected graphs, by the reduction from vertex cover in~\cite{kempe2003maximizing}. Second, the fairness objective lacks the structure that makes classical influence maximization tractable. The expected outreach $\sigma$ is monotone and submodular under the ICM and, for thresholds drawn independently and uniformly from $[0,1]$, under the LTM~\cite{kempe2003maximizing}, which underlies the $(1-1/e)$ approximation guarantee of greedy maximization~\cite{nemhauser1978analysis}. The fairness utility $F(S)=1-\Delta(S)$ has neither property. Consider two isolated nodes $r\in R$ and $b\in B$ with $|R|=|B|=1$, so that under either diffusion model a seed activates only itself. Then $F(\emptyset)=1$, $F(\{r\})=F(\{b\})=0$ and $F(\{r,b\})=1$. Adding $r$ to the empty set decreases $F$, so $F$ is not monotone; and the marginal gain of $b$ is $-1$ with respect to $\emptyset$ but $+1$ with respect to the larger set $\{r\}$, so $F$ is not submodular. Greedy guarantees therefore do not carry over to fairness objectives, in line with the inapproximability of deterministic maximin fair influence maximization~\cite{becker2022fairness}. These observations motivate reporting the outreach--disparity trade-off of each measure rather than a single scalarised score.

%% =====================================================================
\section{Analytical Analysis}\label{sec:theory}
%% =====================================================================

This section gives two lemmas that explain the main empirical patterns of Section~\ref{sec:results}. Lemma~\ref{lem:bounds} bounds disparity from above and from below in terms of outreach, community size, and the structure of the community boundary, and Lemma~\ref{lem:homophily} relates the community boundary to the homophily parameter $h$ of the two generative models. We first recall a known property of the ICM and introduce the two quantities that appear in the bounds.

\paragraph{Monotonicity of the ICM.} Under the ICM and for a fixed seed set $S$, the group outreach $\sigma_j(S)$ of every community, and hence $\sigma(S)$, is non-decreasing in every arc probability $p_{uv}$, in particular in $\pin$ and $\pout$. This follows from the live-arc representation~\cite{kempe2003maximizing}: drawing one uniform random number per arc couples two settings with probabilities $\mathbf p\le\mathbf p'$ so that every arc live in the first is live in the second. Since a non-edge is equivalent to two arcs with probability $0$, adding edges cannot decrease $\sigma_j(S)$ and removing edges cannot increase it. The property concerns expected values, so estimates from independent simulations can deviate from it within the Monte Carlo error. It has no analog for the LTM, because adding an edge at $v$ reduces the weight $1/k_v$ with which each existing neighbor of $v$ counts towards $\phi_v$. For example, if a seed $s$ is the only neighbour of $v$ and $w$ belongs to a triangle $\{w,a,b\}$ without further edges, then, with thresholds uniform on $[0,1]$, adding the edge $\{v,w\}$ lowers the expected outreach of $\{s\}$ from $2$ to $23/12$. Disparity need not be monotone in the probabilities even under the ICM, because all groups gain outreach but not necessarily by the same normalized amount (Section~\ref{sec:sens}).

\paragraph{Two auxiliary quantities.} Recall from Section~\ref{sec:notation} that $e_j$ is the number of edges with exactly one endpoint in community $C_j$ (the size of its cut). Under the ICM with probabilities~\eqref{eq:puv}, each of these edges transmits influence across the boundary of $C_j$ with probability $\pout$ in each direction, and we write
\begin{equation}\label{eq:beta}
\beta_j=1-(1-\pout)^{e_j}\;\le\;\min\{1,\pout e_j\}
\end{equation}
for the probability that at least one of the $e_j$ arcs that cross the boundary of $C_j$ in a given direction (into or out of $C_j$) is live. Thus $\beta_j$ is small when the community has few boundary edges or when transmission across the boundary is weak, and $\beta_j\to0$ as $\pout\to0$. When all seeds lie in a single community $C_i$, we also write $\rho_i^{\circ}(S)$ for the expected normalized outreach that $S$ would achieve if the diffusion were confined to $C_i$, i.e.\ under the ICM with probability $\pin$ on the subgraph $G[C_i]$ induced by $C_i$; it depends only on the internal structure of $C_i$ and on $\pin$.

\begin{lemma}[Bounds on disparity]\label{lem:bounds}
Let $S$ be a seed set.
\begin{enumerate}[label=(\alph*)]
\item \emph{Outreach bound.} For every diffusion model, $\Delta(S)\le\max_l\rho_l(S)\le\sigma(S)/\min_l|C_l|$.
\item \emph{Boundary bound.} Under the ICM with probabilities~\eqref{eq:puv}, every community $C_j$ that contains no seed ($S\cap C_j=\emptyset$) satisfies $\rho_j(S)\le\beta_j$.
\item \emph{Two-sided bound.} Under the ICM with probabilities~\eqref{eq:puv}, if there are $q\ge2$ communities and all seeds lie in one community $C_i$, then
\[
\rho_i^{\circ}(S)-\beta_i\;\le\;\Delta(S)\;\le\;\rho_i^{\circ}(S)+\beta_i .
\]
\end{enumerate}
The outer bound in (a) and the two-sided bound in (c) also hold for the run-averaged disparity, $\mathbb{E}[\Delta^{(t)}]\le\sigma(S)/\min_l|C_l|$ and $\rho_i^{\circ}(S)-\beta_i\le\mathbb{E}[\Delta^{(t)}]\le\rho_i^{\circ}(S)+\beta_i$, and (a) also holds for the estimates, $\widehat\Delta\le\widehat\sigma/\min_l|C_l|$.
\end{lemma}
\begin{proof}
(a) Since $\rho_l\ge0$ for every $l$, $\Delta=\max_l\rho_l-\min_l\rho_l\le\max_l\rho_l$, and $\rho_l=\sigma_l/|C_l|\le\sigma/\min_m|C_m|$. The same argument gives $\Delta^{(t)}\le|A^{(t)}|/\min_l|C_l|$ in every Monte Carlo run, and taking means gives the outer bound for $\mathbb{E}[\Delta^{(t)}]$ and $\widehat\Delta$.

(b) We use the live-arc description of Section~\ref{sec:diffusion}. Since $S\cap C_j=\emptyset$, every path of live arcs from $S$ to a node of $C_j$ enters $C_j$ through an arc $(u,v)$ with $u\notin C_j$ and $v\in C_j$. There are exactly $e_j$ such arcs, one for each boundary edge of $C_j$, and each is live independently with probability $\pout$. With probability $(1-\pout)^{e_j}=1-\beta_j$, none of them is live, and no node of $C_j$ becomes active; otherwise, at most all $|C_j|$ nodes do. Hence $\rho_j(S)\le\beta_j$. The inequality in~\eqref{eq:beta} follows from Bernoulli's inequality $(1-\pout)^{e_j}\ge1-\pout e_j$.

(c) Let $R^{\circ}$ be the set of nodes reachable from $S$ along live arcs inside $G[C_i]$. Its distribution is that of the final active set of the ICM with probability $\pin$ on $G[C_i]$, so $\mathbb{E}|R^{\circ}|=|C_i|\,\rho_i^{\circ}(S)$, and $R^{\circ}\subseteq A_\infty(S)$. Let $L$ be the event that none of the $e_i$ arcs leaving $C_i$ is live; $\Pr(L^c)=\beta_i$. On $L$ no live path leaves $C_i$, so $A_\infty(S)=R^{\circ}$, every other community has realised normalised outreach $0$, and the realised disparity equals $|R^{\circ}|/|C_i|$. Since all realised normalised quantities lie in $[0,1]$, it follows that $\rho_i^{\circ}\le\rho_i\le\rho_i^{\circ}+\beta_i$, that $\rho_l\le\beta_i$ for every $l\ne i$, and that $|\mathbb{E}[\Delta^{(t)}]-\rho_i^{\circ}|\le\beta_i$. Finally, choosing any $l\ne i$ (which exists because $q\ge2$), $\Delta\ge\rho_i-\rho_l\ge\rho_i^{\circ}-\beta_i$, and $\Delta\le\max_m\rho_m\le\rho_i^{\circ}+\beta_i$.
\end{proof}

Part~(a) shows that low disparity can be an artifact of low outreach. In the HBA networks, with $N=10{,}000$ nodes and a minority of $2{,}000$ nodes, for example, a seed set that activates $50$ nodes on average cannot have a disparity above $0.025$, regardless of how unequally these nodes are distributed. We call such seed sets \emph{trivially fair} and always read disparity together with outreach. Parts~(b) and~(c) explain the effect of attenuated transmission from $\ICM{1}$ to $\ICM{3}$. A community without seeds can be reached only through its boundary, so its outreach is limited by the number of boundary edges and by $\pout$. When the seeds are concentrated in one community, disparity is pinned to the reach of the seeds inside their own community, up to an error $\beta_i$ that vanishes as $\pout\to0$; attenuation therefore concentrates influence in the community where diffusion starts. The bound in~(b) and~(c) is numerically informative only when $\pout e_j$ is at most of order one, i.e.\ for communities with at most a few tens of boundary edges under $\ICM{3}$. The boundaries of the real networks are far larger ($e_R=28{,}373$ on Facebook and $22{,}849$ on DBLP, Table~\ref{tab:realnets}); there the lemma describes the mechanism rather than giving a useful number, and attenuation acts by weakening each cross-community path rather than by closing the boundary. The LTM has no transmission probability that could be attenuated. For thresholds drawn independently and uniformly from $[0,1]$, the LTM with weights $1/k_v$ is equivalent to a live-edge process in which every node $v$ with $k_v\ge1$ selects exactly one neighbor, each with probability $1/k_v$, and the final active set consists of the nodes from which $S$ can be reached by repeatedly following selected neighbours~\cite{kempe2003maximizing}. A community $C_j$ without seeds is therefore reached only if one of its nodes $v$ selects an outside neighbor, which for a given $v$ happens with probability $k_v^{\mathrm{out}}/k_v$. Cross-community reach under the LTM is thus determined by the graph alone, in particular by the fractions $k_v^{\mathrm{out}}/k_v$ at the community boundaries, and not by a tunable transmission parameter.

\begin{lemma}[Homophily and inter-community edges]\label{lem:homophily}
\begin{enumerate}[label=(\alph*)]
\item In the HBA model~\cite{karimi2018homophily}, a new node of group $g$ selects an existing node $u$ with probability proportional to $h\cdot k_u$ if $u\in g$ and to $(1-h)\,k_u$ otherwise, where $h\in[0,1]$. For a given set of candidate nodes, let $D_s>0$ and $D_o>0$ be the total degrees of the candidates in the same and in the other group. Then the probability that the selection creates an inter-group edge is
\[
\pi(h)=\frac{(1-h)D_o}{hD_s+(1-h)D_o},\qquad \frac{\mathrm{d}\pi}{\mathrm{d}h}=-\frac{D_sD_o}{\bigl(hD_s+(1-h)D_o\bigr)^2}<0 ,
\]
i.e.\ it strictly decreases with $h$.
\item In the HICH-BA model~\cite{saxena2023fairness}, node additions never create inter-community edges, and, for a given network state, an ordinary edge-creation step creates one with probability $(1-h)\,a\le1-h$. Here $a\in[0,1]$ is the probability that, once the step has turned to the other communities, a candidate target is found and the edge is accepted; it depends on the network state, $p_{PA}$ and $\mathbf r$, but not on $h$.
\end{enumerate}
\end{lemma}
\begin{proof}
(a) Normalizing the attachment weights over the candidates gives $\pi(h)$; its denominator is at least $\min\{D_s, D_o\}>0$ for $h\in[0,1]$, and the quotient rule gives the derivative. (b) A new node is attached to a node of its own community. An ordinary edge-creation step considers only candidates from the source's own community with probability $h$, so in that case it cannot create an inter-community edge; with probability $1-h$ it considers candidates from the other communities and then adds the edge only if a candidate exists and passes the acceptance test (Section~\ref{sec:networks}), which happens with probability $a$. Neither the choice of the source nor the acceptance test involves $h$.
\end{proof}

Lemmas~\ref{lem:bounds} and~\ref{lem:homophily} together connect disparity with the homophily parameter. Higher homophily lowers the probability that an attachment step creates an inter-community edge, so the community boundaries tend to grow more slowly. By Lemma~\ref{lem:bounds}(b), fewer boundary edges $e_j$ mean a smaller $\beta_j$ and hence a tighter limit on the outreach of communities without seeds, and by Lemma~\ref{lem:bounds}(c) disparity is then pinned more tightly to the reach inside the seeds' community. The connection is qualitative for two reasons: the reach $\rho_i^{\circ}$ itself depends on $h$, and an exact closed form for the expected cut size over the whole network evolution is not available, because $D_s$ and $D_o$ themselves depend on $h$ and, in HICH-BA, triadic closure can create edges of both types.

The lemmas lead to predictions that we test in Section~\ref{sec:results}. Lemma~\ref{lem:bounds}(a) implies that measures with small cascades have small disparity regardless of how their seeds are distributed; on the synthetic networks under the LTM this applies to the clustering coefficient, PC and LASPN (Section~\ref{sec:res:ltm}), which we therefore treat separately from the cascade-triggering measures. Lemma~\ref{lem:bounds}(c) implies that, when all seeds lie in one community, attenuating cross-community transmission drives disparity towards the reach of the seeds inside that community. Consistent with the monotonicity of the ICM, every measure reaches more nodes at $\pin\in\{0.2,0.3\}$ than at $\pin=0.1$ (Section~\ref{sec:sens}; on HICH-BA the sensitivity runs use $k=100$ instead of $k=50$). In the robustness experiments, the seed sets are kept fixed, so under $\ICM{2}$ (Section~\ref{sec:robust}) edge addition cannot decrease and edge removal cannot increase the expected outreach; accordingly, edge addition raises and edge removal lowers the estimated outreach of almost every measure. The two exceptions, PC on Facebook for $k=50$ (about $4\%$ more outreach after edge removal) and the clustering coefficient on DBLP for $k=10$ (about $4\%$ less after adding $1\%$ of the edges), are small and we report them as observed. What the lemmas do not predict is which centrality measures concentrate their seeds in which community; this depends on the structural positions a measure rewards and is the subject of the empirical comparison.

%% =====================================================================
\section{Methodology}\label{sec:method}
%% =====================================================================

To evaluate the fairness of centrality-based influence propagation systematically, we combine a broad set of centrality measures with two diffusion models and networks of different structures. Algorithm~\ref{alg:pipeline} states the evaluation procedure, which is the same for every network. The procedure \textsc{Evaluate} (lines~1--6) estimates the outreach, disparity, and IBR of a given seed set on a given network by $T$ independent Monte Carlo simulations, using the estimators of Section~\ref{sec:metrics}. In the main experiment (lines~7--15), each network is preprocessed and labeled with its groups (Section~\ref{sec:networks}), each of the $25$ centrality measures of Section~\ref{sec:centralities} is computed once, and the $k$ highest-ranked nodes are selected as seeds for every budget $k$. The seed difference is recorded for the two-group networks, and every seed set is evaluated under $\ICM{1}$, $\ICM{2}$, $\ICM{3}$ and the LTM. The resulting points are compared in the outreach--disparity plane (Section~\ref{sec:formulation}). Because the seed sets do not depend on the diffusion parameters, the same seed sets are reused in the two stability analyses. In the sensitivity analysis (lines~16--17), the ICM experiments are repeated with higher activation probabilities (Section~\ref{sec:sens}). In the robustness analysis (lines~18--21), diffusion is simulated on $M$ randomly perturbed copies of each real-world network, while the centrality scores and seed sets are kept from the original network (Section~\ref{sec:robust}). The components of the procedure are described in detail below.

\begin{algorithm}[!t]
\caption{Fairness evaluation of centrality-based seeding}\label{alg:pipeline}
\small\raggedright
\begin{algorithmic}[1]
\Require networks $\mathcal{G}$ with node partitions $\mathcal{C}=\{C_1,\dots,C_q\}$; centrality measures $\mathcal{M}$; budgets $K$; diffusion settings $\mathcal{D}$; number of runs $T$ (values in Table~\ref{tab:settings})
\Ensure $\widehat\sigma$, $\widehat\Delta$ and $\widehat{\IBR}$ (mean and standard deviation) and $\mathrm{SD}$ for every network, measure, budget and setting
\Procedure{Evaluate}{$G,\mathcal{C},S,D,T$}
  \For{$t=1,\dots,T$}
    \State $A^{(t)}\gets$ final active set of one run of $D$ on $G$ from $S$ \Comment{LTM: new $\theta_v$ per run}
    \State $\rho_i^{(t)}\gets|A^{(t)}\cap C_i|/|C_i|$ for $i=1,\dots,q$
    \State $\Delta^{(t)}\gets\max_i\rho_i^{(t)}-\min_i\rho_i^{(t)}$;\quad $\IBR^{(t)}\gets$ ratio of Section~\ref{sec:metrics}
  \EndFor
  \State \Return mean and standard deviation of $|A^{(t)}|$, $\Delta^{(t)}$ and $\IBR^{(t)}$ over the $T$ runs
\EndProcedure
\Statex \textit{Main experiment}
\ForAll{$(G,\mathcal{C})\in\mathcal{G}$}
  \State $G\gets\textsc{Preprocess}(G)$ \Comment{simple, undirected, group labels}
  \ForAll{$c\in\mathcal{M}$}
    \State compute $c(v)$ for all $v\in V$ \Comment{$\mathcal{C}$ used by community-/fairness-aware $c$}
    \ForAll{$k\in K$}
      \State $S_c^{k}\gets$ the $k$ nodes with the largest values of $c$
      \State $\mathrm{SD}\gets|S_c^{k}\cap B|-|S_c^{k}\cap R|$ \Comment{two-group networks}
      \ForAll{$D\in\mathcal{D}$}
        \State \textsc{Evaluate}$(G,\mathcal{C},S_c^{k},D,T)$
      \EndFor
    \EndFor
  \EndFor
\EndFor
\Statex \textit{Sensitivity analysis} (same seed sets)
\ForAll{$\pin\in\{0.2,0.3\}$ and $\pout\in\{\pin,\pin/2,\pin/4\}$}
  \State repeat the ICM evaluations with these probabilities
\EndFor
\Statex \textit{Robustness analysis} (real-world networks, seed sets from the original $G$)
\ForAll{perturbation type $\in\{\text{removal},\text{addition}\}$ and level $\eta\in\{1\%,2\%,3\%\}$}
  \For{$\ell=1,\dots,M$}
    \State $G'_\ell\gets G$ with $\lfloor\eta|E|\rfloor$ random edges removed or added
    \State \textsc{Evaluate}$(G'_\ell,\mathcal{C},S_c^{k},D,T)$ for all $c$, $k$ and $D\in\{\ICM{2},\mathrm{LTM}\}$
  \EndFor
\EndFor
\end{algorithmic}
\end{algorithm}

\subsection{Centrality measures}\label{sec:centralities}

Centrality measures rank the nodes of a network by their structural importance, and their usefulness is commonly assessed through downstream tasks such as influence maximization, influence minimization, or network robustness~\cite{singh2022centrality,saxena2020centrality}. We select seeds as the top-$k$ nodes of a measure, the most widely used centrality-based formulation of influence maximization~\cite{kundu2011new,dey2021survey}. Recent work considers the dual objective of maximizing influence and spreading it equitably across communities, and in this setting the identity of the top-$k$ nodes is critical, because they drive the diffusion process. To examine how structurally informed seed selection affects fairness, we evaluate $25$ centrality measures. Following~\cite{saxena2020centrality}, we group them into local, semi-local, and global measures according to the extent of structural information they use, and we additionally distinguish \emph{community-aware} (modular) measures, which use the partition $\mathcal{C}$, and \emph{fairness-aware} measures, which use a protected group $R$ and a fairness parameter. Table~\ref{tab:centralities} summarises the classification; the class of a measure also determines its marker shape in all figures.

\subsubsection{Local centrality measures}
Local measures quantify a node's importance using only its one-hop neighborhood. They are cheap to compute and highly scalable, but they ignore the network's structure beyond the immediate neighbors.
\begin{enumerate}
\item \textbf{Degree centrality (DC)}~\cite{freeman2002centrality}. The degree of a node is the number of its direct neighbors, $DC(v)=k_v=|N(v)|$. A node with a high degree can be considered more popular than other nodes, and in homophilic networks with preferential attachment the highest-degree nodes are concentrated in the majority~\cite{karimi2018homophily}.

\item \textbf{Clustering coefficient (CC)}~\cite{watts1998collective}. The clustering coefficient measures how strongly the neighbors of a node are interconnected,
\begin{equation*}
CC(v)=\frac{2\,\mathrm{tri}(v)}{k_v(k_v-1)}\quad(k_v\ge2),\qquad CC(v)=0\quad(k_v<2),
\end{equation*}
where $\mathrm{tri}(v)$ is the number of triangles that contain $v$. Ranking nodes by their clustering coefficient places seeds in tightly connected groups, which favors local reinforcement but often limits the reach of a cascade.

\item \textbf{$H$-index}~\cite{hirsch2005index,lu2016h}. Originally a bibliometric indicator, the $H$-index of a node is the largest integer $h$ such that the node has at least $h$ neighbors of degree at least $h$. With the operator $\mathcal{H}(f,v)=\max\{h\in\mathbb{N}:|\{u\in N(v):f(u)\ge h\}|\ge h\}$, defined for any node function $f$, we have $H(v)=\mathcal{H}(k,v)$. The $H$-index identifies nodes whose neighbors are themselves well-connected, which helps a message spread beyond the immediate neighborhood.

\item \textbf{$H^2$-index}~\cite{lu2016h}. The second-order $H$-index applies the $H$-operator a second time, now to the $H$-indices of the neighbours instead of their degrees, $H^2(v)=\mathcal{H}(H,v)$ (the index $h^{(2)}$ of~\cite{lu2016h}, not the bibliometric $h(2)$-index of Kosmulski~\cite{kosmulski2006new}). It rewards nodes embedded in neighborhoods of well-connected nodes and is less sensitive to single high-degree neighbors.

\item \textbf{Z-score}~\cite{guimera2005functional}. The Z-score compares the connectivity of a node within its own community with that of the other members of the community,
\begin{equation*}
Z(v)=\frac{k_v^{\mathrm{in}}-\mu_{C(v)}}{s_{C(v)}},
\end{equation*}
where $k_v^{\mathrm{in}}$ is the within-community degree of $v$, and $\mu_{C(v)}$ and $s_{C(v)}$ are the mean and standard deviation of the within-community degrees in $C(v)$; we set $Z(v)=0$ if $s_{C(v)}=0$. Because it standardizes connectivity separately within each community, the Z-score is community-aware, and its top-ranked nodes can come from every community, including small ones.
\end{enumerate}

\subsubsection{Semi-local centrality measures}
Semi-local measures use information from the two- or three-hop neighborhood of a node. They aim at a compromise between the low cost of local measures and the structural coverage of global ones.
\begin{enumerate}
\item \textbf{Chen et al.\ (CLSZZ)}~\cite{chen2012identifying}. This measure, called local centrality in~\cite{chen2012identifying} and named here after the initials of its authors, approximates the behavior of global measures at a cost close to that of local ones by aggregating the nearest and next-nearest neighbors of the neighbors of a node. With $|\Gamma_2(w)|$ the number of nodes within distance two of $w$,
\begin{equation*}
Q(u)=\sum_{w\in N(u)}|\Gamma_2(w)|,\qquad CLSZZ(v)=\sum_{u\in N(v)}Q(u).
\end{equation*}

\item \textbf{Local structural centrality (LSC)}~\cite{gao2014ranking}. LSC combines the size of the two-hop neighborhood of each neighbor with the clustering coefficients of that neighbor's neighbors, thus taking into account both the number and the interconnection of the nodes in the neighborhood,
\begin{equation*}
LSC(v)=\sum_{u\in N(v)}\Bigl[\omega\,|\Gamma_2(u)|+(1-\omega)\sum_{w\in N(u)}CC(w)\Bigr],\qquad\omega\in[0,1].
\end{equation*}

\item \textbf{LASPN}~\cite{xiao2024new} (local average shortest path with extended neighborhood). LASPN is a weighted sum of a local, a semi-local, and a structural term,
\begin{equation*}
LASPN(v)=a_1\,I_{\mathrm{loc}}(v)+a_2\,I_{\mathrm{semi}}(v)+a_3\,I_{\mathrm{LRASP}}(v),\qquad a_1,a_2,a_3\ge0 .
\end{equation*}
The local term is the normalized degree $k_v/n$. The semi-local term aggregates the influence of the nodes within a few hops of $v$, weighted by the degrees along the connecting paths and attenuated with distance. The local relative average shortest-path term $I_{\mathrm{LRASP}}(v)$ is obtained from the subgraph formed by $v$ and its extended neighborhood: it compares the average shortest-path length and the size of this subgraph with those obtained after removing $v$, and the larger the change, the more the local structure depends on $v$. Xiao et al.~\cite{xiao2024new} use equal weights $a_1=a_2=a_3$.
\end{enumerate}

\subsubsection{Global centrality measures}
Global measures use the entire network. They capture long-range structure and the role of a node in carrying information across the network, at a higher computational cost.
\begin{enumerate}
\item \textbf{$k$-core}~\cite{seidman1983network}. The $k$-core $G_k$ of a network is the maximal subgraph in which every node has at least $k$ neighbors within the subgraph; it is obtained by repeatedly removing all nodes of degree less than $k$. The core number (or $k$-shell index) $\mathrm{core}(v)=\max\{k\in\mathbb{N}:v\in G_k\}$ is the largest $k$ for which $v$ belongs to $G_k$. Nodes with a high core number lie in densely connected, structurally central regions of the network.

\item \textbf{$k$-shell with gravity (KSG)}~\cite{yang2021improved}. KSG (called KSGC in~\cite{yang2021improved} and labeled ``K-shell'' in the figures) combines the $k$-shell decomposition with a gravity model in which nodes attract each other in proportion to their degrees and inversely to the square of their distance,
\begin{equation*}
KSG(v)=\sum_{u\in\Gamma_r(v)}c_{vu}\,\frac{k_v\,k_u}{d(v,u)^2},\qquad
c_{vu}=\exp\Bigl(\frac{\mathrm{core}(v)-\mathrm{core}(u)}{\mathrm{core}_{\max}-\mathrm{core}_{\min}}\Bigr),
\end{equation*}
where $r$ is the distance cut-off (Table~\ref{tab:settings}) and the attraction coefficient $c_{vu}$, with $\mathrm{core}_{\max}$ and $\mathrm{core}_{\min}$ the largest and smallest core numbers in the network, favors nodes that lie in deeper shells than the nodes around them.

\item \textbf{Katz centrality}~\cite{katz1953new}. Katz centrality counts all walks that end at a node and penalizes longer walks,
\begin{equation*}
Katz(v)=\sum_{p=1}^{\infty}\sum_{u\in V}\alpha^{p}\,(A^{p})_{uv},\qquad 0<\alpha<1/\lambda_1 ,
\end{equation*}
where $(A^p)_{uv}$ is the number of walks of length $p$ from $u$ to $v$ and $\alpha$ is the attenuation factor. Nodes with high Katz centrality can reach many others through short and medium-length walks.

\item \textbf{Betweenness centrality (BC)}~\cite{freeman1977set}. Betweenness counts the shortest paths that pass through a node,
\begin{equation*}
BC(v)=\sum_{s\neq t;\ s,t\neq v}\frac{\sigma_{st}(v)}{\sigma_{st}},
\end{equation*}
where $\sigma_{st}$ is the number of shortest $s$--$t$ paths and $\sigma_{st}(v)$ the number of those passing through $v$; pairs that are not connected contribute zero. In networks with several communities, nodes with high betweenness often act as bridges between them.

\item \textbf{Closeness centrality}~\cite{freeman2002centrality}. Closeness is the reciprocal of the average distance of a node to all other nodes, $CLC(v)=(n-1)/\sum_{u\neq v}d(v,u)$, which requires a connected network. Nodes with high closeness can reach the rest of the network quickly, which gives them a strategic advantage in communication and influence propagation.

\item \textbf{Harmonic centrality}~\cite{opsahl2010node,boldi2014axioms}. Harmonic centrality replaces the reciprocal of the sum of distances by the sum of reciprocal distances, $HC(v)=\sum_{u\neq v}1/d(v,u)$ with $1/d(v,u)=0$ for unreachable $u$, and therefore remains well-defined on disconnected networks.

\item \textbf{Eigenvector centrality}~\cite{bonacich1972factoring}. Eigenvector centrality assigns a high score to nodes connected to other high-scoring nodes, $EC(v)=\lambda_1^{-1}\sum_{u\in N(v)}EC(u)$, i.e.\ the scores are the entries of the principal eigenvector of $A$. Unlike degree, it rewards connections to influential nodes rather than the number of connections.

\item \textbf{PageRank}~\cite{brin1998anatomy}. PageRank models a random walker who, at every step, follows a uniformly chosen edge of the current node with probability $d$ and jumps to a uniformly random node with probability $1-d$,
\begin{equation*}
PR(v)=\frac{1-d}{n}+d\sum_{u\in N(v)}\frac{PR(u)}{k_u},
\end{equation*}
where $d\in(0,1)$ is the damping factor. Originally developed to rank web pages, it ranks individuals in a social network by their links to other well-connected individuals.

\item \textbf{Flow closeness (FC) and flow betweenness (FB)}~\cite{brandes2005centrality}. These are the current-flow (random-walk) counterparts of closeness and betweenness, defined for connected networks, in which the network is viewed as an electrical circuit with unit resistances. With $L$ the graph Laplacian, $L^{+}$ its pseudo-inverse and $\Omega(u,v)=L^{+}_{uu}+L^{+}_{vv}-2L^{+}_{uv}$ the effective resistance between $u$ and $v$,
\begin{equation*}
FC(v)=\frac{n-1}{\sum_{u\neq v}\Omega(v,u)},\qquad
FB(v)=\frac{1}{(n-1)(n-2)}\sum_{s\neq t;\ s,t\neq v}\tau_{st}(v),
\end{equation*}
where $\tau_{st}(v)=\frac12\sum_{u\in N(v)}|\psi^{st}_v-\psi^{st}_u|$ is the current through $v$ when a unit current is injected at $s$ and extracted at $t$, and $\boldsymbol{\psi}^{st}=L^{+}(\mathbf{e}_s-\mathbf{e}_t)$ are the resulting node potentials. FC coincides with information centrality. These measures take all paths into account rather than only the shortest ones, and therefore favor nodes that lie on many redundant routes.

\item \textbf{Bridging centrality}~\cite{hwang2006bridging,rajeh2021investigating}. Bridging centrality is the product of betweenness and the bridging coefficient, which relates the inverse degree of a node to the inverse degrees of its neighbors,
\begin{equation*}
BRC(v)=BC(v)\cdot\frac{1/k_v}{\sum_{u\in N(v)}1/k_u}.
\end{equation*}
It favors nodes that lie on many shortest paths and connect densely connected regions through low-degree positions. In this form, with betweenness computed on the whole network, bridging centrality does not use the partition.
\end{enumerate}

\subsubsection{Community-aware (modular) centrality measures}
Modular centrality measures use the community structure and distinguish intra- from inter-community connections~\cite{ghalmane2019centrality,rajeh2021investigating}. They decompose the importance of a node into its influence within its own community and its ability to bridge communities, which is particularly relevant in homophilic networks, where intra-community edges dominate.
\begin{enumerate}
\item \textbf{Participation coefficient (PC)}~\cite{guimera2005functional,rajeh2021investigating}. Nodes whose links are spread evenly over many communities score highly,
\begin{equation*}
PC(v)=1-\sum_{i=1}^{q}\Bigl(\frac{k_{v,C_i}}{k_v}\Bigr)^2\quad(k_v\ge1),\qquad PC(v)=0\quad(k_v=0).
\end{equation*}

\item \textbf{Community-based mediator (CBM)}~\cite{tulu2018identifying,rajeh2021investigating}. CBM weights a node's degree share by the entropy of its intra- and inter-community connections. With $\pi_v^{\mathrm{in}}=k_v^{\mathrm{in}}/k_v$, $\pi_v^{\mathrm{out}}=k_v^{\mathrm{out}}/k_v$ and $0\log0=0$,
\begin{equation*}
\mathcal{E}_v=-\pi_v^{\mathrm{in}}\log\pi_v^{\mathrm{in}}-\pi_v^{\mathrm{out}}\log\pi_v^{\mathrm{out}},\qquad
CBM(v)=\mathcal{E}_v\,\frac{k_v}{\sum_{u\in V}k_u}.
\end{equation*}
Nodes with many connections both inside and outside their community are ranked higher. For two groups, this is the only possible form; with $q>2$ communities, the inter-community term could also be split over the individual communities.

\item \textbf{Modulus modular betweenness (MMB)}~\cite{ghalmane2019centrality}. The network is split into a \emph{local} network $G^{\ell}$ containing only intra-community edges and a \emph{global} network $G^{g}$ containing only inter-community edges. The local component measures the role of $v$ within its community and the global component its role in connecting communities; the score is the Euclidean norm of both,
\begin{gather*}
\beta_L(v)=\sum_{\substack{s,t\in C(v)\\ s\neq t;\ s,t\neq v}}\frac{\sigma^{\ell}_{st}(v)}{\sigma^{\ell}_{st}},\qquad
\beta_G(v)=\sum_{\substack{s,t\in \mathcal{S}^{g}(v)\\ s\neq t;\ s,t\neq v}}\frac{\sigma^{g}_{st}(v)}{\sigma^{g}_{st}},\\
MMB(v)=\sqrt{\beta_L(v)^2+\beta_G(v)^2},
\end{gather*}
where $\sigma^{\ell}_{st}$ and $\sigma^{g}_{st}$ count shortest paths in $G^{\ell}$ and $G^{g}$, and $\mathcal{S}^{g}(v)$ is the connected component of $G^{g}$ that contains $v$. A node without inter-community edges has $\beta_G(v)=0$.

\item \textbf{Modulus modular closeness (MMC)}~\cite{ghalmane2019centrality}. Analogously, with $d^{\ell}$ and $d^{g}$ the distances in $G^{\ell}$ and $G^{g}$,
\begin{gather*}
\gamma_L(v)=\frac{1}{\sum_{u\in C(v)}d^{\ell}(v,u)},\qquad
\gamma_G(v)=\frac{1}{\sum_{u\in \mathcal{S}^{g}(v)}d^{g}(v,u)},\\
MMC(v)=\sqrt{\gamma_L(v)^2+\gamma_G(v)^2},
\end{gather*}
where the sums run over the nodes reachable from $v$, and a component is set to $0$ if no other node is reachable.
\end{enumerate}
The Z-score, which we list among the local measures because it uses only a node's within-community degree, is also community-aware (Table~\ref{tab:centralities}).

\subsubsection{Fairness-aware centrality measures}
Both fairness-aware measures were proposed by Tsioutsiouliklis et al.~\cite{tsioutsiouliklis2021fairness}. They take a protected group $R$ and a fairness parameter $\varphi\in[0,1]$, the target share of PageRank mass for $R$. Let $P$ be the random-walk transition matrix ($P_{uv}=1/k_u$ for $v\in N(u)$) and $d$ the damping factor.
\begin{enumerate}
\item \textbf{Fairness-Sensitive PageRank (FSPR)} keeps the transition matrix $P$ and replaces the uniform jump vector of PageRank with a fairness-aware jump vector $\mathbf{v}_F$,
\begin{equation*}
\mathbf{p}_F^{\top}=d\,\mathbf{p}_F^{\top}P+(1-d)\,\mathbf{v}_F^{\top},
\end{equation*}
where the probability vector $\mathbf{v}_F$ is chosen to minimise the distance $\|\mathbf{p}_F-\mathbf{p}\|_2$ to the ordinary PageRank vector $\mathbf{p}$, subject to $\sum_{v\in R}p_F(v)=\varphi$; this is a convex optimisation problem. The network structure and the transition probabilities remain unchanged; only the teleportation favors the protected group.

\item \textbf{Locally Fair PageRank (LFPR)} enforces fairness locally, in the neighborhood of every node. In the neighborhood variant LFPR$_N$ of~\cite{tsioutsiouliklis2021fairness}, every node sends a fraction $\varphi$ of its mass to its neighbours in $R$ and $1-\varphi$ to its neighbours in $B$,
\begin{equation*}
\widetilde{P}_{uv}=\begin{cases}\varphi/|N(u)\cap R|, & v\in N(u)\cap R,\\ (1-\varphi)/|N(u)\cap B|, & v\in N(u)\cap B,\\ 0,&\text{otherwise,}\end{cases}\qquad
\mathbf{p}_L^{\top}=d\,\mathbf{p}_L^{\top}\widetilde{P}+(1-d)\,\mathbf{v}_N^{\top},
\end{equation*}
with the $\varphi$-fair jump vector $\mathbf{v}_N$, whose entries are $\varphi/|R|$ for nodes in $R$ and $(1-\varphi)/|B|$ for nodes in $B$. A node $u$ without neighbours in $R$ distributes the share $\varphi$ uniformly over all nodes of $R$, i.e.\ $\widetilde{P}_{uv}=\varphi/|R|$ for $v\in R$, and analogously for $B$. Tsioutsiouliklis et al.~\cite{tsioutsiouliklis2021fairness} also propose residual-based variants (LFPR$_U$, LFPR$_P$, LFPR$_O$).
\end{enumerate}

\begin{table}[!t]
\centering
\caption{Summary of the $25$ centrality measures by category. CA: community-aware (uses the partition $\mathcal{C}$); FA: fairness-aware (uses a protected group and the parameter $\varphi$). The last column gives the marker shape used in all figures.}
\label{tab:centralities}
\footnotesize
\renewcommand{\arraystretch}{1.1}
\begin{tabularx}{\linewidth}{@{}l>{\raggedright\arraybackslash}Xccl@{}}
\toprule
Category & Measure & CA & FA & Marker \\
\midrule
Local (5) & Degree~\cite{freeman2002centrality} & & & star \\
 & Clustering coefficient (CC)~\cite{watts1998collective} & & & star \\
 & $H$-index~\cite{hirsch2005index,lu2016h} & & & star \\
 & $H^2$-index~\cite{lu2016h} & & & star \\
 & Z-score~\cite{guimera2005functional} & \checkmark & & star \\
\midrule
Semi-local (3) & CLSZZ~\cite{chen2012identifying} & & & triangle \\
 & Local structural centrality (LSC)~\cite{gao2014ranking} & & & triangle \\
 & LASPN~\cite{xiao2024new} & & & triangle \\
\midrule
Global (11) & $k$-core~\cite{seidman1983network}, KSG (``K-shell'' in the figures)~\cite{yang2021improved}, Katz~\cite{katz1953new}, betweenness~\cite{freeman1977set}, closeness~\cite{freeman2002centrality}, harmonic~\cite{opsahl2010node,boldi2014axioms}, eigenvector~\cite{bonacich1972factoring}, PageRank~\cite{brin1998anatomy}, flow closeness (FC) and flow betweenness (FB)~\cite{brandes2005centrality}, bridging~\cite{hwang2006bridging,rajeh2021investigating} & & & square \\
\midrule
Global, community-aware (4) & Participation coefficient (PC)~\cite{guimera2005functional}, community-based mediator (CBM)~\cite{tulu2018identifying}, modulus modular betweenness (MMB) and closeness (MMC)~\cite{ghalmane2019centrality} & \checkmark & & diamond \\
\midrule
Global, fairness-aware (2) & Fairness-Sensitive PageRank (FSPR), Locally Fair PageRank (LFPR)~\cite{tsioutsiouliklis2021fairness} & & \checkmark & circle \\
\bottomrule
\end{tabularx}
\end{table}

\subsection{Networks}\label{sec:networks}

We use two synthetic generative models and two real-world networks. Both generative models combine homophily with preferential attachment but produce different structures, which allows us to separate the effect of community-size imbalance from that of clustering and of the number of communities.

\subsubsection{Synthetic networks}

\paragraph{A. Homophily Barab\'asi--Albert (HBA) model} 
The HBA model~\cite{avin2015homophily,karimi2018homophily,lee2019homophily} extends the Barab\'asi--Albert model~\cite{barabasi1999emergence}. Preferential attachment alone produces scale-free networks but ignores the tendency of nodes to connect with similar others, which is widely observed in real social networks. The HBA model incorporates this tendency by making the attachment probability depend on both group membership and degree. A network of $N$ nodes is generated in which a fraction $f$ of randomly chosen nodes form the minority, and the remaining nodes form the majority. Starting from $m$ initial nodes, the network grows one node at a time, and every new node $v$ creates edges to $m$ distinct existing nodes. An existing node $u$ is chosen with probability proportional to $h\cdot k_u$ if $u$ and $v$ belong to the same group and to $(1-h)\,k_u$ otherwise, where $h\in[0,1]$ is the homophily parameter; values $h<0.5$ produce heterophilic and values $h>0.5$ homophilic networks. We use $f=0.2$, i.e.\ a minority of $20\%$, which yields a structurally imbalanced two-community network, and vary $h\in\{0.2,0.4,0.6,0.8\}$.

\paragraph{B. High-Clustering Homophily Barab\'asi--Albert (HICH-BA) model} To study fairness in a more realistic network with several communities and high clustering, we use the HICH-BA model~\cite{saxena2023fairness}. It combines scale-free degrees, community structure, homophily-driven edge creation, and clustering through triadic closure, and it has six parameters: the number of nodes $n$; the node-addition probability $p_N$; the community probabilities $\mathbf{r}=(r_1,\dots,r_q)$, where $r_i$ is the probability that a new node joins community $i$; the homophily parameter $h\in[0,1]$; the triadic-closure probability $p_T$; and the preferential-attachment probability $p_{PA}$. The network starts with one node per community and grows until it has $n$ nodes. Unless stated otherwise, every random choice of a node is preferential (with probability proportional to degree plus one) with probability $p_{PA}$ and uniform otherwise. In each iteration, a new node is added with probability $p_N$; it is assigned to a community according to $\mathbf{r}$ and connected to an existing node of the same community. With probability $1-p_N$, an edge is created instead. With probability $p_T$, this is done by triadic closure: a node $u$ with at least two neighbors is selected, a neighbor $w$ of $u$ is chosen uniformly, and $w$ is connected to another neighbor of $u$ that is not yet adjacent to $w$. Otherwise, an ordinary edge is created: a source node $u\in C_i$ is selected, and the target $w$ is chosen among the nodes not adjacent to $u$, from $C_i$ with probability $h$ and from the other communities with probability $1-h$; an inter-community edge to $w\in C_j$ is accepted only with probability $\min\{1,r_j/r_i\}$, so that edges from larger to smaller communities are thinned. An iteration in which a candidate set is empty is skipped. The model thus offers separate control over community structure, homophily, degree heterogeneity, and clustering. We use $q=4$ communities with $\mathbf{r}=(0.5,0.3,0.15,0.05)$, $p_N=0.4$, $p_T=0.3$ and $p_{PA}=0.8$, and vary $h\in\{0.2,0.4,0.6,0.8\}$.

%\paragraph{Comparability of the two models.} 

%The homophily parameter does not have the same meaning in the two models. In HBA, it re-weights preferential attachment between two groups, whereas in HICH-BA, it is the probability that an ordinary edge-creation step considers the node's own community among four communities. Networks with the same nominal value of $h$ are therefore not structurally equivalent, and we compare the two models through the direction of the trends in $h$ rather than through values at equal $h$.

\subsubsection{Real-world networks}\label{sec:realnets}

We use two real-world networks in which a binary attribute defines a majority and a minority group (Table~\ref{tab:realnets}). Facebook is the combined ego-network collection from the SNAP repository~\cite{snapnets}, with $4{,}039$ nodes and $88{,}234$ edges; $3{,}232$ nodes belong to the majority and $807$ to the minority. DBLP is a co-authorship network~\cite{fairPRrec} with $16{,}501$ nodes and $66{,}613$ edges and groups of $12{,}259$ and $4{,}242$ nodes. In both networks, the groups are derived from the gender attribute of the node metadata. Table~\ref{tab:realnets} also compares the observed fraction of inter-group edges with the value $2f_B(1-f_B)$ expected under random mixing, where $f_B$ is the majority fraction. On Facebook, the two coincide ($0.322$ versus $0.320$), so the partition is essentially non-homophilic; in DBLP, the observed fraction is about $10\%$ lower ($0.343$ versus $0.382$), indicating mild homophily. The two networks also differ strongly in density, with average degrees of $43.69$ and $8.07$.

\begin{table}[!t]
\centering
\caption{Real-world networks. $f_B$: majority fraction; $e_{RB}$: number of inter-group edges; $e_{RB}/|E|$: observed fraction of inter-group edges; $2f_B(1-f_B)$: expected fraction under random mixing.}
\label{tab:realnets}
\footnotesize
\begin{tabular}{@{}lrrrrrrr@{}}
\toprule
Network & $n$ & $|E|$ & $\langle k\rangle$ & $f_B$ & $e_{RB}$ & $e_{RB}/|E|$ & $2f_B(1-f_B)$ \\
\midrule
Facebook & $4{,}039$ & $88{,}234$ & $43.69$ & $0.800$ & $28{,}373$ & $0.322$ & $0.320$ \\
DBLP & $16{,}501$ & $66{,}613$ & $8.07$ & $0.743$ & $22{,}849$ & $0.343$ & $0.382$ \\
\bottomrule
\end{tabular}
\end{table}

\subsubsection{Data preprocessing}\label{sec:preprocessing}
All networks are treated as simple, unweighted, and undirected graphs: edge directions are ignored, and self-loops and multiple edges are removed. Every node carries exactly one group label. In real networks, the label is the binary attribute described above, and the same partition is used for both the fairness metrics and the community partition required by the community-aware measures. In the synthetic networks, the partition is the one produced by the generative model. The fairness-aware measures use the minority group as the protected group $R$.

\subsection{Experimental settings}\label{sec:settings}

Table~\ref{tab:settings} lists all parameters of the study. The intra-community activation probability $\pin=0.1$ is the higher of the two uniform probabilities used by Kempe et al.~\cite{kempe2003maximizing} and the highest level of the trivalency model of Chen et al.~\cite{chen2010scalable}. It is a modeling convention rather than an estimate: methods that learn edge-specific probabilities from action logs~\cite{goyal2010learning} require diffusion traces, which are not available for generative networks or for the static real-world networks used here. The inter-community probability is set to $\pin$, $\pin/2$ and $\pin/4$, motivated by evidence that transmission across community boundaries is weaker than within communities~\cite{gupta2016modeling}; the factors $1/2$ and $1/4$ are our choice, and on the real networks the attenuated variants act as stress tests of fairness under constrained cross-group transmission. At higher probabilities, cascades reach a large fraction of the network for most measures, and the disparities among measures overlap so strongly that a fair seeding strategy can no longer be identified. Section~\ref{sec:sens} examines how the results change for $\pin\in\{0.2,0.3\}$.

\begin{table}[!t]
\centering
\caption{Experimental parameters.}
\label{tab:settings}
\footnotesize
\begin{tabularx}{\linewidth}{@{}l>{\raggedright\arraybackslash}X@{}}
\toprule
Parameter & Value \\
\midrule
HBA & $n=10{,}000$; $m=2$ ($|E|=19{,}996$ edges); minority fraction $f=0.2$; $h\in\{0.2,0.4,0.6,0.8\}$ \\
HICH-BA & $n=10{,}000$; $|E|=21{,}442$ edges; $q=4$; $\mathbf{r}=(0.5,0.3,0.15,0.05)$; $p_N=0.4$; $p_T=0.3$; $p_{PA}=0.8$; $h\in\{0.2,0.4,0.6,0.8\}$ \\
ICM & $\pin=0.1$; $\pout\in\{0.1,0.05,0.025\}$ ($\ICM{1},\ICM{2},\ICM{3}$); sensitivity: $\pin\in\{0.2,0.3\}$ with $\pout/\pin\in\{1,0.5,0.25\}$ \\
LTM & thresholds drawn from $[0,1]$; neighbour weights $1/k_v$ \\
Seed budgets & $k\in\{5,10,50,100\}$ \\
Monte Carlo runs & $T=1{,}000$ (main experiments); $T=1{,}000$ per perturbed network (robustness) \\
Centrality parameters & damping $d=0.6$ (PageRank), $0.6$ (FSPR), $0.5$ (LFPR); fairness parameter $\varphi\in\{0.25,0.5\}$, depending on the network; Katz $\alpha=0.001$; LSC $\omega=0.5$, LASPN weights $(a_1=\frac{1}{3},a_2=\frac{1}{3},a_3=\frac{1}{3})$, KSG $r=0.5\,\hat{d}$ \\
Perturbation & random addition or removal of a fraction $\eta\in\{1\%,2\%,3\%\}$ of the edges; $M=20$ perturbed networks per setting \\
\bottomrule
\end{tabularx}
\end{table}

%% =====================================================================
\section{Experimental Results}\label{sec:results}
%% =====================================================================

Each scatter plot shows one point per centrality measure in the outreach--disparity or outreach--IBR plane; favorable points lie in the bottom-right. Marker shapes encode the class of a measure (Table~\ref{tab:centralities}), and axes are scaled separately for each panel, so the values quoted below are approximate readings. The synthetic networks are reported with $k=100$ (HBA) and $k=50$ (HICH-BA) seeds under the ICM and with $k=5$ seeds under the LTM; the effect of the budget is examined in Section~\ref{sec:res:seed}.

\subsection{Independent Cascade Model on the HBA network}\label{sec:res:hba}

\begin{figure}
    \centering
    \includegraphics[width=\linewidth]{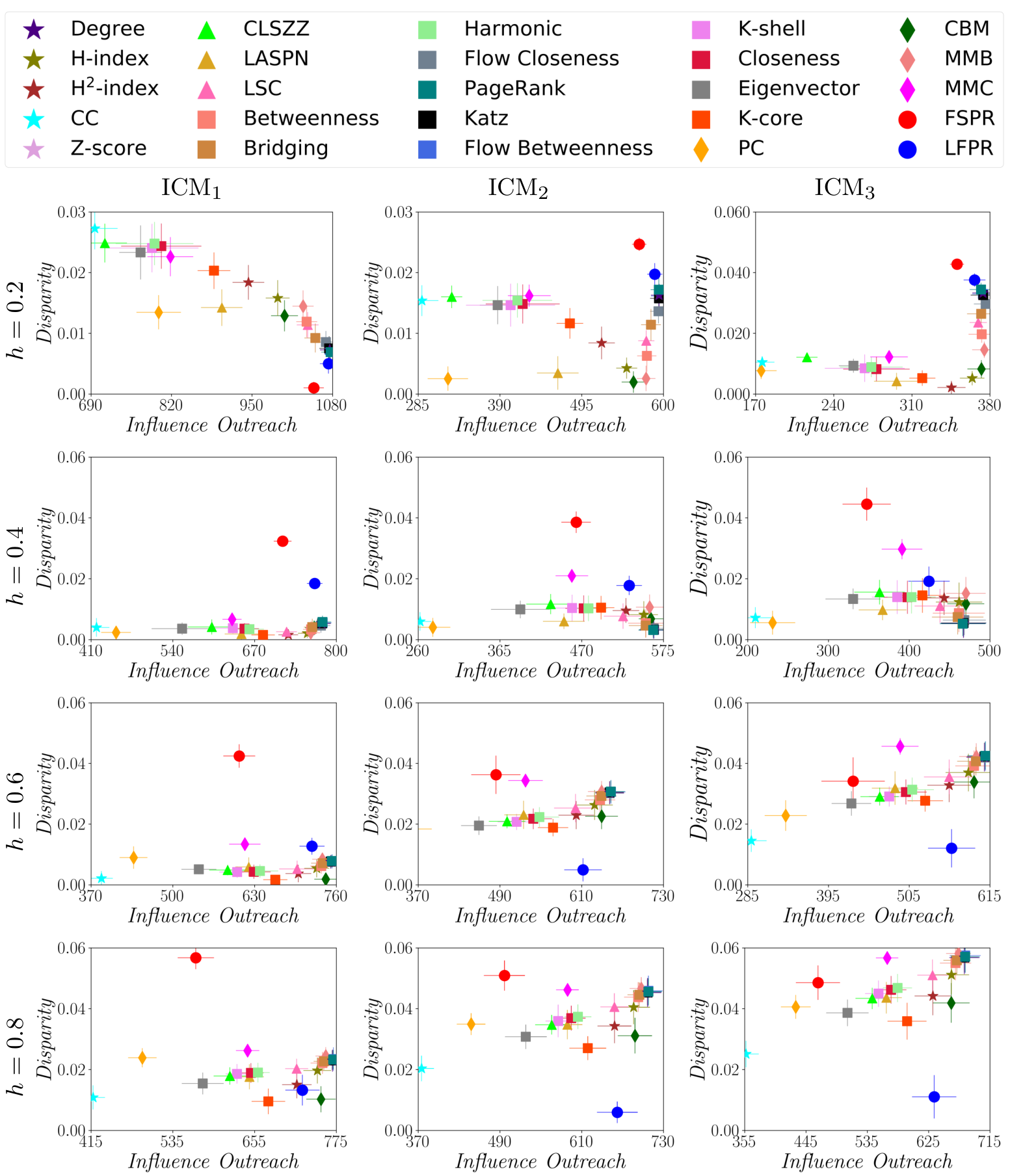}
    \caption{Disparity versus influence outreach on the HBA network ($k=100$, $\pin=0.1$). Rows: homophily $h$; columns: diffusion model. Error bars show one standard deviation.}
\label{fig:hba_disp}
\end{figure}

\begin{figure}
    \centering
    \includegraphics[width=\linewidth]{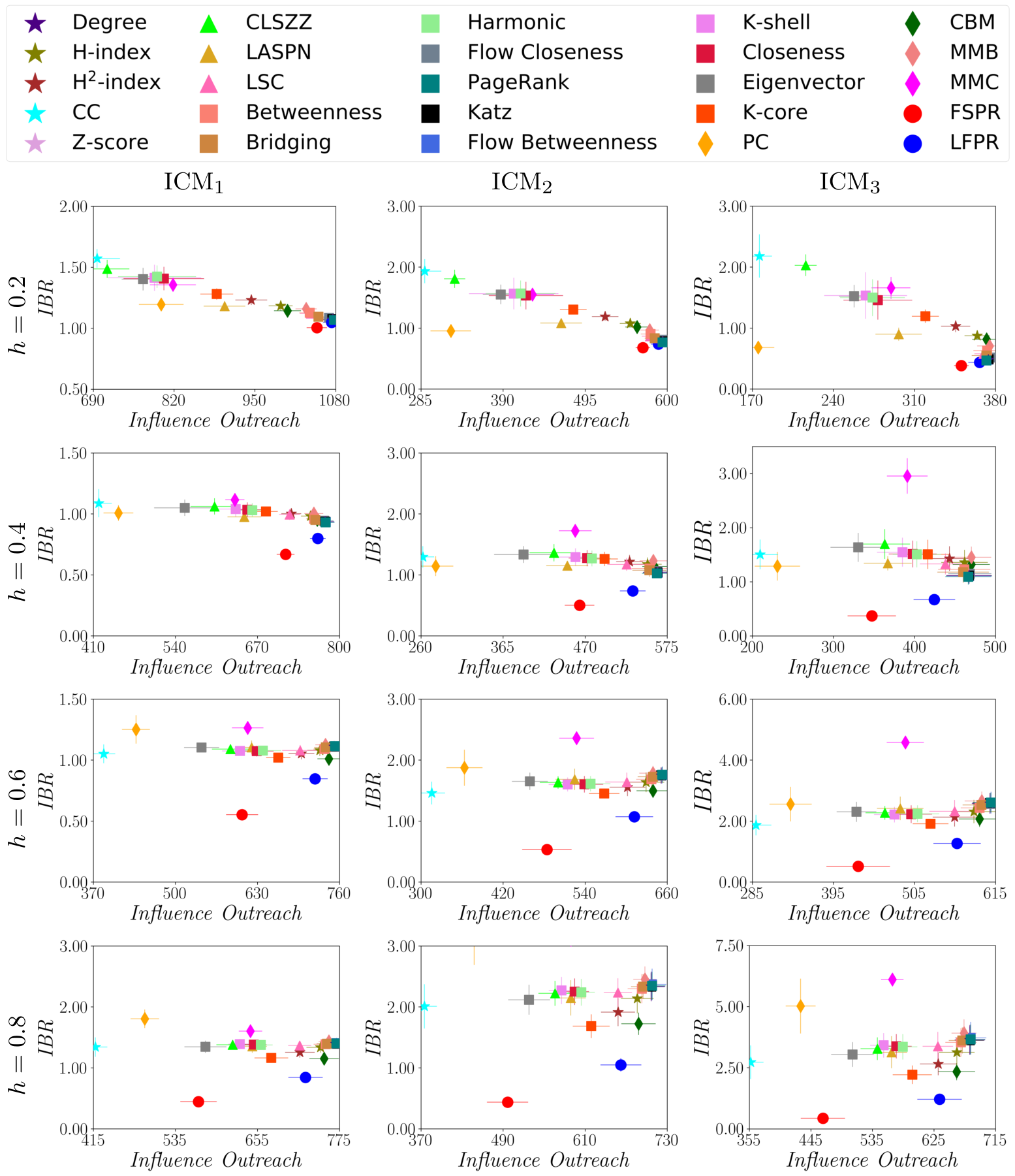}
\caption{Influence balance ratio $\IBR=\rho_B/\rho_R$ versus influence outreach on the HBA network ($k=100$). Values above $1$ favor the majority and values below $1$ the minority.}
\label{fig:hba_ibr}
\end{figure}

%\partitl{Effect of attenuation.} 
Weakening inter-community transmission reduces the outreach of every measure, as the monotonicity of the ICM implies (Section~\ref{sec:theory}), and the reduction is largest in heterophilic networks and smallest in homophilic ones. In homophilic networks, attenuation also raises disparity; at $h=0.8$ the disparity of most measures more than doubles from $\ICM{1}$ to $\ICM{3}$. Since the majority holds most high-centrality seeds, it receives a growing share of the reach, as Lemma~\ref{lem:bounds}(c) predicts in the limiting case in which all seeds lie in one community. The IBR (Fig.~\ref{fig:hba_ibr}) confirms that the gap favors the majority: at $h=0.8$ it grows from below $2$ under $\ICM{1}$ to as much as $6$ under $\ICM{3}$, with MMC and PC at the upper end.

\partitle{\textbf{Heterophilic networks.}} At $h=0.2$, a different pattern emerges. Under uniform transmission, measures with high outreach are also fair, and FSPR and LFPR are close to the best on both criteria, while the clustering coefficient reaches the fewest nodes with the highest disparity. Under attenuated transmission, the high-outreach measures (PageRank, Katz, flow closeness, FSPR and LFPR) reach the minority proportionally \emph{more} than the majority ($\IBR<1$, down to $\approx0.4$ for FSPR under $\ICM{3}$). In heterophilic networks, minority nodes attract many links from the majority and become hubs~\cite{karimi2018homophily}, so hub-seeking measures select many minority seeds (Fig.~\ref{fig:seeddiff}), whose cascades stay in the minority when cross-group transmission is weak. The lowest disparity at near-maximal outreach is obtained by measures with balanced seed sets, in particular CBM and the $H$-index, together with MMB under $\ICM{2}$ and the $H^2$-index under $\ICM{3}$, whose disparity stays below about $0.01$ under attenuated transmission. LASPN and PC also have low disparity, but at clearly lower outreach, so their fairness is partly trivial in the sense of Lemma~\ref{lem:bounds}(a).

\partitle{\textbf{Homophilic networks.}} From $h=0.6$ onward, LFPR separates from all other measures under attenuated transmission. Under $\ICM{2}$, its disparity is several times lower than that of any other measure, at more than $90\%$ of the maximal outreach and with an IBR close to $1$; under $\ICM{3}$ it is again the fairest measure. FSPR behaves very differently: its IBR lies between about $0.4$ and $0.7$ in every configuration except uniform transmission at $h=0.2$, i.e.\ it reaches the minority up to about $2.5$ times as intensively as the majority. FSPR therefore over-corrects, and because disparity is symmetric, this over-correction appears as unfairness; whether it should be penalized is a normative choice that the IBR makes visible. Among the other measures, CBM and $k$-core are comparatively fair under uniform transmission at $h=0.8$, whereas MMC is among the least fair measures in homophilic networks because it selects its seeds only from the majority (Fig.~\ref{fig:seeddiff}); under $\ICM{3}$ it reaches the majority up to six times as intensively as the minority.

\begin{figure}
    \centering
    \includegraphics[width=\linewidth]{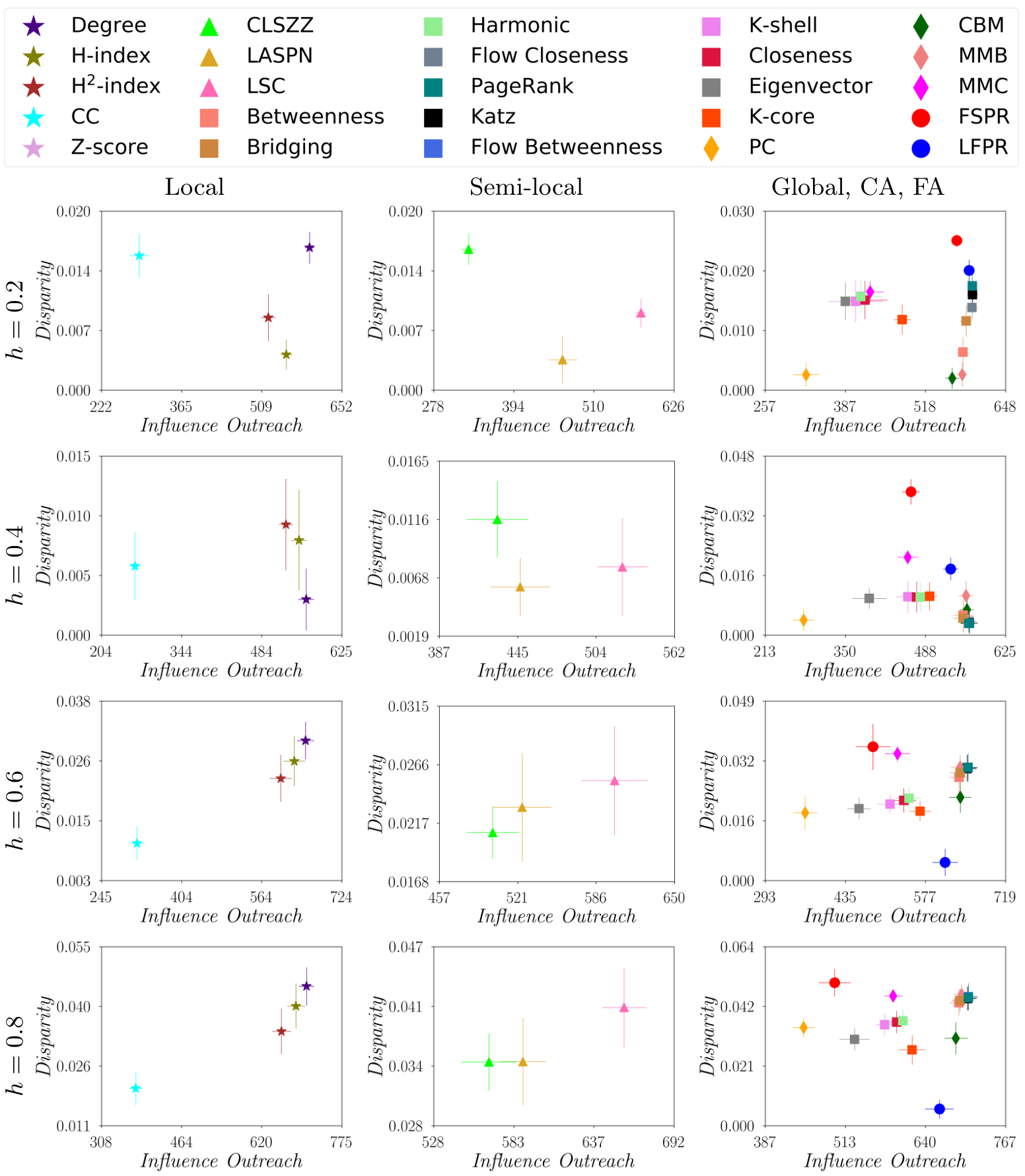}
\caption{Class-wise view of the $\ICM{2}$ column of Fig.~\ref{fig:hba_disp}: local measures (left), semi-local measures (middle), and global (squares), community-aware (CA, diamonds), and fairness-aware (FA, circles) measures (right). Axes are rescaled per panel.}
\label{fig:hba_class}
\end{figure}

\partitle{\textbf{The class of a measure does not determine its fairness.}} The class-wise view in Fig.~\ref{fig:hba_class} shows that the spread of disparity within each class is as large as the spread between classes; at $h=0.8$ the same panel contains both the fairest measure (LFPR) and the least fair one (FSPR). What matters is which positions a measure rewards. Betweenness favors boundary-spanning nodes and is comparatively fair at low homophily. However, path-based measures become biased towards the majority as homophily grows, because dense intra-majority links make the majority the main carrier of shortest paths. Closeness and harmonic centrality penalize the peripheral minority; eigenvector centrality favors the dense majority core at every homophily level, whereas PageRank does so only in homophilic networks; and the $H$-indices and KSG favor dense regions. Class-wise views of the HICH-BA network under $\ICM{2}$ are shown in Section~\ref{sec:res:hich}, and those of the LTM are shown in Appendix~\ref{app:classwise}.

\subsection{Independent Cascade Model on the HICH-BA network}\label{sec:res:hich}

\begin{figure}
    \centering
    \includegraphics[width=\linewidth]{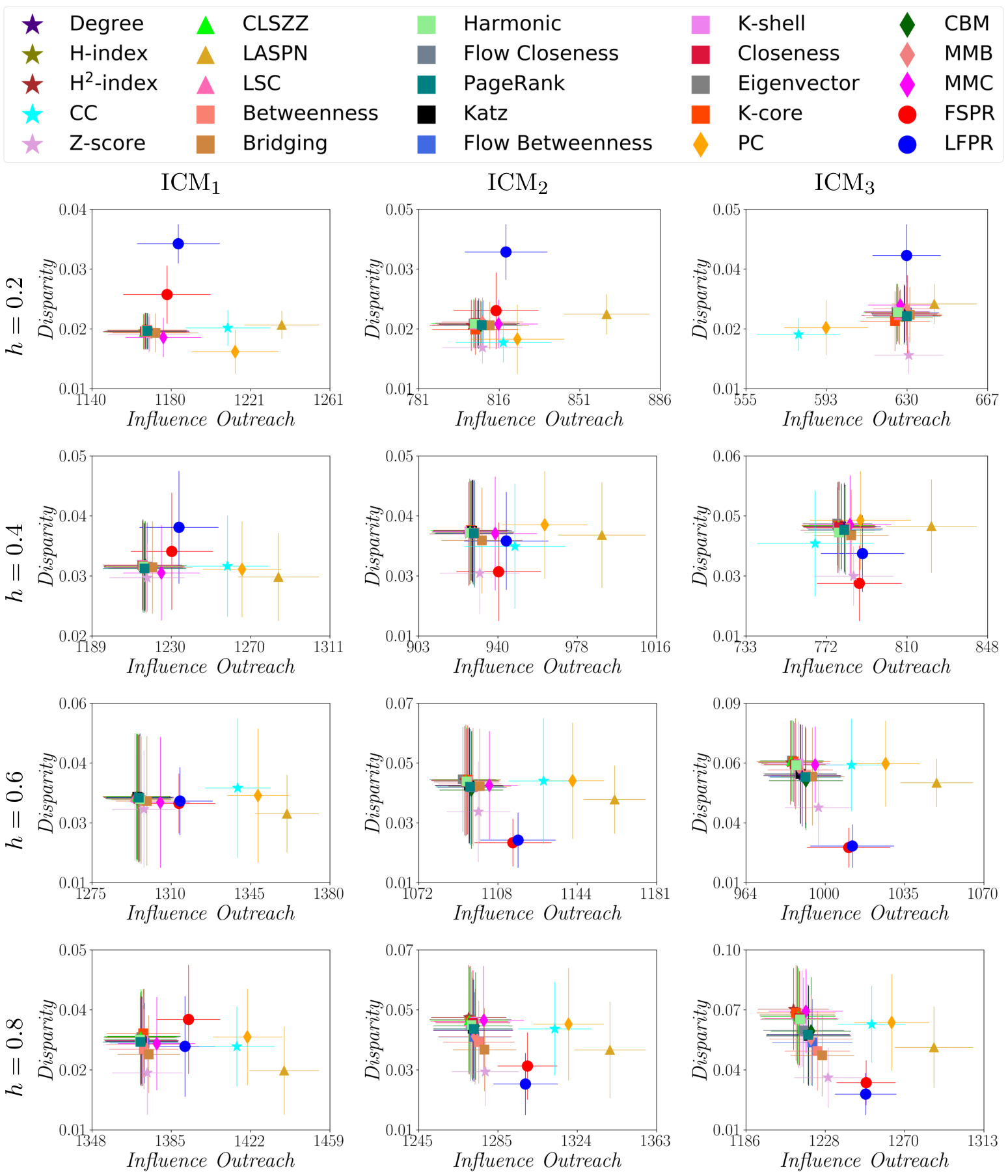}
\caption{Disparity versus influence outreach on the four-community HICH-BA network ($k=50$, $\pin=0.1$).}
\label{fig:hich_disp}
\end{figure}

\begin{figure}
    \centering
    \includegraphics[width=\linewidth]{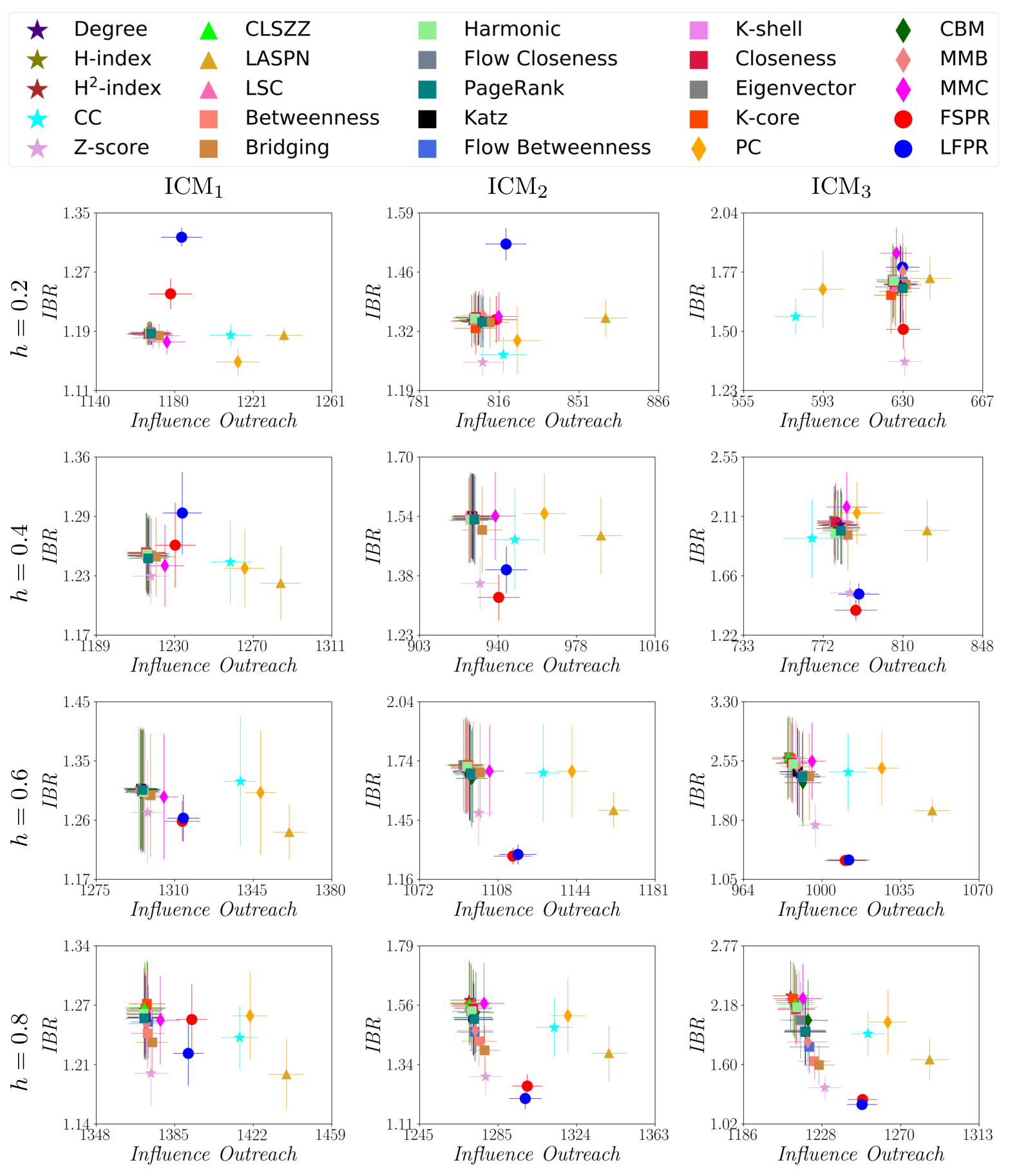}
\caption{IBR $=\max_i\rho_i/\min_i\rho_i$ versus influence outreach on the HICH-BA network ($k=50$).}
\label{fig:hich_ibr}
\end{figure}

In the four-community, highly clustered HICH-BA network (Figs.~\ref{fig:hich_disp} and~\ref{fig:hich_ibr}), outreach \emph{increases} with homophily, because homophilic edge creation and triadic closure produce dense communities in which cascades percolate easily. As in HBA, attenuation costs most outreach at low homophily, and disparity grows with attenuation and, up to $h\approx0.6$, with homophily; the IBR peaks at $h=0.6$ and decreases slightly at $h=0.8$. Most local and global measures presumably select largely the same hubs in the core of the largest community and form a tight cluster with overlapping error bars, so that in a highly clustered multi-community network the choice among conventional measures matters much less than in the two-community model.

Three groups of measures separate from the cluster. LASPN attains the highest outreach in every configuration, and PC and the clustering coefficient usually also exceed the cluster (except under $\ICM{3}$ at low homophily). In clustered networks, hub seeds are redundant, and measures that avoid the core spread their seeds more widely. LASPN, however, is not fairer than the cluster. At $h=0.2$, the lowest disparity is obtained by PC under $\ICM{1}$ and, under attenuated transmission, by the Z-score, which draws seeds from all four communities. The fairness-aware measures switch sides relative to homophily. At $h=0.2$, LFPR has the highest disparity of all measures. From $h=0.6$ onwards, under $\ICM{2}$ and $\ICM{3}$, LFPR and FSPR have the lowest or near-lowest disparity, well below that of the cluster, together with slightly higher outreach and the lowest IBR, so that they dominate the cluster on both criteria.

\begin{figure}
    \centering
    \includegraphics[width=\linewidth]{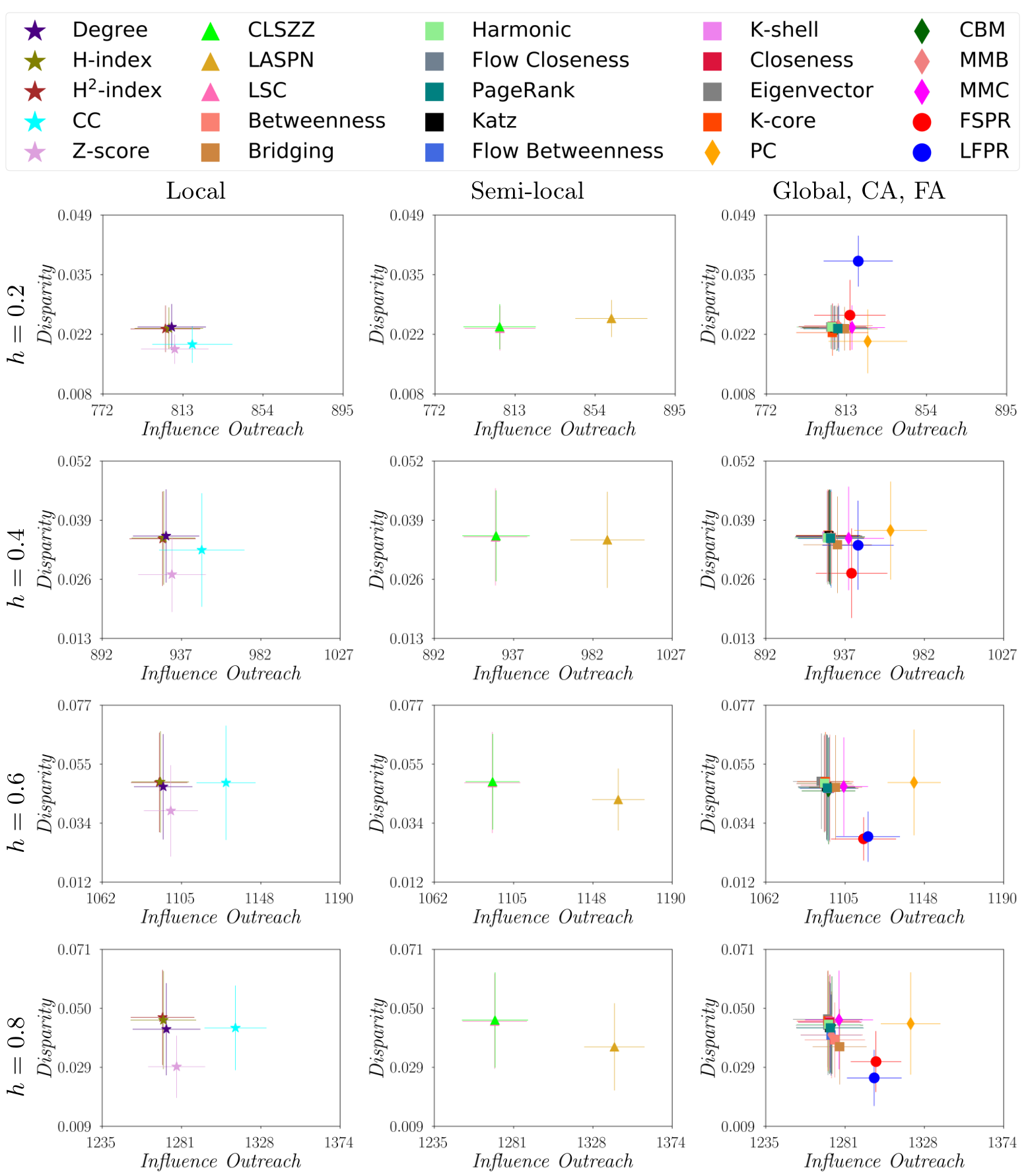}
\caption{Class-wise view of the $\ICM{2}$ results on the HICH-BA network ($k=50$).}
\label{fig:cls_hich_icm2}
\end{figure}

The class-wise view of the $\ICM{2}$ results in Fig.~\ref{fig:cls_hich_icm2} makes these patterns visible. Among the local measures, the Z-score has the lowest disparity at both homophily levels (at $h=0.2$ only marginally below the clustering coefficient), and the clustering coefficient reaches more nodes than degree and the $H$-indices. Among the semi-local measures, LASPN clearly reaches more nodes than CLSZZ and LSC, which are almost identical. In the panel of global measures, most points overlap in a tight cluster, from which three measures separate: at $h=0.2$, LFPR lies far above the cluster and PC slightly below it, whereas at $h=0.8$, LFPR and FSPR lie clearly below the cluster at slightly higher outreach, and PC reaches the most nodes at cluster-level disparity.

\subsection{Linear Threshold Model on the synthetic networks}\label{sec:res:ltm}

\begin{figure}
    \centering
    \includegraphics[width=\linewidth]{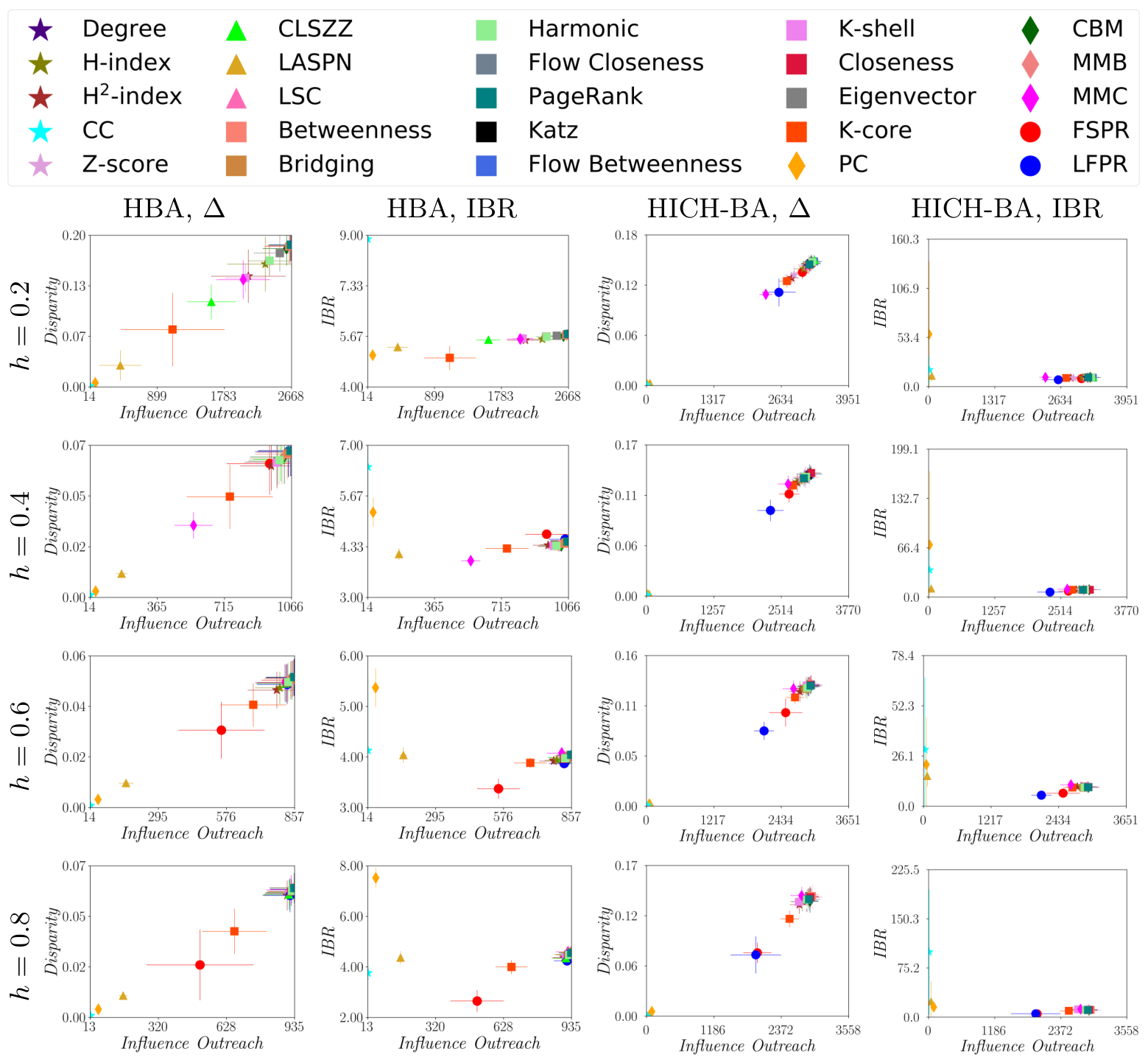}
\caption{Disparity and IBR versus influence outreach under the LTM ($k=5$) on the HBA (left two columns) and HICH-BA (right two columns) networks.}
\label{fig:ltm}
\end{figure}

Under the LTM (Fig.~\ref{fig:ltm}), the clustering coefficient, PC, and LASPN select nodes that cannot trigger the reinforcement required for activation and reach at most a few hundred nodes; their low disparity is trivial fairness. The other measures trigger cascades of intermediate to large size: in HBA at $h=0.2$ and in HICH-BA at every homophily level, five seeds reach more than twice as many nodes as $100$ and $50$ seeds, respectively, under $\ICM{1}$. Once triggered, a cascade spreads through the densely connected region of the seeds, so disparity grows with outreach. Disparity is also higher than under the ICM, markedly so in heterophilic HBA networks (up to $\approx0.19$ at $h=0.2$) and in HICH-BA, where the best-served community is reached about ten times as intensively as the least-served one; in homophilic HBA networks it is comparable to that under $\ICM{3}$. In HBA, both the largest outreach and the corresponding disparity drop sharply from $h=0.2$ to $h=0.4$ and change little thereafter, presumably because cascades no longer percolate through the whole network; in HICH-BA, outreach decreases only slightly with homophily, and the cluster's disparity shows no clear trend.

The fairness-aware measures again behave differently from each other. In homophilic HBA networks, FSPR offers the most favorable trade-off: at $h=0.8$ its disparity is less than half that of the high-outreach cluster, at the price of about $45\%$ less outreach, whereas LFPR behaves like the other hub-based measures. Because FSPR places most of its seeds in the minority, it starts cascades inside the minority itself, which majority cascades rarely reach when activation requires reinforcement from several neighbors. In HICH-BA, LFPR has the lowest disparity among the cascade-triggering measures for $h\ge0.4$ (at $h=0.2$ it is tied with MMC), and at $h=0.8$ LFPR and FSPR both halve the disparity and the IBR of the cluster while reaching about two-thirds of its outreach. $k$-core lies between the fairness-aware measures and the cluster.

\subsection{Effect of homophily and network structure}\label{sec:res:structure}

Homophily affects the disparity differently in the two models. In HBA, it is non-monotone in $h$: it is lowest near neutral mixing ($h=0.4$) and increases in both directions because hub-seeking measures over-serve the minority under heterophily ($\IBR<1$) and the majority under homophily ($\IBR>1$). In HICH-BA, where new nodes always attach within their own community and therefore, heterophily does not turn members of small communities into hubs, the disparity increases with $h$ and levels off for $h\ge0.6$. Structural isolation alone does not determine which measures are fair: in the two-community model the group composition of the seed set is decisive (Section~\ref{sec:res:sd}), whereas in the clustered multi-community model the redundancy of hub seeds in a dense core determines which measures separate from the majority-biased cluster.

\subsection{Impact of the seed budget}\label{sec:res:seed}

\begin{figure}
    \centering
    \includegraphics[width=\linewidth]{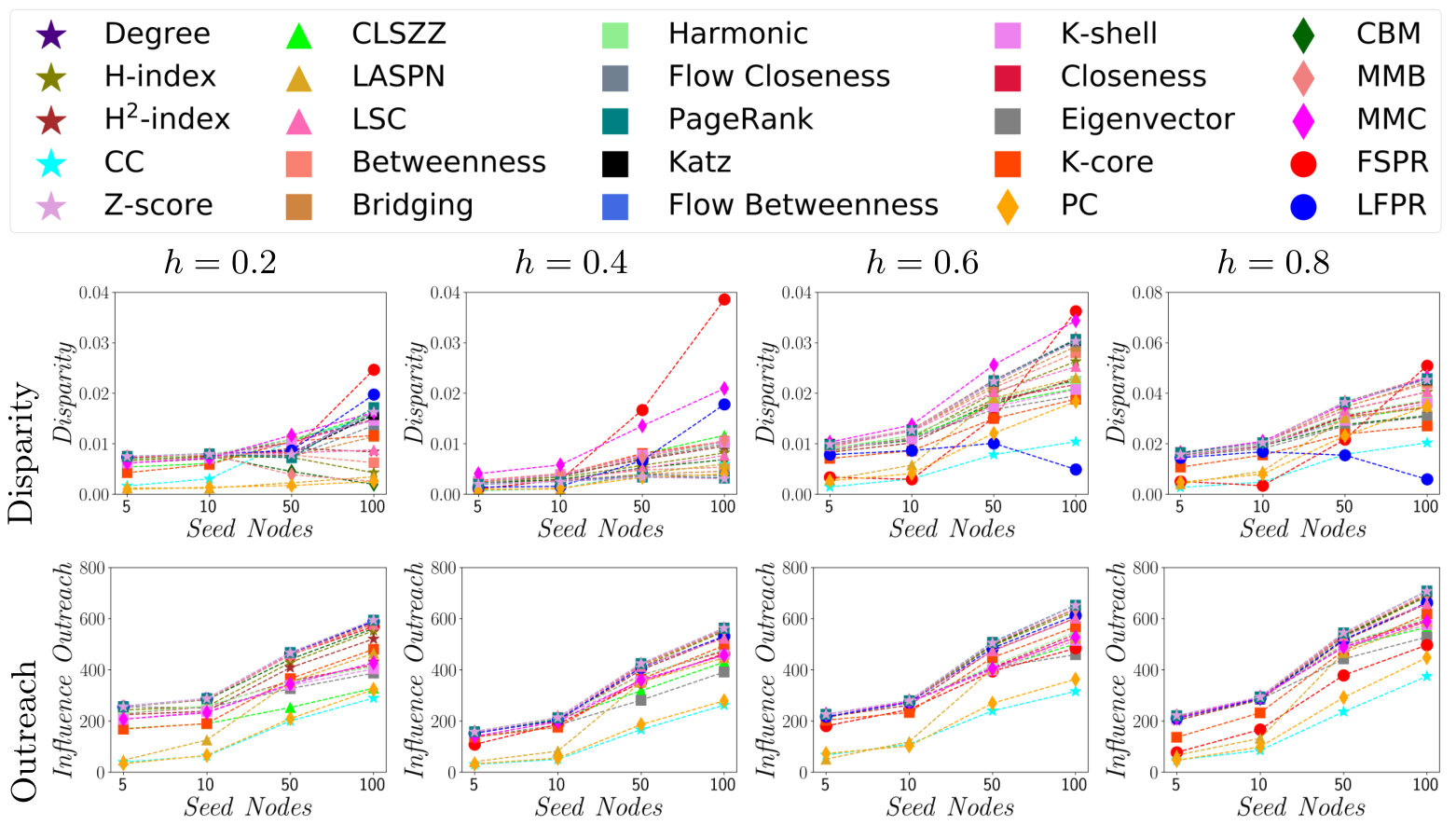}
\caption{Disparity (top) and influence outreach (bottom) versus the seed budget $k\in\{5,10,50,100\}$ on the HBA network under $\ICM{2}$.}
\label{fig:seed_hba}
\end{figure}

\begin{figure}
    \centering
    \includegraphics[width=\linewidth]{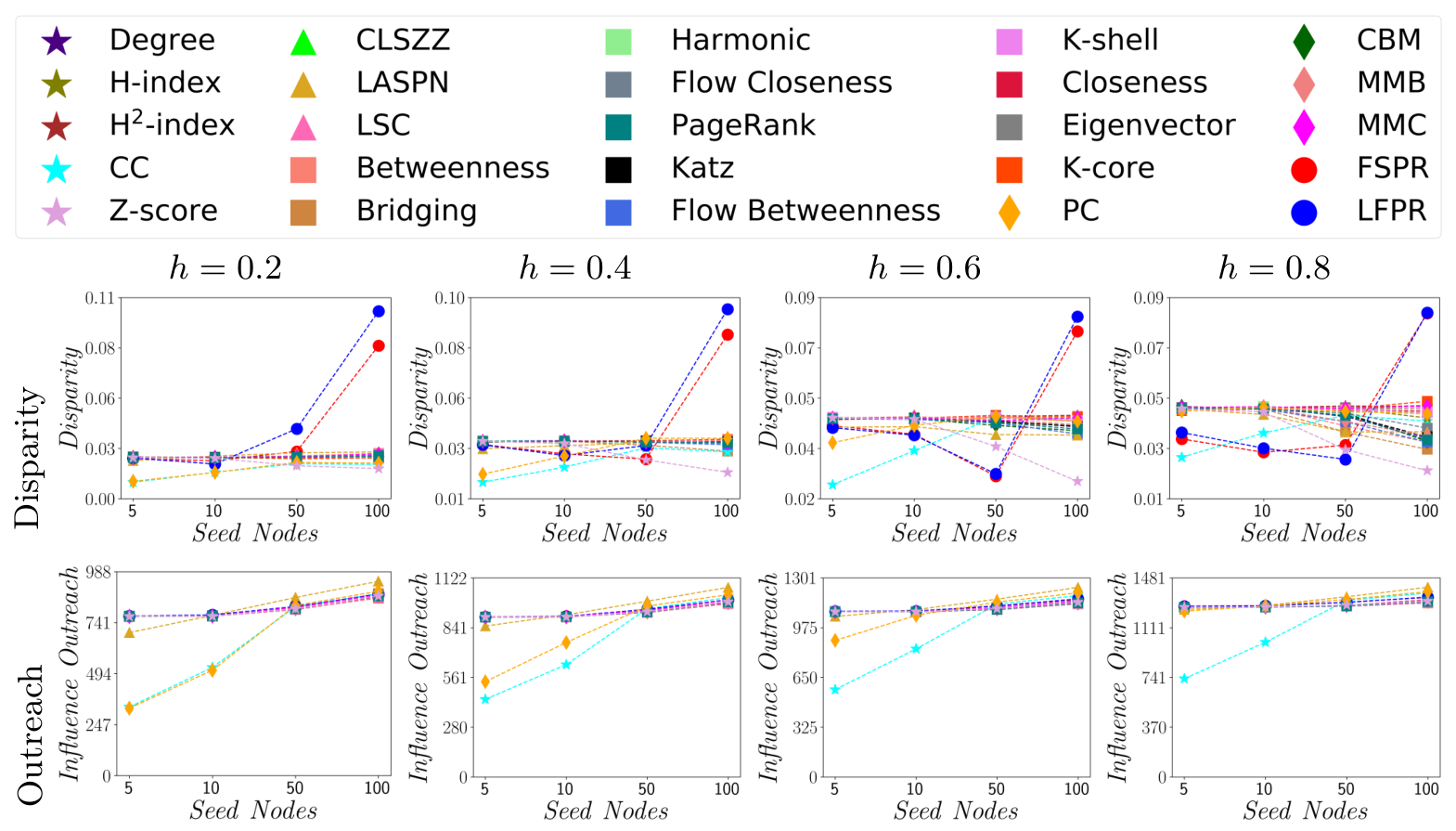}
\caption{Disparity (top) and influence outreach (bottom) versus the seed budget on the HICH-BA network under $\ICM{2}$.}
\label{fig:seed_hich}
\end{figure}

\partitle{\textbf{Cascade diffusion.}} Under $\ICM{2}$ on HBA (Fig.~\ref{fig:seed_hba}), outreach grows with the budget, and disparity is small and similar across measures for small budgets but diverges for larger ones. In homophilic networks, disparity increases with $k$ for almost all measures, because additional seeds come mostly from the majority. LFPR is the exception: its disparity drops sharply at $k=100$, where its seed set is roughly balanced. FSPR shows the steepest increase at every level of homophily, consistent with a growing over-allocation of seeds to the minority. For $k\le10$, the lowest disparity is generally achieved by the clustering coefficient, PC, and LASPN, all of which also have the lowest outreach; from $h=0.4$ onward, FSPR is also among the fairest measures at $k=5$ and $k=10$, with reduced outreach. On HICH-BA (Fig.~\ref{fig:seed_hich}), the outreach of almost all measures converges for $k\ge50$, so the choice of measure matters less as the budget grows. The fairness-aware measures are fairer than the cluster up to $k=50$ in homophilic networks but become the least fair at $k=100$, indicating that their advantage depends on the budget; the Z-score, in contrast, becomes fairer as $k$ grows.

\begin{figure}
    \centering
    \includegraphics[width=\linewidth]{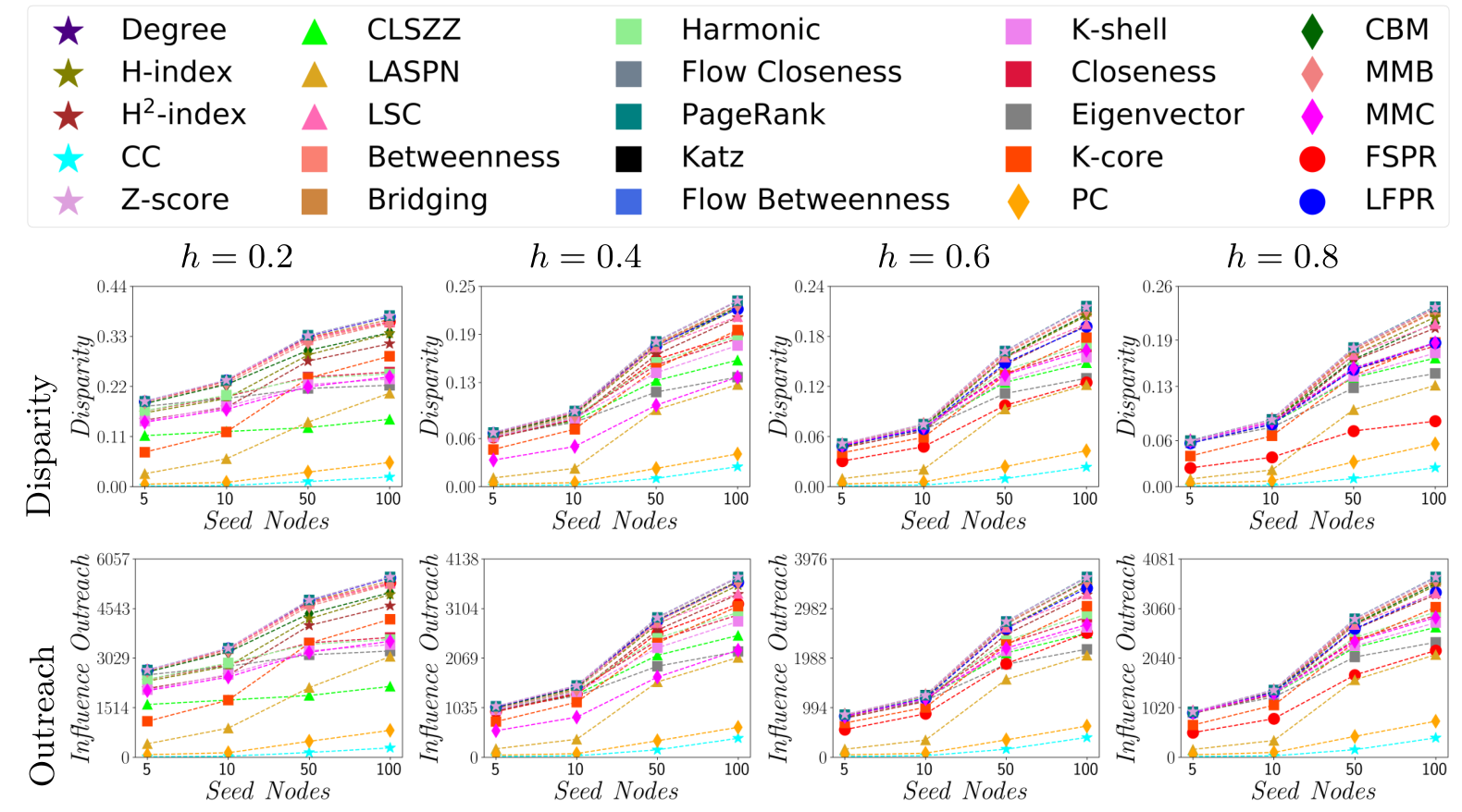}
\caption{Disparity (top) and influence outreach (bottom) versus the seed budget on the HBA network under the LTM.}
\label{fig:seed_ltm_hba}
\end{figure}

\begin{figure}
    \centering
    \includegraphics[width=\linewidth]{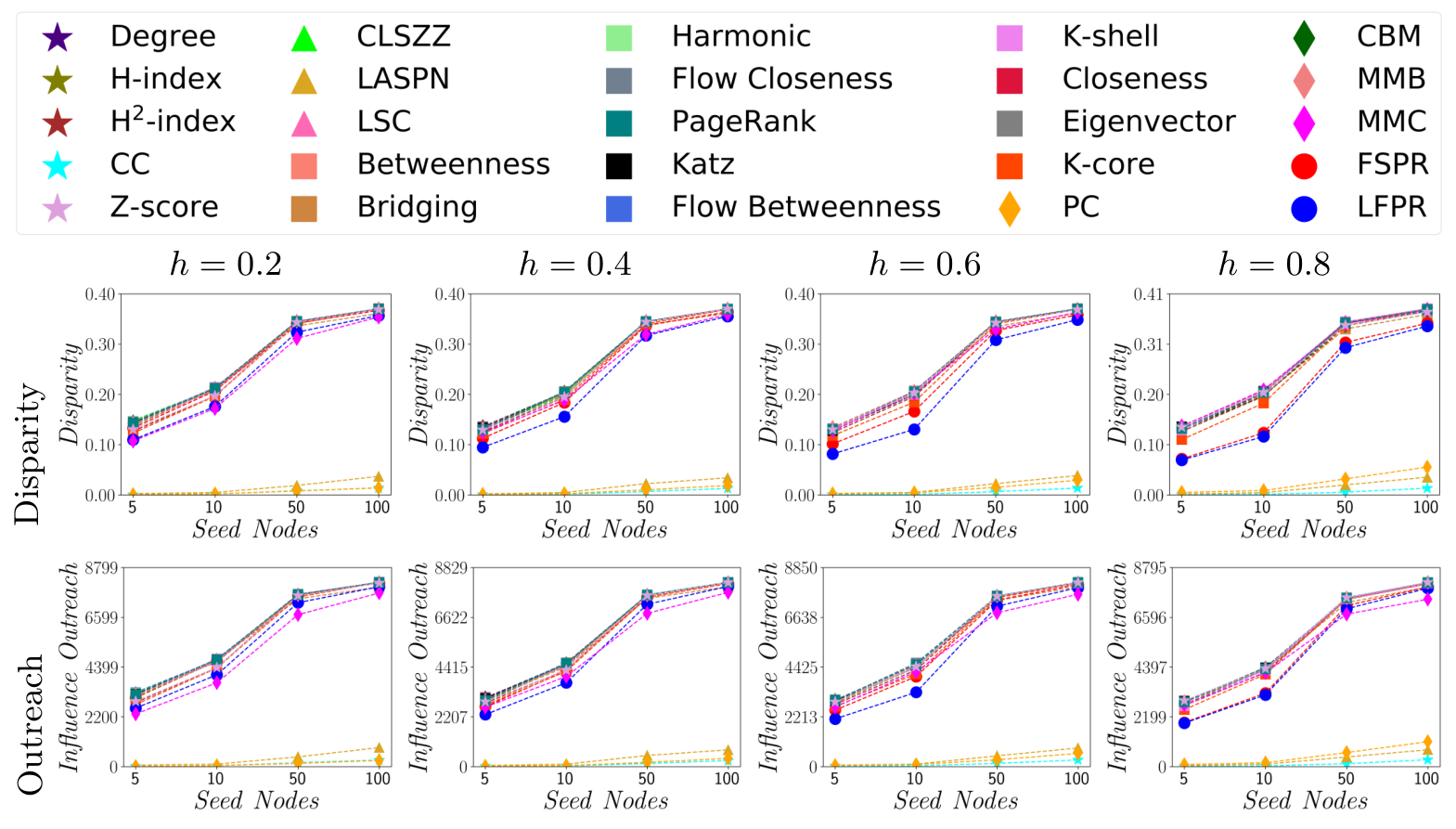}
\caption{Disparity (top) and influence outreach (bottom) versus the seed budget on the HICH-BA network under the LTM.}
\label{fig:seed_ltm_hich}
\end{figure}

\partitle{\textbf{Threshold diffusion.}} Under the LTM (Figs.~\ref{fig:seed_ltm_hba} and~\ref{fig:seed_ltm_hich}) outreach and disparity both increase with the budget for all cascade-triggering measures; larger budgets push the majority closer to saturation while the minority lags behind. In HBA, disparity at $h=0.2$ roughly doubles from $k=5$ to $k=100$; in HICH-BA it more than doubles at every homophily level, mostly between $k=10$ and $k=50$. In homophilic networks, the fairness-aware measures remain among the fairest triggering measures across all budgets (FSPR in HBA, LFPR and FSPR in HICH-BA). In HICH-BA, their advantage is largest at small budgets, where it also costs the most outreach, and almost vanishes at $k=100$. The positions of other measures change with the budget; for example, in HBA $k$-core is among the fairest triggering measures at $k=5$ but only mid-ranked at $k=100$.

\subsection{Fairness in seed selection}\label{sec:res:sd}

\begin{figure}
    \centering
    \includegraphics[width=\linewidth]{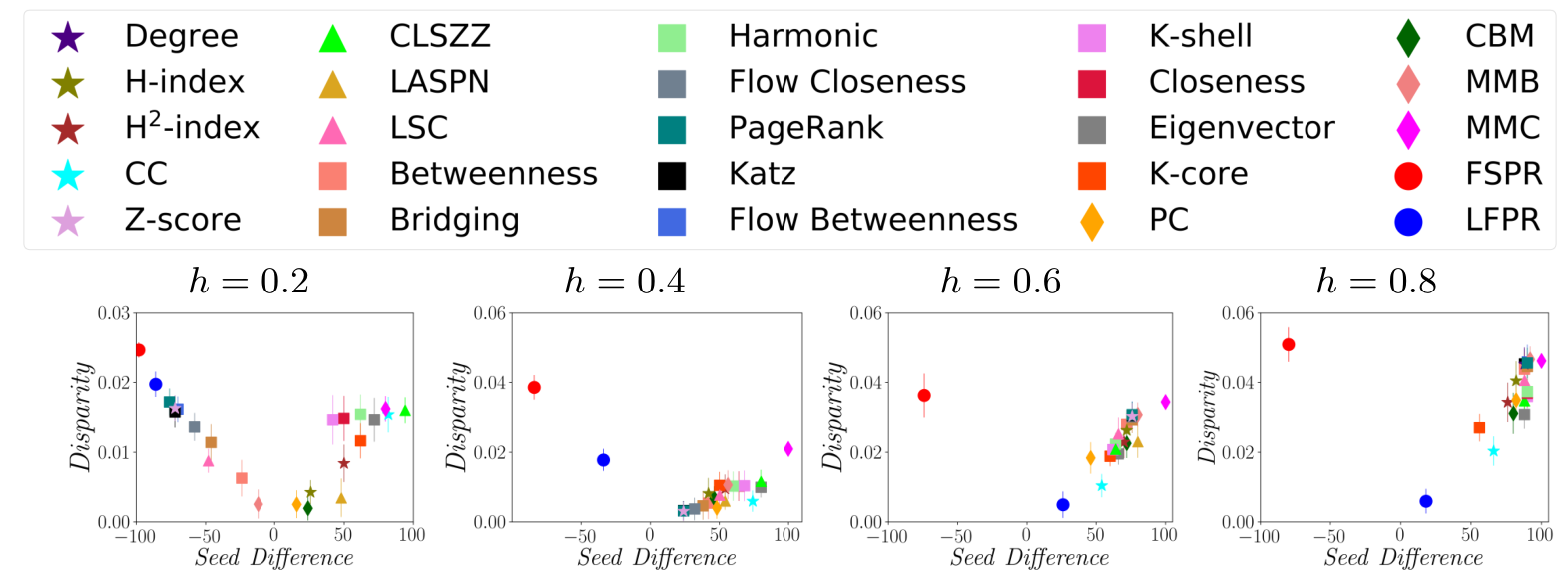}
\caption{Disparity versus seed difference $\mathrm{SD}=|S\cap B|-|S\cap R|$ on the HBA network under $\ICM{2}$ ($k=100$). Proportional representation corresponds to $\mathrm{SD}^\star=60$.}
\label{fig:seeddiff}
\end{figure}

Fig.~\ref{fig:seeddiff} relates the group composition of the seed set to disparity on HBA. At $h=0.2$, the relation is V-shaped: disparity is lowest for nearly balanced seed sets (MMB, PC and CBM, followed by the $H$-index) and grows in both directions, for minority-heavy seed sets (FSPR, LFPR, PageRank and Katz) and for majority-heavy ones (eigenvector centrality, MMC, CLSZZ). At $h=0.4$, most measures select seed sets with a clear majority surplus yet maintain low disparity, whereas FSPR and, to a lesser extent, LFPR select more minority than majority seeds. In homophilic networks ($h\ge0.6$), almost all measures select majority-dominated seed sets, and disparity increases roughly linearly with the seed difference. MMC selects only majority seeds; FSPR remains strongly minority-biased with high disparity, confirming the over-correction observed in Section~\ref{sec:res:hba}; and LFPR is the only measure with a roughly balanced seed set and attains the lowest disparity. Disparity in the two-community model is therefore largely a question of seed allocation. The disparity-minimizing allocation is, however, not proportional: in all four settings the lowest mean disparity is attained when the $20\%$ minority receives about $37$--$41\%$ of the seeds. Cascades started from majority seeds spill over into the minority less than the reverse, so the minority needs proportionally more seeds to obtain proportionally equal reach.

\subsection{Real-world networks}\label{sec:res:real}

\begin{figure}
    \centering
    \includegraphics[width=\linewidth]{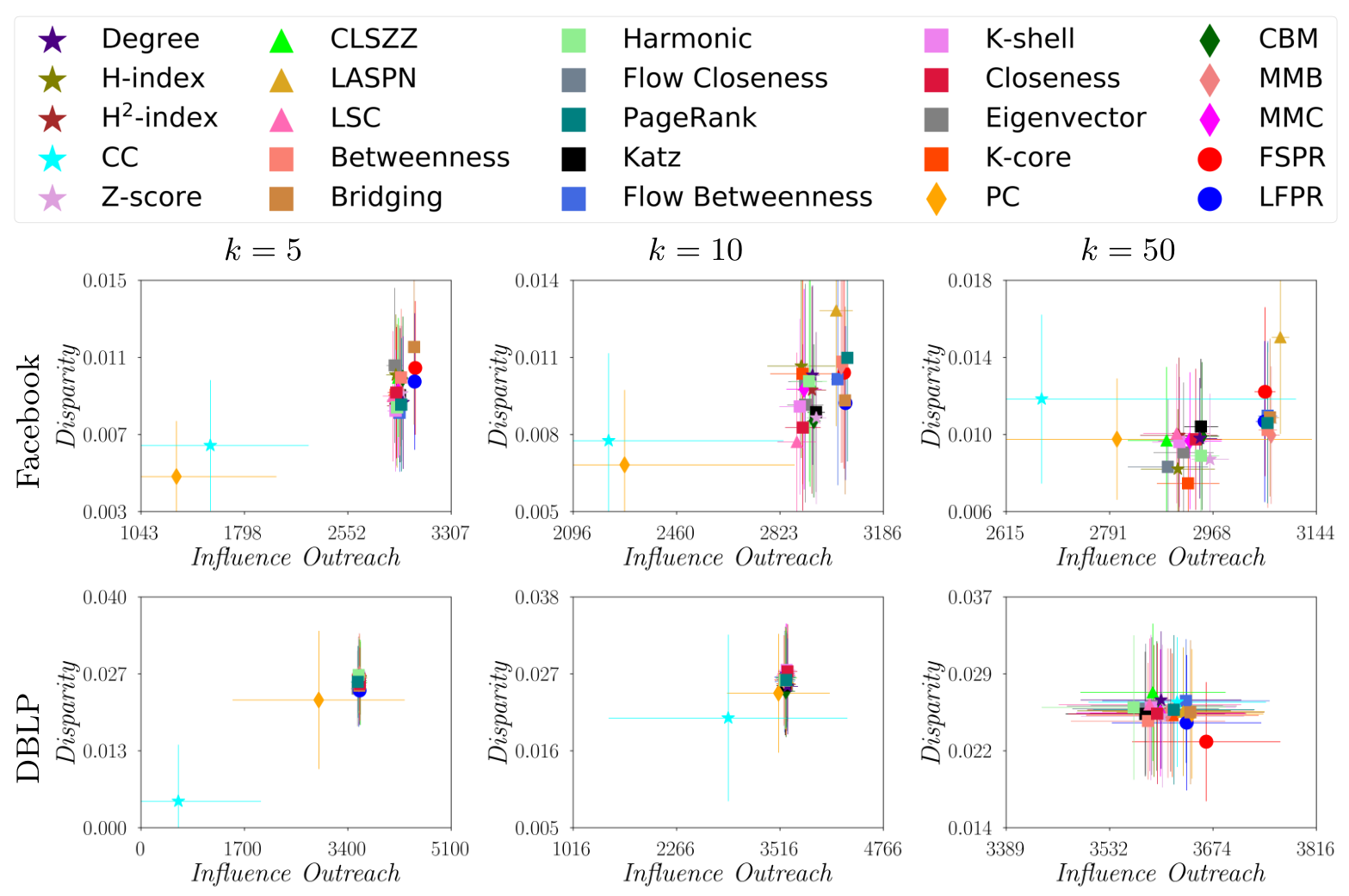}
\caption{Disparity versus influence outreach on the Facebook and DBLP networks under $\ICM{1}$.}
\label{fig:real_icm1}
\end{figure}

\begin{figure}
    \centering
    \includegraphics[width=\linewidth]{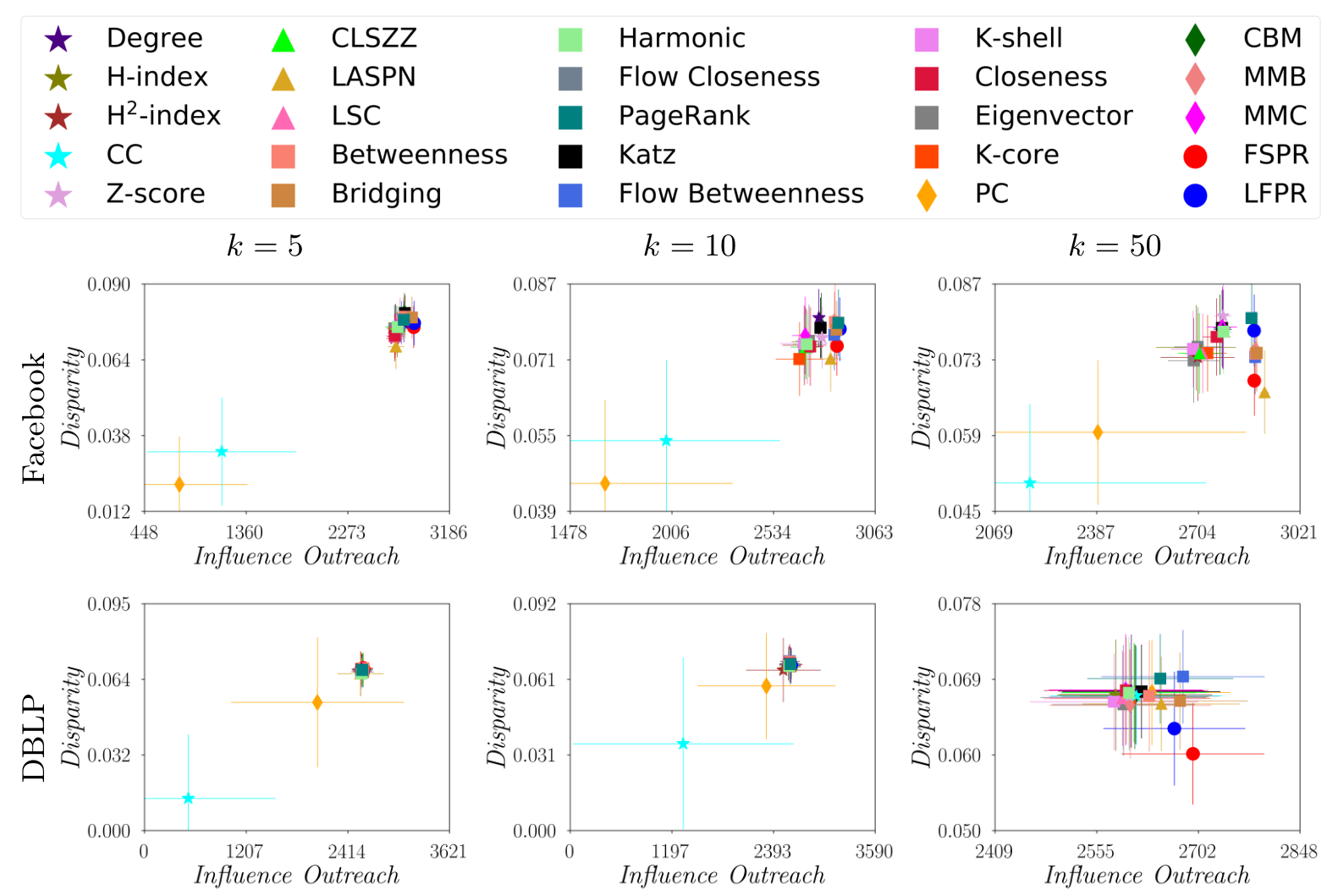}
\caption{Disparity versus influence outreach on the Facebook and DBLP networks under $\ICM{2}$.}
\label{fig:real_icm2}
\end{figure}

\begin{figure}
    \centering
    \includegraphics[width=\linewidth]{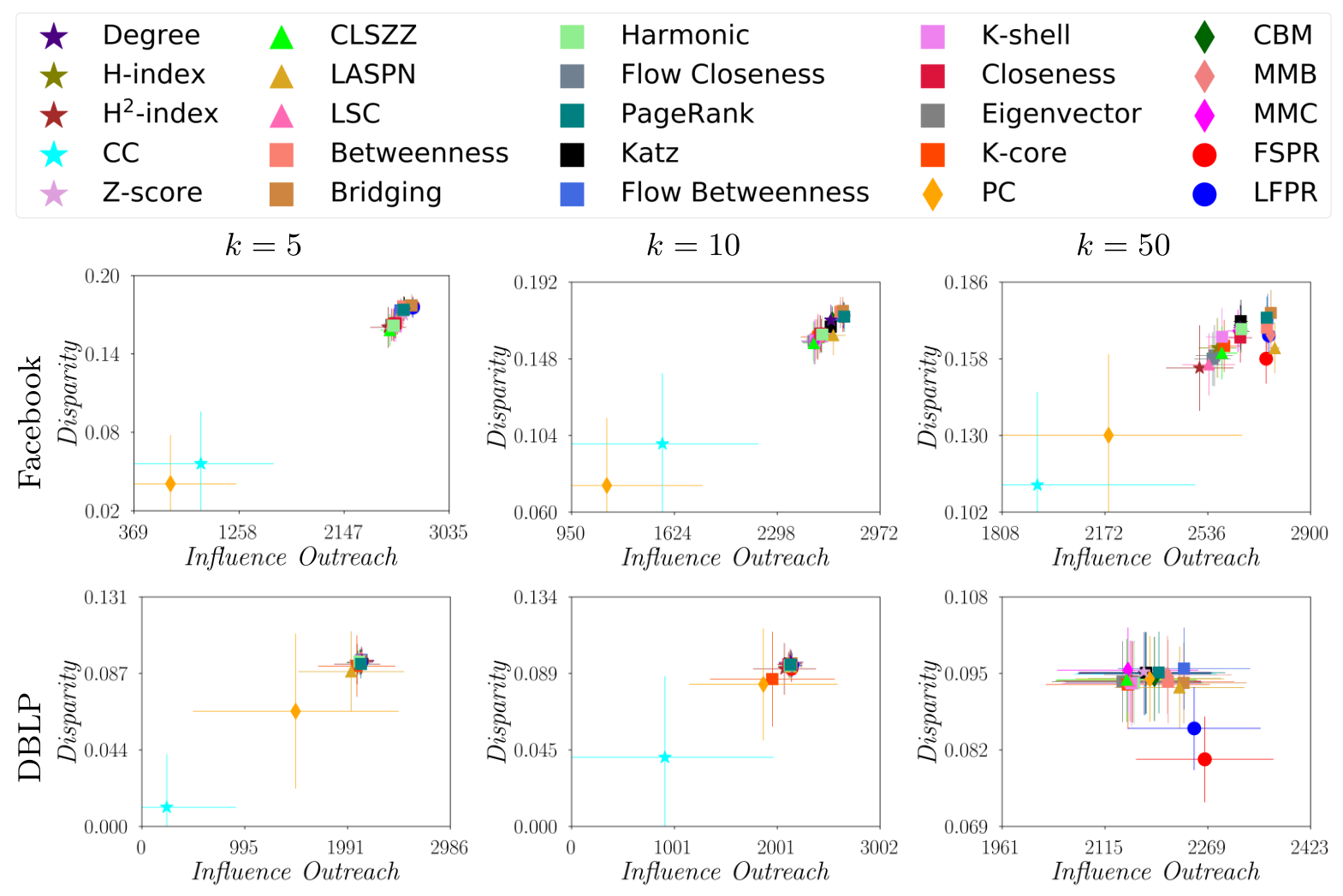}
\caption{Disparity versus influence outreach on the Facebook and DBLP networks under $\ICM{3}$.}
\label{fig:real_icm3}
\end{figure}

\begin{figure}
    \centering
    \includegraphics[width=\linewidth]{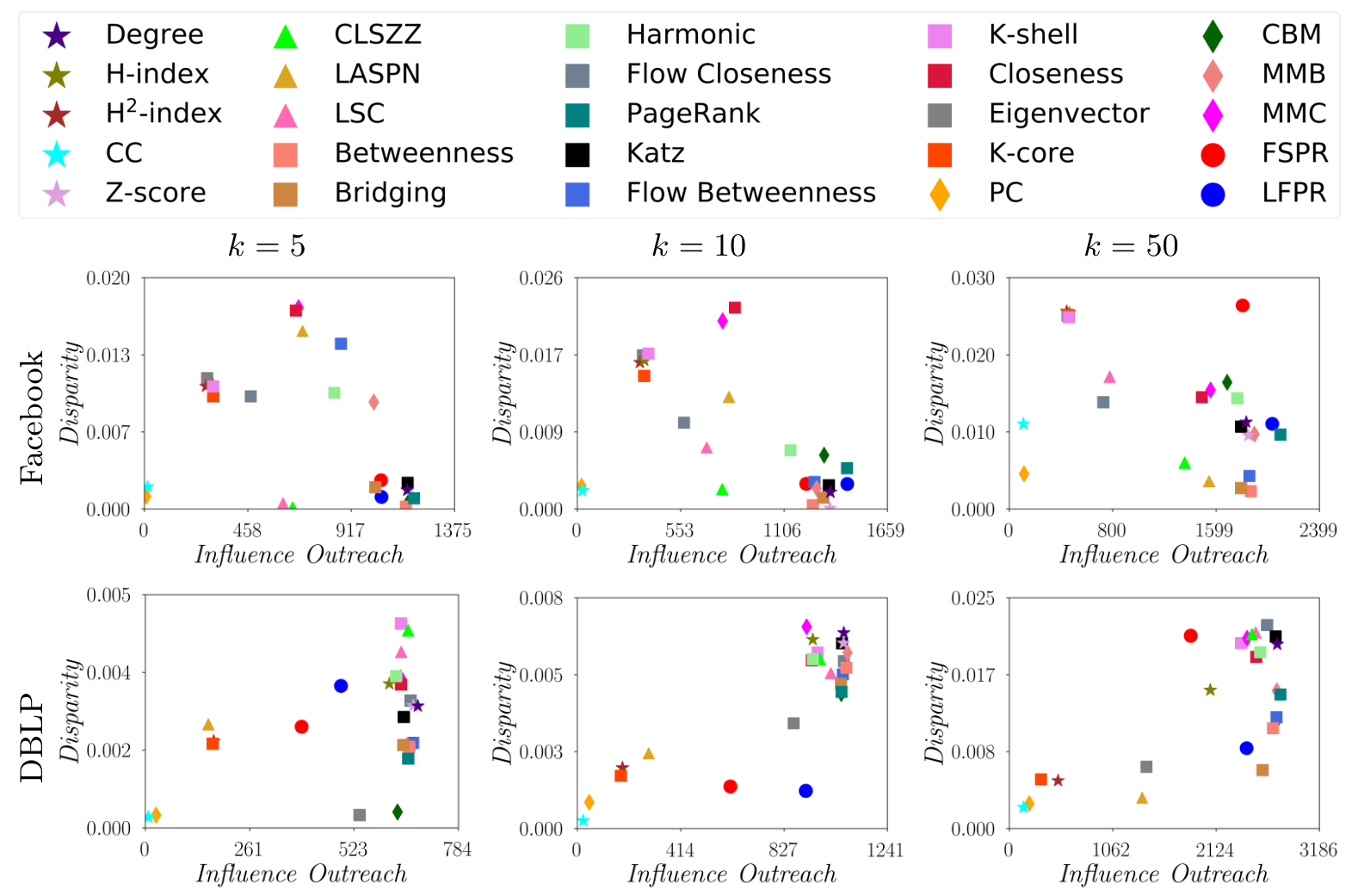}
\caption{Disparity versus influence outreach on the Facebook and DBLP networks under the LTM.}
\label{fig:real_ltm}
\end{figure}

Figs.~\ref{fig:real_icm1}--\ref{fig:real_ltm} report the Facebook and DBLP networks for $k\in\{5,10,50\}$; results for $k=100$ (not shown) are very similar to those for $k=50$, and the IBR plots mirror the disparity plots and are omitted. Facebook, in particular, is much denser than the synthetic networks.

\partitle{\textbf{ICM diffusion model.}} Under uniform transmission ($\ICM{1}$), the outreach of most measures is close to its maximum already with five seeds, and disparity is low: at most about $0.015$ on Facebook, whose group partition is essentially non-homophilic, and about $0.025$ on DBLP. On Facebook, for $k\ge10$, a group of PageRank-, path- and flow-based measures together with MMB, FSPR, LFPR and LASPN reaches a few per cent more nodes than the others, at similar or slightly higher (still small) disparity; at $k=5$ only FSPR, LFPR and bridging stand out in this way. Because the error bars of all measures overlap, no measure is clearly fairer than the others on Facebook under $\ICM{1}$. The clustering coefficient and PC reach markedly fewer nodes at small budgets on both networks. On DBLP, FSPR and LFPR have both the highest outreach and the lowest disparity for $k=50$. Attenuating cross-group transmission then sharply increases disparity on both networks, while outreach decreases (Figs.~\ref{fig:real_icm2} and~\ref{fig:real_icm3}): under $\ICM{3}$, it reaches $0.16$--$0.17$ on Facebook and about $0.09$ on DBLP. That this happens on the non-homophilic Facebook network indicates that community-size imbalance alone suffices to create substantial disparity. Under random mixing, about $80\%$ of the edges of a minority node but only $20\%$ of the edges of a majority node cross the group boundary, so attenuating cross-group transmission penalizes the smaller group far more. For $k=50$ under attenuated transmission, the fairness-aware measures, FSPR in particular, are again the fairest high-outreach measures on DBLP; on Facebook, FSPR shares the lowest disparity with LASPN under $\ICM{2}$ and has the lowest disparity among the high-outreach measures under $\ICM{3}$, whereas LFPR is not consistently fairer than the other measures.

\partitle{\textbf{LTM diffusion model.}} Under the LTM (Fig.~\ref{fig:real_ltm}) outreach is much lower than under the ICM, particularly for small budgets, and grows strongly with $k$, while disparity remains small (at most $\approx0.026$). The two networks differ: on Facebook disparity tends to decrease with outreach, whereas on DBLP it increases. On Facebook, core-type measures ($k$-core, $k$-shell, the $H$-indices and eigenvector centrality) reach only a few hundred nodes but have much higher disparity than the hub measures, while hub measures such as PageRank and betweenness reach the most nodes at low disparity; on DBLP, $k$-core and the $H^2$-index reach few nodes with low disparity. The clustering coefficient and PC barely spread on either network. In contrast to the synthetic networks, FSPR does not perform well under threshold diffusion on real networks: at $k=50$ it has the highest disparity of all measures on Facebook and one of the highest on DBLP. LFPR fares better: on Facebook, it is among the two measures with the highest outreach for $k\ge10$, with low to moderate disparity, and on DBLP, at $k=50$, it has the second-lowest disparity among the high-outreach measures.

\partitle{\textbf{Comparison of the two networks.}} The two networks differ in the two properties that, according to Section~\ref{sec:theory}, govern cross-group reach. Facebook is dense and essentially non-homophilic, so each minority node has many edges into the majority, and cascades that started anywhere reach both groups; disparity is then small under uniform transmission and, once cross-group transmission is attenuated, arises mainly from community-size imbalance. DBLP is about five times sparser and mildly homophilic, so a larger share of the reach depends on which group the seeds belong to; this is consistent with the observation that the fairness-aware measures, which are designed to favor the protected group, offer a clearer advantage on DBLP than on Facebook under cascade diffusion. Under threshold diffusion, the relationship between outreach and disparity also differs: it decreases on Facebook and increases on DBLP. A plausible explanation is density: on the dense Facebook network, large-threshold cascades eventually reach both groups, whereas on the sparser, mildly homophilic DBLP network, they grow mainly within the group in which they start.

\subsection{Robustness to network perturbations}\label{sec:robust}

Centrality scores are computed on an observed snapshot of a network, which may contain missing or spurious edges. To assess the effect of such errors, we keep the seed sets computed on the original network $G=(V,E)$ and simulate diffusion on a perturbed network $G'=(V,E')$. The centrality scores and seed sets are not recomputed, thereby isolates the effect of the structural change on the diffusion from its effect on seed selection. For a perturbation level $\eta\in\{1\%,2\%,3\%\}$ we consider two types of perturbation. For edge removal, a set $E_{\mathrm{del}}\subseteq E$ with $|E_{\mathrm{del}}|=\lfloor\eta|E|\rfloor$ is drawn uniformly at random and $E'=E\setminus E_{\mathrm{del}}$. For edge addition, a set $E_{\mathrm{add}}\subseteq\binom{V}{2}\setminus E$ of non-edges with $|E_{\mathrm{add}}|=\lfloor\eta|E|\rfloor$ is drawn uniformly at random and $E'=E\cup E_{\mathrm{add}}$. The perturbed network thus has about $(1-\eta)|E|$ or $(1+\eta)|E|$ edges. To account for the randomness of the perturbation, we generate $M=20$ independent perturbed networks for every perturbation type and level and run $T=1{,}000$ Monte Carlo simulations on each, under $\ICM{2}$ ($\pin=0.1$, $\pout=0.05$) and under the LTM. Figs.~\ref{fig:rob_icm_3} and~\ref{fig:rob_ltm_3} show $\eta=3\%$; the other levels are reported in Appendix~\ref{app:robust}.

\begin{figure}
    \centering
    \includegraphics[width=\linewidth]{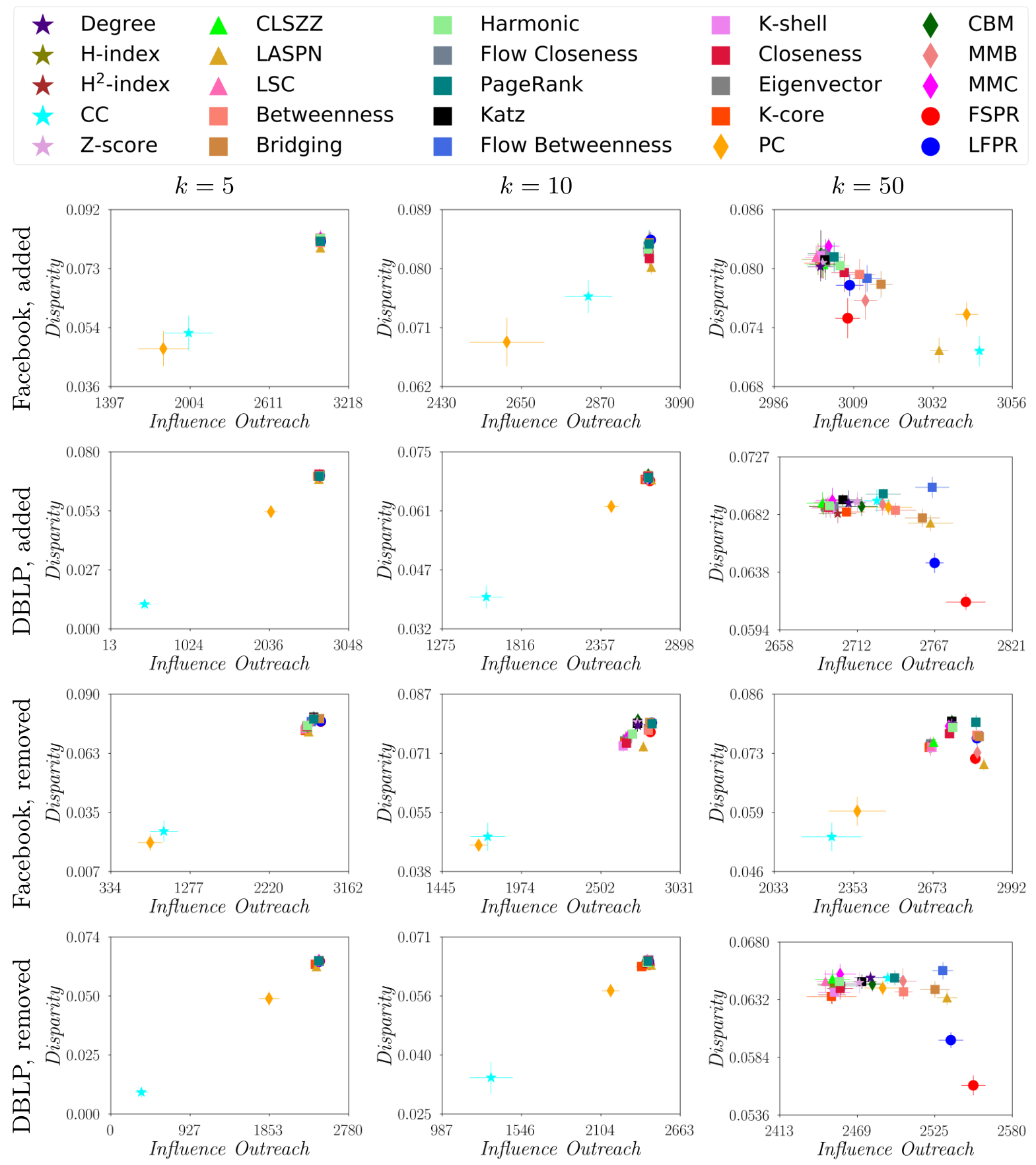}
\caption{{\textbf{Robustness, ICM, $\eta=3\%$:} disparity versus influence outreach under $\ICM{2}$ ($\pin=0.1$) on Facebook and DBLP after randomly adding (top two rows) or removing (bottom two rows) $3\%$ of the edges; seed sets computed on the original networks; averages over $M=20$ perturbed networks.}}
\label{fig:rob_icm_3}
\end{figure}

\partitle{\textbf{ICM diffusion model.}} Under $\ICM{2}$, edge addition shifts outreach to the right and edge removal shifts it to the left, as the monotonicity of the ICM implies for the expected outreach (Section~\ref{sec:theory}), and the shift grows with $\eta$; the only departures are PC on Facebook for $k=50$, whose outreach is about $4\%$ higher after edge removal at every level, and the clustering coefficient on DBLP for $k=10$, which loses about $4\%$ after $1\%$ edge addition. On DBLP, the effect is small and of similar size in both directions; the relative positions of the measures are preserved, and FSPR and LFPR remain the two measures with the lowest disparity and the highest outreach for $k=50$ in every setting. On Facebook, edge removal changes the outreach of the main group of measures by less than $2\%$ and leaves disparity unchanged, whereas edge addition raises it by $4$--$10\%$ and slightly increases disparity. Edge addition also changes the ranking for $k=50$: the clustering coefficient and PC move from low outreach to the highest outreach of all measures, presumably because the random edges give their seeds shortcuts to the rest of the network. Apart from this case, the conclusions drawn under $\ICM{2}$ on the real networks are robust to random errors of up to $3\%$ of the edges.

\begin{figure}
    \centering
    \includegraphics[width=\linewidth]{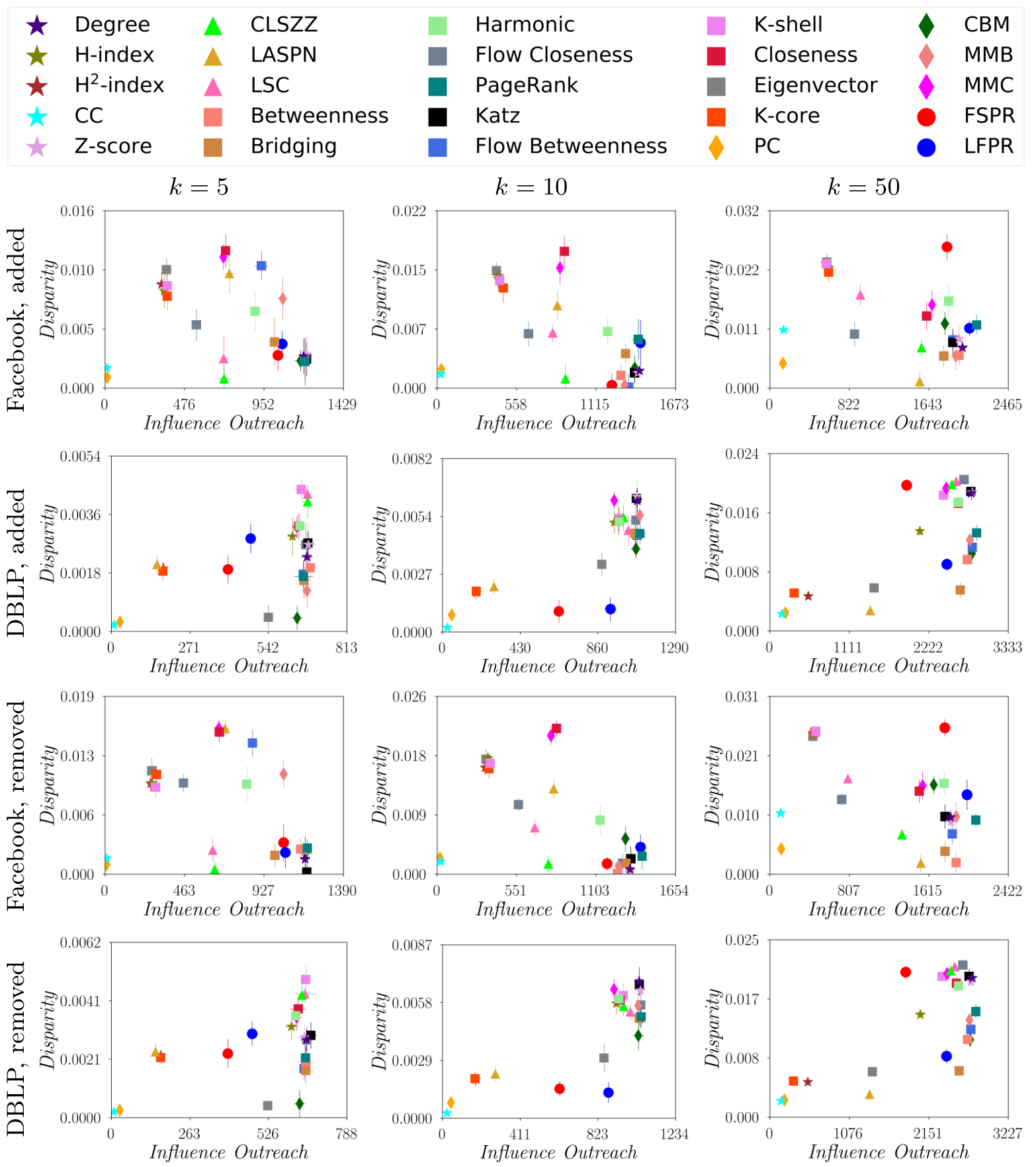}
\caption{{\textbf{Robustness, LTM, $\eta=3\%$:} disparity versus influence outreach under the LTM on Facebook and DBLP after randomly adding (top two rows) or removing (bottom two rows) $3\%$ of the edges; seed sets computed on the original networks; averages over $M=20$ perturbed networks.}}
\label{fig:rob_ltm_3}
\end{figure}

\partitle{\textbf{LTM diffusion model.}} Under the LTM, the results are also stable. Outreach changes by only a few percent; the ordering of the measures by outreach and by disparity is largely preserved on both networks and for all budgets, and the largest disparity at $k=50$ is essentially unchanged. On Facebook, edge addition lifts some low- and mid-outreach measures by $10$--$27\%$ and lowers the largest disparity at small budgets. %The hub measures achieve the highest outreach with near-zero disparity at $k=5$ on Facebook, whereas FSPR achieves the low disparity at $k=50$ on Facebook (and one of the lowest on DBLP). 
LFPR keeps its position as well: on Facebook, it remains one of the two measures with the highest outreach for $k\ge10$, with low to moderate disparity, whereas on DBLP, its outreach remains below that of most other measures, although at $k=50$ it retains the second-lowest disparity among the high-outreach measures.

\subsection{Sensitivity to the activation probability}\label{sec:sens}

To examine whether the conclusions depend on $\pin=0.1$, we repeated the ICM experiments with $\pin\in\{0.2,0.3\}$ and $\pout\in\{\pin,\pin/2,\pin/4\}$ on the synthetic networks ($k=100$; Figs.~\ref{fig:sens_hba_02}--\ref{fig:sens_hich_03}) and on the real networks (Figs.~\ref{fig:sens_real_icm1}--\ref{fig:sens_real_icm3}). For HICH-BA, the main experiments use $k=50$, so where this network is compared with $\pin=0.1$, we state which budget is used. Consistent with the monotonicity of the ICM (Section~\ref{sec:theory}), outreach increases for every measure. Disparity does not respond uniformly. It is comparable or lower at higher $\pin$ on Facebook, on HBA at low homophily, and on HICH-BA at high homophily, presumably because larger cascades cross community boundaries more often. It increases markedly on HBA at high homophily, on HICH-BA at low homophily, and on DBLP under attenuated transmission, presumably because larger cascades saturate the majority first while the minority remains poorly connected to it; on DBLP under uniform transmission it changes little.

The effects of attenuation and homophily persist, but the ranking of the measures changes in three respects. First, on HBA at $h=0.8$, FSPR rather than LFPR becomes the fairest measure under $\ICM{2}$ and $\ICM{3}$ at both activation probabilities and under $\ICM{1}$ at $\pin=0.3$; LFPR remains the fairest only under $\ICM{1}$ at $\pin=0.2$. Presumably, with stronger transmission, cascades started from the many minority seeds of FSPR also reach the majority. Second, on HICH-BA, FSPR and LFPR have the lowest mean disparity under $\ICM{2}$ and $\ICM{3}$ across all homophily levels, although at $\pin=0.3$ the differences are mostly within the error bars. At $\pin=0.1$ they are instead among the least fair measures at $h=0.2$ for $k=50$ (Fig.~\ref{fig:hich_disp}), and under $\ICM{2}$ with the same budget as here, $k=100$, they have the highest disparity of all measures at every homophily level (Fig.~\ref{fig:seed_hich}). Third, on DBLP the differences between measures become very small: for $k=50$, FSPR, closely followed by LFPR, has the lowest disparity under $\ICM{2}$ and $\ICM{3}$ at $\pin=0.2$, but PC has the lowest at $\pin=0.3$. On Facebook, LASPN achieves the highest outreach at $k=50$ with comparatively low disparity across all three variants. At higher probabilities, the error bars of many measures overlap on the synthetic and real networks. This makes it harder to identify a clearly preferable seeding strategy and supports using $\pin=0.1$ in the main experiments.

\section{Discussion}\label{sec:discussion}

\begin{table}[!t]
\centering
\caption{Measures with the most favorable outreach--disparity trade-off, i.e.\ the lowest disparity among the measures with near-maximal outreach (approximate disparity in parentheses). Trivially fair measures (Lemma~\ref{lem:bounds}(a)) are excluded. Where no measure with near-maximal outreach has comparatively low disparity, the fairest cascade-triggering measures are listed with their outreach as a percentage of the maximum; ``none'' marks settings in which no measure is clearly preferable, with the reason given in the cell. Synthetic networks: $k=100$ (HBA) and $k=50$ (HICH-BA) under the ICM, $k=5$ under the LTM; real-world networks: $k=50$.}
\label{tab:summary}
\footnotesize
\begin{tabularx}{\linewidth}{@{}ll>{\raggedright\arraybackslash}X>{\raggedright\arraybackslash}X@{}}
\toprule
Network & Model & Low homophily ($h=0.2$) & High homophily ($h=0.8$) \\
\midrule
HBA & $\ICM{1}$ & FSPR ($0.001$), LFPR ($0.005$) & CBM, $k$-core ($0.010$), LFPR ($0.013$) \\
 & $\ICM{2}$ & CBM ($0.002$), MMB, $H$-index ($\le0.004$) & LFPR ($0.006$) \\
 & $\ICM{3}$ & $H^2$-index ($0.002$), $H$-index ($0.005$), CBM ($0.008$) & LFPR ($0.011$) \\
 & LTM & none; disparity grows with outreach & FSPR ($0.024$; $55\%$), $k$-core ($0.040$; $72\%$) \\
\midrule
HICH-BA & $\ICM{1}$ & PC ($0.016$) & Z-score ($0.022$), LASPN ($0.023$) \\
 & $\ICM{2}$ & Z-score ($0.018$), CC, PC ($0.019$--$0.020$) & LFPR ($0.025$), Z-score ($0.030$), FSPR ($0.031$) \\
 & $\ICM{3}$ & Z-score ($0.017$) & LFPR ($0.028$), FSPR ($0.034$) \\
 & LTM & LFPR ($0.109$; $77\%$), MMC ($0.107$; $70\%$) & LFPR ($0.071$), FSPR ($0.073$); $\approx65\%$ \\
\midrule
 & & Facebook & DBLP \\
\cmidrule(l){3-4}
Real & $\ICM{1}$ & none; all within $0.007$--$0.015$ (overlapping error bars) & FSPR ($0.023$), LFPR ($0.024$) \\
 & $\ICM{2}$ & FSPR, LASPN ($0.068$) & FSPR ($0.060$), LFPR ($0.062$) \\
 & $\ICM{3}$ & FSPR ($0.157$), LASPN ($0.163$) & FSPR ($0.080$), LFPR ($0.085$) \\
 & LTM & betweenness ($0.002$; $89\%$), bridging ($0.003$; $86\%$) & bridging ($0.006$; $93\%$), LFPR ($0.009$; $88\%$) \\
\bottomrule
\end{tabularx}
\end{table}

\paragraph{Centrality, network structure, and diffusion dynamics.}
Table~\ref{tab:summary} summarizes the measures offering the most favorable outreach--disparity trade-offs. No measure or class consistently performs best. Fairness-aware measures are most effective in homophilic networks and real networks under the ICM model. At low homophily with attenuated transmission, however, CBM, MMB and the $H$-indices on HBA networks, and the Z-score, clustering coefficient and PC (HICH-BA), often perform better. Fairness-aware corrections may over-serve the minority or increase disparity. Community awareness alone is insufficient: MMC performs poorly in homophilic HBA networks, whereas CBM and MMB are effective at low levels of homophily. Fairness therefore depends on the structural positions prioritized by each measure and their distribution across groups.

Diffusion dynamics further influence these outcomes. Attenuated cross-community transmission reduces outreach and increases disparity, consistent with Lemma~\ref{lem:bounds}(b) and~(c). Community-size imbalance alone can generate disparity when cross-group transmission is weak (Section~\ref{sec:res:real}). On synthetic networks, LTM diffusion generally produces greater disparity than ICM diffusion and distinguishes measures that trigger cascades from those that do not. LFPR balances outgoing mass locally, producing approximately balanced seed sets in homophilic networks (Section~\ref{sec:res:sd}). It performs consistently under attenuated cascades at $\pin=0.1$ and outperforms FSPR under the LTM on real networks. Conversely, FSPR concentrates seeds in the protected group, potentially over-correcting under weak cascades but benefiting from reinforcement in homophilic HBA networks. Its high LTM disparity on Facebook and DBLP, however, shows that these benefits do not generalize. Fairness assessments must therefore account for the diffusion model.

These observations extend previous findings. Karimi et al.~\cite{karimi2018homophily} showed that minority degree rankings depend on homophily, explaining why hub-based measures shift from minority over-service at $h=0.2$ to majority over-service at $h=0.8$ in HBA networks. Esp\'in-Noboa et al.~\cite{espin2022inequality} established the dependence of PageRank inequalities on homophily and group size; we demonstrate that attenuated transmission amplifies these inequalities in influence propagation. Consistent with Stoica et al.~\cite{stoica:2020}, balanced seeding can improve both fairness and outreach, although the disparity-minimizing allocation need not be proportional to group size. Moreover, the low disparity at low outreach (Lemma~\ref{lem:bounds}(a)) highlights why gap-based fairness measures~\cite{fish2019gaps} must be evaluated alongside total outreach.

\paragraph{Outreach--fairness trade-offs and seed allocation.}
Under ICM diffusion, improving fairness often incurs little outreach loss. In homophilic HBA networks, LFPR retains over $90\%$ of maximal outreach while reducing disparity several-fold; in homophilic HICH-BA networks, LFPR and FSPR can even outperform majority-biased measures in outreach. On DBLP, both achieve the highest outreach and lowest disparity under uniform transmission at $k=50$. Under LTM diffusion, however, the trade-off becomes substantial on synthetic networks: the fairest cascade-triggering measures achieve approximately $55\%$ of maximal outreach in homophilic HBA and $65\%$ in homophilic HICH-BA networks. This may reflect reinforcement within densely connected regions, where seeds in smaller, less connected groups trigger weaker cascades. The trade-off is milder on real networks, where betweenness (Facebook) and bridging (DBLP) achieve low disparity with approximately $90\%$ of maximal outreach.

Seed allocation strongly influences disparity. In the two-community model, the disparity-minimizing allocation assigns the minority approximately twice its population share of seeds. Increasing the seed budget does not necessarily improve fairness: disparity increases with budget under the LTM, while fairness-aware measures exhibit the highest disparity under $\ICM{2}$ on HICH-BA at $k=100$. On real-world networks, the findings remain largely robust to random perturbations affecting up to $3\%$ of edges under $\ICM{2}$ and the LTM, except for edge addition on Facebook under $\ICM{2}$ (Section~\ref{sec:robust}).

\paragraph{Practical implications and adversarial robustness.}
Several practical recommendations emerge. First, disparity should be reported alongside outreach and a directional measure such as IBR to distinguish minority under-service from over-correction. Second, structurally balanced measures (CBM or MMB for two communities, Z-score or PC for multiple communities) are preferable in heterophilic or weakly homophilic networks with attenuated cross-community transmission. Third, fairness-aware PageRank variants are preferable under homophily and attenuated transmission: LFPR is generally more reliable at low activation probabilities, whereas FSPR becomes competitive at higher probabilities and can outperform LFPR in homophilic HBA networks (Section~\ref{sec:sens}). Finally, group-specific quotas for selecting top-ranked nodes could make conventional centrality measures fairness-aware, although this approach remains unevaluated. %Since performance varies with network structure, activation probability, and budget, measure selection should be validated on the target network.

Our robustness analysis covers random perturbations, whereas targeted manipulation presents additional risks. By Lemma~\ref{lem:bounds}(b), modifying inter-community edges can directly alter cross-group reach and disparity. Fairness-aware measures also introduce vulnerabilities associated with protected attributes. In LFPR, each node allocates a share $\varphi$ of its mass to protected-group neighbors; an injected protected-labeled node connected to many nodes lacking other protected neighbors could accumulate substantial mass and enter the seed set, constituting a Sybil-type attack~\cite{yu2006sybilguard}. In FSPR, injected protected-labeled nodes instead dilute the protected mass. Misreporting group membership affects community-aware and fairness-aware measures but not structure-only measures. Quantifying vulnerabilities to Sybil injection, strategic edge rewiring, structural poisoning~\cite{waniek2018hiding,zugner2018adversarial}, and label manipulation remains for future work.

\paragraph{Limitations.} The synthetic results are based on one network realization per homophily level, so the error bars reflect the randomness of the diffusion rather than variability across networks. We compare the measures using their mean values and standard deviations without conducting formal statistical tests. Therefore, we conclude only when differences are clearly larger than the reported error bars and avoid ranking conventional measures on HICH-BA when their error bars overlap. %Comparisons rely on means and standard deviations rather than formal tests; we base the main conclusions on differences that clearly exceed the error bars where these are shown, and therefore do not rank the measures within the overlapping cluster of conventional measures on HICH-BA. %, such as the penta-level spreading model \cite{gupta2016modeling} and the Multiple Adoption Linear Threshold model \cite{pham2019multiple},
Both real-world networks have a binary, metadata-derived attribute, so a multi-community real-world evaluation remains to be done. Furthermore, the ICM and LTM are simplistic diffusion models; however, advanced models \cite{saxena2021fake} that capture more complex diffusion processes could be considered in future evaluations. Additionally, the choice of $\pin=0.1$ in the ICM is conventional rather than empirically estimated. 
%The ICM and LTM are simple models; however, there are several extensions \cite{saxena2021fake}, such as the penta-level spreading model \cite{gupta2016modeling} and the Multiple Adoption Linear Threshold model \cite{pham2019multiple},  that better capture complex real-world diffusion. These models can be explored in future work. Additionally, in ICM, $\pin=0.1$ is a conventional choice rather than an estimate. 
%Disparity and IBR capture the differences between the most and least served communities, potentially overlooking inequalities among other communities. Finally, FSPR and LFPR are evaluated using a single value of the fairness parameter $\varphi$ per network, leaving a more comprehensive sensitivity analysis of the parameter for future work.
Disparity and IBR are extreme statistics, dominated by the most- and least-served communities. Finally, FSPR and LFPR are evaluated for a single value of the fairness parameter $\varphi$ per network. 

%% =====================================================================

\section{Conclusion}\label{sec:conclusion}
%% =====================================================================

We evaluated the fairness of centrality-based seeding for influence propagation in networks with imbalanced communities, combining analytical bounds with a systematic comparison of $25$ centrality measures on synthetic and real-world networks under cascade and threshold diffusion. Three findings are most relevant in practice. First, no class of centrality measures is fair by construction, and low disparity is meaningful only when paired with high outreach. Second, the fairness-aware PageRank variants are the best choice in homophilic networks under attenuated cascade and threshold diffusion, but which is preferable depends on the diffusion model, the activation probability, and the network. LFPR is the more reliable under cascade diffusion at the default activation probability, whereas at higher activation probabilities FSPR becomes the fairer of the two in homophilic HBA networks; under threshold diffusion, FSPR helps on homophilic synthetic networks but performs poorly on the real ones. Third, in the two-community model the fairest seed sets give the minority about twice its population share of seeds, which suggests quota-based seed allocation as a simple fairness intervention. On the real networks, the relative performance of the measures is largely stable under random removal or addition of up to $3\%$ of the edges under both $\ICM{2}$ and the LTM, with edge addition on Facebook under $\ICM{2}$ as the main exception. Future work includes an explicit relation between homophily and the expected cut size, allocation-based fair seeding rules, targeted adversarial manipulation, individual-level fairness, and hypergraphs and multilayer networks.

\backmatter

\section*{Declarations}

\paragraph{Funding} The authors declare that no funds, grants, or other support were received during the preparation of this manuscript.

\paragraph{Competing interests} The authors have no relevant financial or non-financial interests to disclose.

\paragraph{Author contributions} All authors contributed to the study conception and design. Problem formulation, analysis of results and manuscript writing were carried out collaboratively by all authors. K.M.\ implemented the code and conducted the experiments. All authors reviewed, revised and approved the final manuscript.

\paragraph{Data availability} The Facebook network is available from the SNAP repository~\cite{snapnets}; the gender-labelled DBLP network is available at~\cite{fairPRrec}. The code for generating the synthetic networks, computing the centrality measures and running the simulations, together with the generated data, will be made publicly available on GitHub upon acceptance.

\paragraph{Use of AI tools} During the preparation of this revision, the authors used a large language model to assist with language editing and rephrasing parts of the text. All scientific content, analyzes, and conclusions are the authors' own; the authors reviewed and edited all AI-assisted text and take full responsibility for the content of the manuscript. %No AI tool is listed as an author.

\bibliography{sn-bibliography}

@misc{snapnets,
  author = {Jure Leskovec and Andrej Krevl},
  title = {{SNAP Datasets}: Stanford Large Network Dataset Collection},
  howpublished = {\url{http://snap.stanford.edu/data}},
  year = {2014}
}

@incollection{saxena2021fake,
  title={Fake news propagation and mitigation techniques: A survey},
  author={Saxena, Akrati and Saxena, Pratishtha and Reddy, Harita},
  booktitle={Principles of Social Networking: The New Horizon and Emerging Challenges},
  pages={355--386},
  year={2021},
  publisher={Springer},
   address={Singapore}
}

@article{kumar2026fairness,
  title={Fairness-Sensitive PageRank Approximation},
  author={Kumar, Mukesh and Dixit, Gaurav and Saxena, Akrati},
  journal={arXiv preprint arXiv:2602.02329},
  year={2026}
}

@article{kumar2026locally,
  title={Locally Fair PageRank: Mean-Field Approximation and One-Step Refinement},
  author={Kumar, Mukesh and Dixit, Gaurav and Saxena, Akrati},
  journal={arXiv preprint arXiv:2609.22923},
  year={2026}
}

@article{stkepien2026fairness,
  title={Fairness in Opinion Dynamics},
  author={St{\k e}pie{\'n}, Stanis{\l}aw and Janik, Michalina and Nurek, Mateusz and Saxena, Akrati and Michalski, Rados{\l}aw},
  journal={arXiv preprint arXiv:2601.03859},
  year={2026}
}

@inproceedings{de2024group,
  title={Group fairness metrics for community detection methods in social networks},
  author={De Vink, Elze and Saxena, Akrati},
  booktitle={International Conference on Complex Networks and Their Applications},
  pages={43--56},
  year={2024},
  organization={Springer}
}

@article{corriera2026individual,
  title={Individual fairness in community detection: quantitative measure and comparative evaluation},
  author={Corriera, Fabrizio and Takes, Frank W and Saxena, Akrati},
  journal={Social Network Analysis and Mining},
  year={2026},
  publisher={Springer}
}

@article{saxena2022nodesim,
  title={Nodesim: node similarity based network embedding for diverse link prediction},
  author={Saxena, Akrati and Fletcher, George and Pechenizkiy, Mykola},
  journal={EPJ Data Science},
  volume={11},
  number={1},
  pages={24},
  year={2022},
  publisher={Springer}
}

@article{saxena2022hm,
  title={HM-EIICT: Fairness-aware link prediction in complex networks using community information},
  author={Saxena, Akrati and Fletcher, George and Pechenizkiy, Mykola},
  journal={Journal of Combinatorial Optimization},
  volume={44},
  number={4},
  pages={2853--2870},
  year={2022},
  publisher={Springer}
}

@article{saxena2025homophily,
  title={Homophily in Complex Networks: Measures, Models, and Applications},
  author={Saxena, Akrati and Kumar, Gaurav and Meena, Chandrakala},
  journal={arXiv preprint arXiv:2509.18289},
  year={2025}
}

@inproceedings{gupta2016modeling,
  title={Modeling memetics using edge diversity},
  author={Gupta, Yayati and Saxena, Akrati and Das, Debarati and Iyengar, SRS},
  booktitle={Complex Networks VII: Proceedings of the 7th Workshop on Complex Networks CompleNet 2016},
  pages={187--198},
  year={2016},
  organization={Springer}
}

@article{freeman2002centrality,
  title   = {Centrality in social networks conceptual clarification},
  author  = {Freeman, Linton C.},
  journal = {Social Networks},
  volume  = {1},
  number  = {3},
  pages   = {215--239},
  year    = {1978},
  doi     = {10.1016/0378-8733(78)90021-7}
}

@article{ghalmane2019centrality,
  title={Centrality in modular networks},
  author={Ghalmane, Zakariya and El Hassouni, Mohammed and Cherifi, Chantal and Cherifi, Hocine},
  journal={EPJ Data Science},
  volume={8},
  number={1},
  pages={15},
  year={2019},
  publisher={Springer}
}

@inproceedings{avin2015homophily,
  title={Homophily and the glass ceiling effect in social networks},
  author={Avin, Chen and Keller, Barbara and Lotker, Zvi and Mathieu, Claire and Peleg, David and Pignolet, Yvonne-Anne},
  booktitle={Proceedings of the 2015 conference on innovations in theoretical computer science},
  pages={41--50},
  year={2015}
}

@article{karimi2018homophily,
  title={Homophily influences ranking of minorities in social networks},
  author={Karimi, Fariba and G{\'e}nois, Mathieu and Wagner, Claudia and Singer, Philipp and Strohmaier, Markus},
  journal={Scientific reports},
  volume={8},
  number={1},
  pages={11077},
  year={2018},
  publisher={Nature Publishing Group UK London}
}

@article{lee2019homophily,
  title={Homophily and minority-group size explain perception biases in social networks},
  author={Lee, Eun and Karimi, Fariba and Wagner, Claudia and Jo, Hang-Hyun and Strohmaier, Markus and Galesic, Mirta},
  journal={Nature human behaviour},
  volume={3},
  number={10},
  pages={1078--1087},
  year={2019},
  publisher={Nature Publishing Group UK London}
}

@article{saxena2023fairness,
  title={Fairness-aware fake news mitigation using counter information propagation},
  author={Saxena, Akrati and Guti{\'e}rrez Bierbooms, Cristina and Pechenizkiy, Mykola},
  journal={Applied Intelligence},
  volume={53},
  number={22},
  pages={27483--27504},
  year={2023},
  doi={10.1007/s10489-023-04928-3}
}

@article{freeman1977set,
  title={A set of measures of centrality based on betweenness},
  author={Freeman, Linton C},
  journal={Sociometry},
  pages={35--41},
  year={1977},
  publisher={JSTOR}
}

@article{bonacich1972factoring,
  title={Factoring and weighting approaches to status scores and clique identification},
  author={Bonacich, Phillip},
  journal={Journal of mathematical sociology},
  volume={2},
  number={1},
  pages={113--120},
  year={1972},
  publisher={Taylor \& Francis}
}

@article{katz1953new,
  title={A new status index derived from sociometric analysis},
  author={Katz, Leo},
  journal={Psychometrika},
  volume={18},
  number={1},
  pages={39--43},
  year={1953},
  publisher={Springer-Verlag}
}

@article{brin1998anatomy,
  title={The anatomy of a large-scale hypertextual web search engine},
  author={Brin, Sergey and Page, Lawrence},
  journal={Computer networks and ISDN systems},
  volume={30},
  number={1-7},
  pages={107--117},
  year={1998},
  publisher={Elsevier}
}

@article{boldi2014axioms,
  title={Axioms for centrality},
  author={Boldi, Paolo and Vigna, Sebastiano},
  journal={Internet Mathematics},
  volume={10},
  number={3-4},
  pages={222--262},
  year={2014},
  publisher={Taylor \& Francis}
}

@article{opsahl2010node,
  title={Node centrality in weighted networks: Generalizing degree and shortest paths},
  author={Opsahl, Tore and Agneessens, Filip and Skvoretz, John},
  journal={Social networks},
  volume={32},
  number={3},
  pages={245--251},
  year={2010},
  publisher={Elsevier}
}

@article{seidman1983network,
  title={Network structure and minimum degree},
  author={Seidman, Stephen B},
  journal={Social networks},
  volume={5},
  number={3},
  pages={269--287},
  year={1983},
  publisher={Elsevier}
}

@inproceedings{zhong2017linear,
  title={On the linear threshold model for diffusion of innovations in multiplex social networks},
  author={Zhong, Yaofeng Desmond and Srivastava, Vaibhav and Leonard, Naomi Ehrich},
  booktitle={2017 IEEE 56th Annual Conference on Decision and Control (CDC)},
  pages={2593--2598},
  year={2017},
  organization={IEEE}
}

@inproceedings{brandes2005centrality,
  title={Centrality measures based on current flow},
  author={Brandes, Ulrik and Fleischer, Daniel},
  booktitle={Annual symposium on theoretical aspects of computer science},
  pages={533--544},
  year={2005},
  organization={Springer}
}

@article{peng2018influence,
  title={Influence analysis in social networks: A survey},
  author={Peng, Sancheng and Zhou, Yongmei and Cao, Lihong and Yu, Shui and Niu, Jianwei and Jia, Weijia},
  journal={Journal of Network and Computer Applications},
  volume={106},
  pages={17--32},
  year={2018},
  publisher={Elsevier}
}

@article{razaghi2022group,
  title={On the Group-Fairness-Aware Influence Maximization in Social Networks},
  author={Razaghi, Behnam and Roayaei, Mehdy and Charkari, Nasrollah Moghadam},
  journal={IEEE Transactions on Computational Social Systems},
  volume={10},
  number={6},
  pages={3406--3414},
  year={2023},
  doi={10.1109/TCSS.2022.3198096}
}

@article{saxena2024fairsna,
  title={{FairSNA}: Algorithmic Fairness in Social Network Analysis},
  author={Saxena, Akrati and Fletcher, George and Pechenizkiy, Mykola},
  journal={ACM Computing Surveys},
  volume={56},
  number={8},
  year={2024},
  doi={10.1145/3653711}
}

@article{chen2012identifying,
  title={Identifying influential nodes in complex networks},
  author={Chen, Duanbing and L{\"u}, Linyuan and Shang, Ming-Sheng and Zhang, Yi-Cheng and Zhou, Tao},
  journal={Physica a: Statistical mechanics and its applications},
  volume={391},
  number={4},
  pages={1777--1787},
  year={2012},
  publisher={Elsevier}
}

@inproceedings{farnad2020unifying,
  title     = {A Unifying Framework for Fairness-Aware Influence Maximization},
  author    = {Farnadi, Golnoosh and Babaki, Behrouz and Gendreau, Michel},
  booktitle = {Companion Proceedings of the Web Conference 2020},
  pages     = {714--722},
  year      = {2020},
  doi       = {10.1145/3366424.3383555}
}

@article{ali2021fairness,
  title   = {On the Fairness of Time-Critical Influence Maximization in Social Networks},
  author  = {Ali, Junaid and Babaei, Mahmoudreza and Chakraborty, Abhijnan and Mirzasoleiman, Baharan and Gummadi, Krishna P. and Singla, Adish},
  journal = {IEEE Transactions on Knowledge and Data Engineering},
  volume  = {35},
  number  = {3},
  pages   = {2875--2886},
  year    = {2023},
  doi     = {10.1109/TKDE.2021.3120561}
}

@inproceedings{stoica2019fairness,
  title={Fairness in social influence maximization},
  author={Stoica, Ana-Andreea and Chaintreau, Augustin},
  booktitle={Companion Proceedings of The 2019 World Wide Web Conference},
  pages={569--574},
  year={2019}
}

@article{hirsch2005index,
  title={An index to quantify an individual's scientific research output},
  author={Hirsch, Jorge E},
  journal={Proceedings of the National academy of Sciences},
  volume={102},
  number={46},
  pages={16569--16572},
  year={2005},
  publisher={National Academy of Sciences}
}

@techreport{hwang2006bridging,
  title       = {Bridging Centrality: Identifying Bridging Nodes in Scale-free Networks},
  author      = {Hwang, Woochang and Cho, Young-rae and Zhang, Aidong and Ramanathan, Murali},
  institution = {Department of Computer Science and Engineering, University at Buffalo, The State University of New York},
  number      = {2006-05},
  year        = {2006}
}

@article{tulu2018identifying,
  title={Identifying influential nodes based on community structure to speed up the dissemination of information in complex network},
  author={Tulu, Muluneh Mekonnen and Hou, Ronghui and Younas, Talha},
  journal={IEEE access},
  volume={6},
  pages={7390--7401},
  year={2018},
  publisher={IEEE}
}

@article{guimera2005functional,
  title={Functional cartography of complex metabolic networks},
  author={Guimera, Roger and Nunes Amaral, Lu{\'\i}s A},
  journal={nature},
  volume={433},
  number={7028},
  pages={895--900},
  year={2005},
  publisher={Nature Publishing Group UK London}
}

@inproceedings{tsioutsiouliklis2021fairness,
  title={Fairness-aware pagerank},
  author={Tsioutsiouliklis, Sotiris and Pitoura, Evaggelia and Tsaparas, Panayiotis and Kleftakis, Ilias and Mamoulis, Nikos},
  booktitle={Proceedings of the Web Conference 2021},
  pages={3815--3826},
  year={2021}
}

@article{yang2021improved,
  title={An improved gravity model to identify influential nodes in complex networks based on k-shell method},
  author={Yang, Xuan and Xiao, Fuyuan},
  journal={Knowledge-Based Systems},
  volume={227},
  pages={107198},
  year={2021},
  doi={10.1016/j.knosys.2021.107198}
}

@article{gao2014ranking,
  title={Ranking the spreading ability of nodes in complex networks based on local structure},
  author={Gao, Shuai and Ma, Jun and Chen, Zhumin and Wang, Guanghui and Xing, Changming},
  journal={Physica A: Statistical Mechanics and its Applications},
  volume={403},
  pages={130--147},
  year={2014},
  publisher={Elsevier}
}

@inproceedings{hussain2013empirical,
  title={Empirical analysis of seed selection criterion in influence mining for different classes of networks},
  author={Hussain, Owais A and Anwar, Zainab and Saleem, Sajid and Zaidi, Faraz},
  booktitle={2013 International Conference on Cloud and Green Computing},
  pages={348--353},
  year={2013},
  organization={IEEE}
}

@article{xiao2024new,
  title={A new semi-local centrality for identifying influential nodes based on local average shortest path with extended neighborhood},
  author={Xiao, Yi and Chen, Yuan and Zhang, Hongyan and Zhu, Xinghui and Yang, Yimin and Zhu, Xiaoping},
  journal={Artificial Intelligence Review},
  volume={57},
  number={5},
  pages={115},
  year={2024},
  doi={10.1007/s10462-024-10725-2}
}

@article{saxena2020centrality,
  title={Centrality measures in complex networks: A survey},
  author={Saxena, Akrati and Iyengar, Sudarshan},
  journal={arXiv preprint arXiv:2011.07190},
  year={2020}
}

@incollection{singh2022centrality,
  title={Centrality Measures: A Tool to Identify Key Actors in Social Networks},
  author={Singh, Rishi Ranjan},
  booktitle={Principles of Social Networking: The New Horizon and Emerging Challenges},
  series={Smart Innovation, Systems and Technologies},
  volume={246},
  pages={1--27},
  publisher={Springer Singapore},
  year={2022},
  doi={10.1007/978-981-16-3398-0_1}
}

@inproceedings{kundu2011new,
  title={A new centrality measure for influence maximization in social networks},
  author={Kundu, Suman and Murthy, CA and Pal, Sankar K},
  booktitle={Pattern Recognition and Machine Intelligence: 4th International Conference, PReMI 2011, Moscow, Russia, June 27-July 1, 2011. Proceedings 4},
  pages={242--247},
  year={2011},
  organization={Springer}
}

@article{jalali2023fairness,
  title   = {Fairness of Information Flow in Social Networks},
  author  = {Jalali, Zeinab S. and Chen, Qilan and Srikanta, Shwetha M. and Wang, Weixiang and Kim, Myunghwan and Raghavan, Hema and Soundarajan, Sucheta},
  journal = {ACM Transactions on Knowledge Discovery from Data},
  volume  = {17},
  number  = {6},
  pages   = {79:1--79:26},
  year    = {2023},
  doi     = {10.1145/3578268}
}

@inproceedings{rajeh2021investigating,
  title={Investigating centrality measures in social networks with community structure},
  author={Rajeh, Stephany and Savonnet, Marinette and Leclercq, Eric and Cherifi, Hocine},
  booktitle={Proceedings of the Ninth International Conference on Complex Networks and Their Applications 2020},
  pages={211--222},
  year={2021},
  organization={Springer}
}

@article{lu2016h,
  title={The H-index of a network node and its relation to degree and coreness},
  author={L{\"u}, Linyuan and Zhou, Tao and Zhang, Qian-Ming and Stanley, H Eugene},
  journal={Nature communications},
  volume={7},
  number={1},
  pages={10168},
  year={2016},
  publisher={Nature Publishing Group UK London}
}

@article{kosmulski2006new,
  title={A new Hirsch-type index saves time and works equally well as the original h-index},
  author={Kosmulski, Marek and others},
  journal={ISSI newsletter},
  volume={2},
  number={3},
  pages={4--6},
  year={2006}
}

@article{de2014role,
  title={Role of centrality for the identification of influential spreaders in complex networks},
  author={De Arruda, Guilherme Ferraz and Barbieri, Andr{\'e} Luiz and Rodriguez, Pablo Martin and Rodrigues, Francisco A and Moreno, Yamir and Costa, Luciano da Fontoura},
  journal={Physical Review E},
  volume={90},
  number={3},
  pages={032812},
  year={2014},
  publisher={APS}
}

@article{dey2021survey,
  title   = {A Survey on the Role of Centrality as Seed Nodes for Information Propagation in Large Scale Network},
  author  = {Dey, Paramita and Bhattacharya, Subhayan and Roy, Sarbani},
  journal = {ACM/IMS Transactions on Data Science},
  volume  = {2},
  number  = {3},
  pages   = {24:1--24:25},
  year    = {2021},
  doi     = {10.1145/3465374}
}

@article{li2018influence,
  title={Influence maximization on social graphs: A survey},
  author={Li, Yuchen and Fan, Ju and Wang, Yanhao and Tan, Kian-Lee},
  journal={IEEE Transactions on Knowledge and Data Engineering},
  volume={30},
  number={10},
  pages={1852--1872},
  year={2018},
  publisher={IEEE}
}

@article{wang2022information,
  title={Information access equality on generative models of complex networks},
  author={Wang, Xindi and Varol, Onur and Eliassi-Rad, Tina},
  journal={Applied Network Science},
  volume={7},
  number={1},
  pages={54},
  year={2022},
  publisher={Springer}
}

@inproceedings{anwar2021balanced,
  title={Balanced influence maximization in the presence of homophily},
  author={Anwar, Md Sanzeed and Saveski, Martin and Roy, Deb},
  booktitle={Proceedings of the 14th ACM International Conference on Web Search and Data Mining},
  pages={175--183},
  year={2021}
}

@inproceedings{tsang2019group,
  title={Group-Fairness in Influence Maximization},
  author={Tsang, Alan and Wilder, Bryan and Rice, Eric and Tambe, Milind and Zick, Yair},
  booktitle={Proceedings of the Twenty-Eighth International Joint Conference on Artificial Intelligence (IJCAI-19)},
  pages={5997--6005},
  year={2019},
  doi={10.24963/ijcai.2019/831}
}

@inproceedings{stoica:2018,
  title={Algorithmic glass ceiling in social networks: The effects of social recommendations on network diversity},
  author={Stoica, Ana-Andreea and Riederer, Christopher and Chaintreau, Augustin},
  booktitle={Proceedings of the 2018 World Wide Web Conference},
  pages={923--932},
  year={2018}
}

@inproceedings{stoica:2020,
  title={Seeding network influence in biased networks and the benefits of diversity},
  author={Stoica, Ana-Andreea and Han, Jessy Xinyi and Chaintreau, Augustin},
  booktitle={Proceedings of The Web Conference 2020},
  pages={2089--2098},
  year={2020}
}

@inproceedings{kempe2003maximizing,
  title={Maximizing the spread of influence through a social network},
  author={Kempe, David and Kleinberg, Jon and Tardos, {\'E}va},
  booktitle={Proceedings of the ninth ACM SIGKDD international conference on Knowledge discovery and data mining},
  pages={137--146},
  year={2003}
}

@article{dey2019influence,
  title   = {Influence maximization in online social network using different centrality measures as seed node of information propagation},
  author  = {Dey, Paramita and Chaterjee, Agneet and Roy, Sarbani},
  journal = {S{\=a}dhan{\=a}},
  volume  = {44},
  pages   = {205},
  year    = {2019},
  doi     = {10.1007/s12046-019-1189-7}
}

@article{ganguly2024influence,
  title={Influence maximization in community-structured social networks: a centrality-based approach},
  author={Ganguly, Maitreyee and Dey, Paramita and Roy, Sarbani},
  journal={The Journal of Supercomputing},
  volume={80},
  number={13},
  pages={19898--19941},
  year={2024},
  publisher={Springer}
}

@article{jiang2025overlapping,
  title={Overlapping community-based fair influence maximization under a multi-transformation optimization algorithm},
  author={Jiang, Chunfeng and Niu, Zegang and Qu, Jingru and Zhao, Yulan and Rezaeipanah, Amin},
  journal={Journal of Big Data},
  volume={12},
  number={1},
  pages={133},
  year={2025},
  publisher={Springer}
}

@article{gong2023influence,
  title={Influence maximization considering fairness: A multi-objective optimization approach with prior knowledge},
  author={Gong, Hao and Guo, Chunxiang},
  journal={Expert Systems with Applications},
  volume={214},
  pages={119138},
  year={2023},
  publisher={Elsevier}
}

@article{lin2023fair,
  title={Fair influence maximization in large-scale social networks based on attribute-aware reverse influence sampling},
  author={Lin, Mingkai and Sun, Lintan and Yang, Rui and Liu, Xusheng and Wang, Yajuan and Li, Ding and Li, Wenzhong and Lu, Sanglu},
  journal={Journal of Artificial Intelligence Research},
  volume={76},
  pages={925--957},
  year={2023}
}

@inproceedings{feng2023influence,
  title={Influence maximization with fairness at scale},
  author={Feng, Yuting and Patel, Ankitkumar and Cautis, Bogdan and Vahabi, Hossein},
  booktitle={Proceedings of the 29th ACM SIGKDD Conference on Knowledge Discovery and Data Mining},
  pages={4046--4055},
  year={2023}
}

@article{barabasi1999emergence,
  title={Emergence of scaling in random networks},
  author={Barab{\'a}si, Albert-L{\'a}szl{\'o} and Albert, R{\'e}ka},
  journal={science},
  volume={286},
  number={5439},
  pages={509--512},
  year={1999},
  publisher={American Association for the Advancement of Science}
}

@article{mcpherson2001birds,
  title={Birds of a feather: Homophily in social networks},
  author={McPherson, Miller and Smith-Lovin, Lynn and Cook, James M},
  journal={Annual review of sociology},
  volume={27},
  number={1},
  pages={415--444},
  year={2001},
  publisher={Annual Reviews 4139 El Camino Way, PO Box 10139, Palo Alto, CA 94303-0139, USA}
}

@article{newman2004fast,
  title={Fast algorithm for detecting community structure in networks},
  author={Newman, Mark EJ},
  journal={Physical review E},
  volume={69},
  number={6},
  pages={066133},
  year={2004},
  publisher={APS}
}

@inproceedings{PSSS:2025,
  title={Fairness-aware Analysis of Centrality Measures in Social Networks},
author={Mohapatra, Kirtidev and Singh, Harsh Vardhan and Singh, Rishi Ranjan and Saxena, Akrati},
booktitle={Proceedings of the 14th International Conference on Computational Data and Social Networks (CSoNet)},
year={2025}
}

@inproceedings{zhao2024fair,
  title={Fair influence maximization in social networks: a group-fairness-aware multi-objective grey wolf optimizer},
  author={Zhao, Ziying and Li, Weihua and Ma, Jing and Jiang, Jianhua and Bai, Quan},
  booktitle={2024 IEEE International Conference on Agents (ICA)},
  pages={88--93},
  year={2024},
  organization={IEEE}
}

@inproceedings{wang2025faim,
  title={FAIM-RL: A Reinforcement Learning Approach for Fairness-Aware Adaptive Influence Maximization},
  author={Wang, Fangzheng and Zhou, Yi and Wang, Yanhao and Karras, Panagiotis and Li, Yuchen},
  booktitle={2025 IEEE International Conference on Data Mining (ICDM)},
  pages={783--792},
  year={2025},
  organization={IEEE}
}

@article{zhao2026fairness,
  title={Fairness-constrained influence maximisation via multi-objective optimisation},
  author={Zhao, Ziying and Li, Weihua and Ma, Jing and Jiang, Jianhua and Bai, Quan and Gu, Wen},
  journal={Swarm and Evolutionary Computation},
  volume={106},
  pages={102428},
  year={2026},
  publisher={Elsevier}
}

@article{osmani2026balancing,
  title={Balancing fairness and influence spread in social networks: a multi-objective evolutionary approach},
  author={Osmani, Amjad and Tafakori, Laleh and Jalili, Mahdi and Hassani, Hossein},
  journal={Artificial Intelligence Review},
  year={2026},
  doi={10.1007/s10462-026-11627-1}
}

@article{guo2026higher,
  title   = {Higher-order learning for fairness: a reinforcement learning framework for equitable and efficient influence maximization},
  author  = {Guo, Peijun and Li, Huan and Mo, Xinyue},
  journal = {Journal of King Saud University Computer and Information Sciences},
  volume  = {38},
  pages   = {403},
  year    = {2026},
  doi     = {10.1007/s44443-026-00798-0}
}

@article{rui2025scalable,
  title={A Scalable Algorithm for Fair Influence Maximization with Unbiased Estimator},
  author={Rui, Xiaobin and Wang, Zhixiao and Peng, Hao and Chen, Wei and Yu, Philip S.},
  journal={IEEE Transactions on Knowledge and Data Engineering},
  volume={37},
  number={7},
  pages={3881--3895},
  year={2025},
  doi={10.1109/TKDE.2025.3564283}
}

@article{zhang2026fairness,
  title={Fairness-Aware Influence Maximization with Randomized Strategies: A Stochastic {Frank--Wolfe} Framework},
  author={Zhang, Yapu and Chen, Shengminjie and Du, Liman and Zhang, Zhenning and Yang, Wenguo},
  journal={ACM Transactions on Knowledge Discovery from Data},
  volume={20},
  number={7},
  pages={1--26},
  year={2026},
  doi={10.1145/3827606}
}

@inproceedings{dam2026fairgcl,
  title={FairGCL: Embedding Fairness for Influence Maximization with Graph Contrastive Learning},
  author={Dam, Arpan and Roy, Sougata and Mitra, Bivas},
  booktitle={Proceedings of the 18th ACM Web Science Conference 2026},
  pages={227--237},
  year={2026}
}

@article{gao2026robust,
  title   = {Robust Fair Influence Maximization under Multiple Community Partitions},
  author  = {Gao, Tianyou and Ito, Takayuki},
  journal = {Proceedings of the ACM on Management of Data},
  volume  = {4},
  number  = {1},
  pages   = {1--25},
  year    = {2026},
  doi     = {10.1145/3786692}
}

@misc{saxena2026dq4fairim,
  title={{DQ4FairIM}: Fairness-aware Influence Maximization using Deep Reinforcement Learning},
  author={Saxena, Akrati and Yadav, Harshith Kumar and Rutten, Bart and Jha, Shashi Shekhar},
  howpublished={arXiv preprint arXiv:2512.00545},
  year={2025}
}

@article{watts1998collective,
  title={Collective dynamics of `small-world' networks},
  author={Watts, Duncan J. and Strogatz, Steven H.},
  journal={Nature},
  volume={393},
  number={6684},
  pages={440--442},
  year={1998}
}

@inproceedings{fish2019gaps,
  title={Gaps in information access in social networks?},
  author={Fish, Benjamin and Bashardoust, Ashkan and boyd, danah and Friedler, Sorelle and Scheidegger, Carlos and Venkatasubramanian, Suresh},
  booktitle={Proceedings of The World Wide Web Conference (WWW)},
  pages={480--490},
  year={2019}
}

@article{becker2022fairness,
  title={Fairness in influence maximization through randomization},
  author={Becker, Ruben and D'Angelo, Gianlorenzo and Ghobadi, Sajjad and Gilbert, Hugo},
  journal={Journal of Artificial Intelligence Research},
  volume={73},
  pages={1251--1283},
  year={2022}
}

@inproceedings{rahmattalabi2021fair,
  title={Fair influence maximization: A welfare optimization approach},
  author={Rahmattalabi, Aida and Jabbari, Shahin and Lakkaraju, Himabindu and Vayanos, Phebe and Izenberg, Max and Brown, Ryan and Rice, Eric and Tambe, Milind},
  booktitle={Proceedings of the AAAI Conference on Artificial Intelligence},
  volume={35},
  number={13},
  pages={11630--11638},
  year={2021}
}

@article{espin2022inequality,
  title={Inequality and inequity in network-based ranking and recommendation algorithms},
  author={Esp{\'\i}n-Noboa, Lisette and Wagner, Claudia and Strohmaier, Markus and Karimi, Fariba},
  journal={Scientific Reports},
  volume={12},
  pages={2012},
  year={2022}
}

@article{nemhauser1978analysis,
  title={An analysis of approximations for maximizing submodular set functions---{I}},
  author={Nemhauser, George L. and Wolsey, Laurence A. and Fisher, Marshall L.},
  journal={Mathematical Programming},
  volume={14},
  number={1},
  pages={265--294},
  year={1978}
}

@inproceedings{chen2010scalable,
  title={Scalable influence maximization for prevalent viral marketing in large-scale social networks},
  author={Chen, Wei and Wang, Chi and Wang, Yajun},
  booktitle={Proceedings of the 16th ACM SIGKDD International Conference on Knowledge Discovery and Data Mining},
  pages={1029--1038},
  year={2010}
}

@inproceedings{goyal2010learning,
  title={Learning influence probabilities in social networks},
  author={Goyal, Amit and Bonchi, Francesco and Lakshmanan, Laks V. S.},
  booktitle={Proceedings of the Third ACM International Conference on Web Search and Data Mining (WSDM)},
  pages={241--250},
  year={2010}
}

@inproceedings{zugner2018adversarial,
  title={Adversarial attacks on neural networks for graph data},
  author={Z{\"u}gner, Daniel and Akbarnejad, Amir and G{\"u}nnemann, Stephan},
  booktitle={Proceedings of the 24th ACM SIGKDD International Conference on Knowledge Discovery and Data Mining},
  pages={2847--2856},
  year={2018}
}

@inproceedings{yu2006sybilguard,
  title={{SybilGuard}: Defending against sybil attacks via social networks},
  author={Yu, Haifeng and Kaminsky, Michael and Gibbons, Phillip B. and Flaxman, Abraham},
  booktitle={Proceedings of the 2006 Conference on Applications, Technologies, Architectures, and Protocols for Computer Communications (SIGCOMM)},
  pages={267--278},
  year={2006}
}

@article{waniek2018hiding,
  title={Hiding individuals and communities in a social network},
  author={Waniek, Marcin and Michalak, Tomasz P. and Wooldridge, Michael J. and Rahwan, Talal},
  journal={Nature Human Behaviour},
  volume={2},
  number={2},
  pages={139--147},
  year={2018}
}

@article{dong2023fairness,
  title={Fairness in graph mining: A survey},
  author={Dong, Yushun and Ma, Jing and Wang, Song and Chen, Chen and Li, Jundong},
  journal={IEEE Transactions on Knowledge and Data Engineering},
  volume={35},
  number={10},
  pages={10583--10602},
  year={2023}
}

@inproceedings{tsioutsiouliklis2022link,
  title={Link recommendations for {PageRank} fairness},
  author={Tsioutsiouliklis, Sotiris and Pitoura, Evaggelia and Semertzidis, Konstantinos and Tsaparas, Panayiotis},
  booktitle={Proceedings of the ACM Web Conference 2022},
  pages={3541--3551},
  year={2022}
}

@misc{fairPRrec,
  author       = {Tsioutsiouliklis, Sotiris and Semertzidis, Konstantinos and Pitoura, Evaggelia and Tsaparas, Panayiotis},
  title        = {{fairPRrec}: Fair Link Recommendation Project},
  howpublished = {\url{https://github.com/ksemer/fairPRrec}},
  note         = {GitHub repository.},
  year         = {2022}
}

%% =====================================================================
\begin{appendices}

\section{Sensitivity analysis: figures}\label{app:sensitivity}

\begin{figure}
    \centering
    \includegraphics[width=\linewidth]{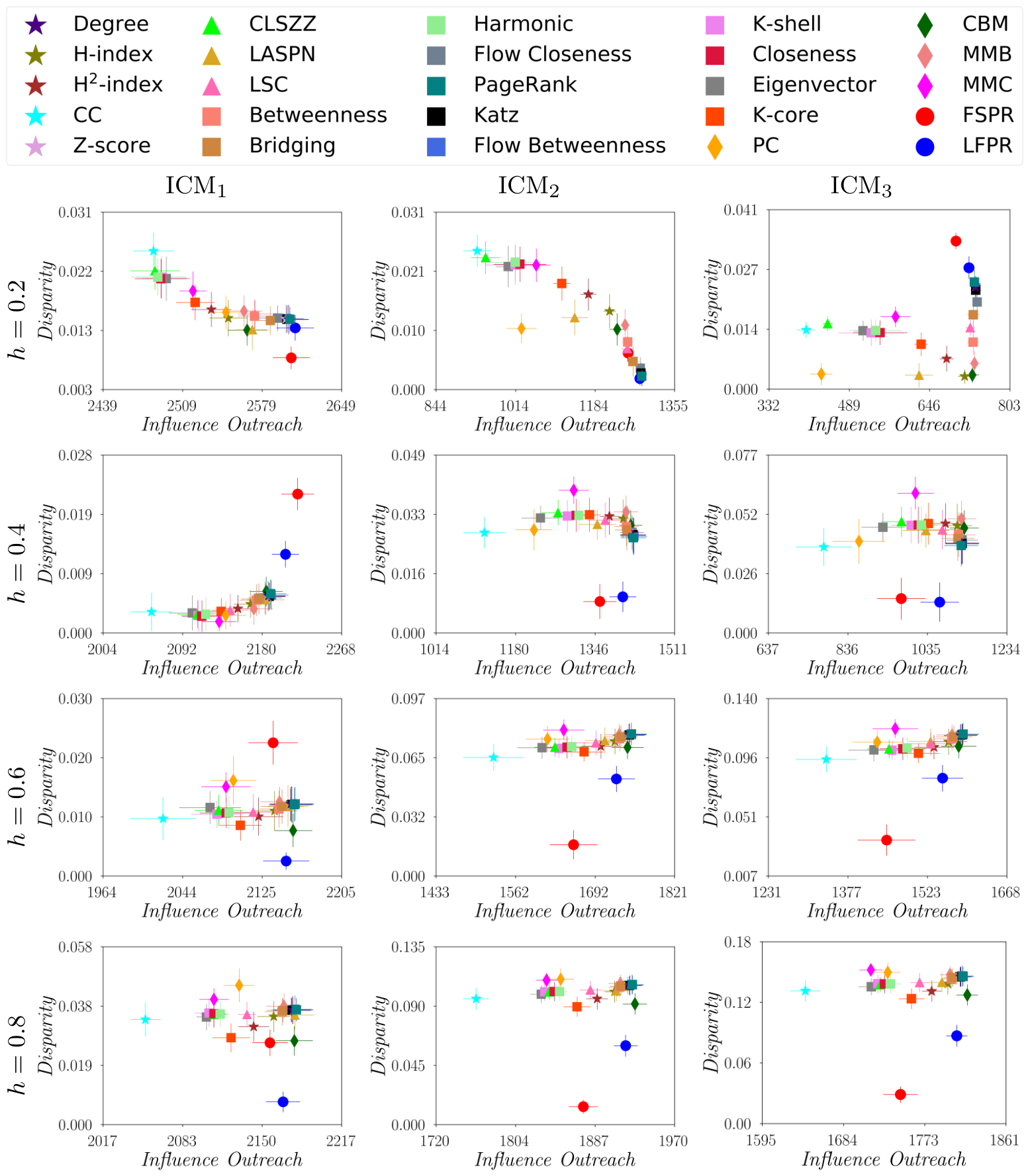}
\caption{{\textbf{Sensitivity to the activation probability:} disparity versus influence outreach on the HBA network with $\pin=0.2$ and $\pout\in\{\pin,\pin/2,\pin/4\}$ ($k=100$).}}
\label{fig:sens_hba_02}
\end{figure}

\begin{figure}
    \centering
    \includegraphics[width=\linewidth]{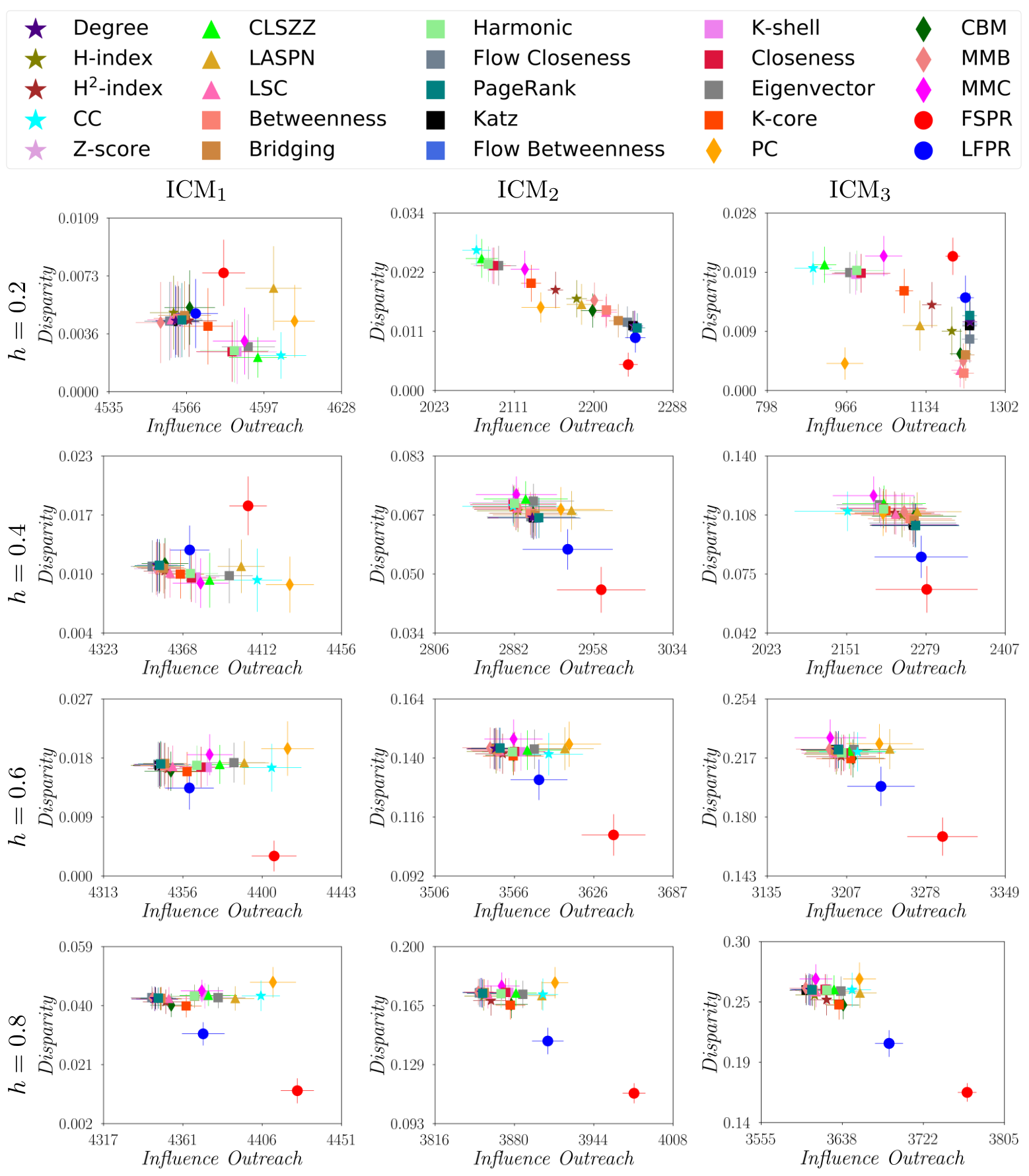}
\caption{{\textbf{Sensitivity to the activation probability:} disparity versus influence outreach on the HBA network with $\pin=0.3$ and $\pout\in\{\pin,\pin/2,\pin/4\}$ ($k=100$).}}
\label{fig:sens_hba_03}
\end{figure}

\begin{figure}
    \centering
    \includegraphics[width=\linewidth]{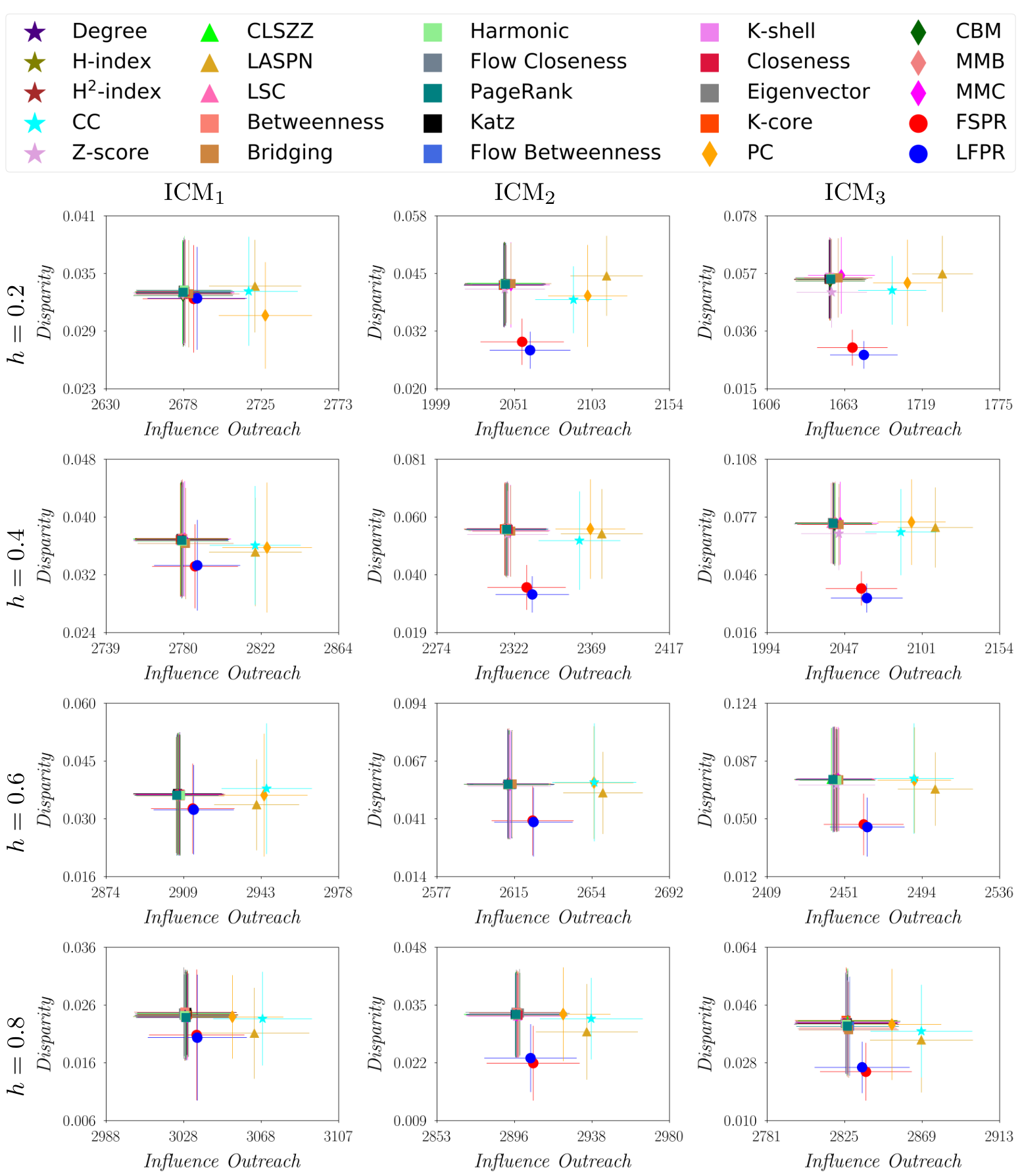}
\caption{{\textbf{Sensitivity to the activation probability:} disparity versus influence outreach on the HICH-BA network with $\pin=0.2$ and $\pout\in\{\pin,\pin/2,\pin/4\}$ ($k=100$).}}
\label{fig:sens_hich_02}
\end{figure}

\begin{figure}
    \centering
    \includegraphics[width=\linewidth]{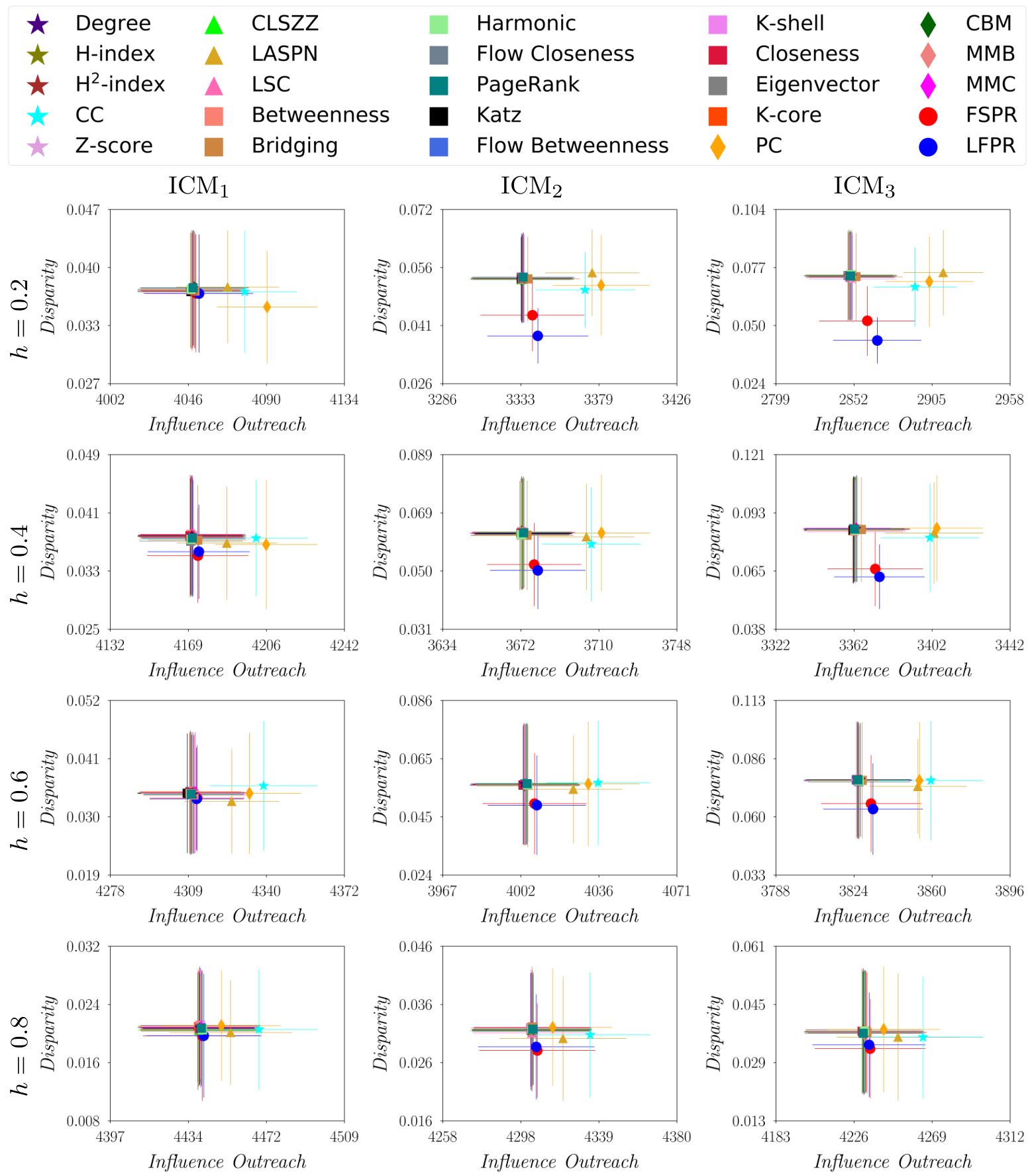}
\caption{{\textbf{Sensitivity to the activation probability:} disparity versus influence outreach on the HICH-BA network with $\pin=0.3$ and $\pout\in\{\pin,\pin/2,\pin/4\}$ ($k=100$).}}
\label{fig:sens_hich_03}
\end{figure}

\begin{figure}
    \centering
    \includegraphics[width=\linewidth]{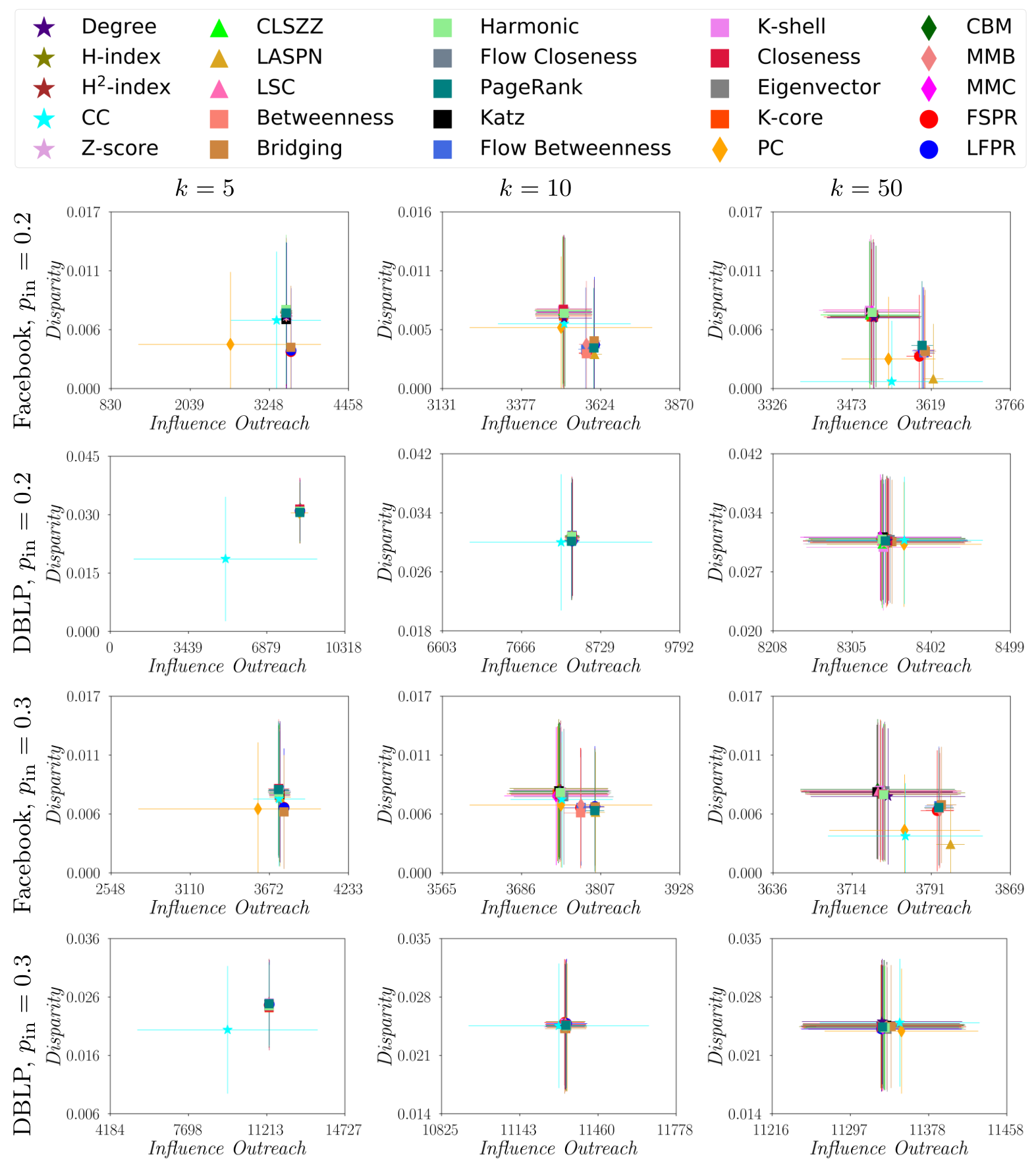}
\caption{{\textbf{Sensitivity to the activation probability, real-world networks:} disparity versus influence outreach under $\ICM{1}$ on Facebook and DBLP with $\pin=0.2$ (top two rows) and $\pin=0.3$ (bottom two rows).}}
\label{fig:sens_real_icm1}
\end{figure}

\begin{figure}
    \centering
    \includegraphics[width=\linewidth]{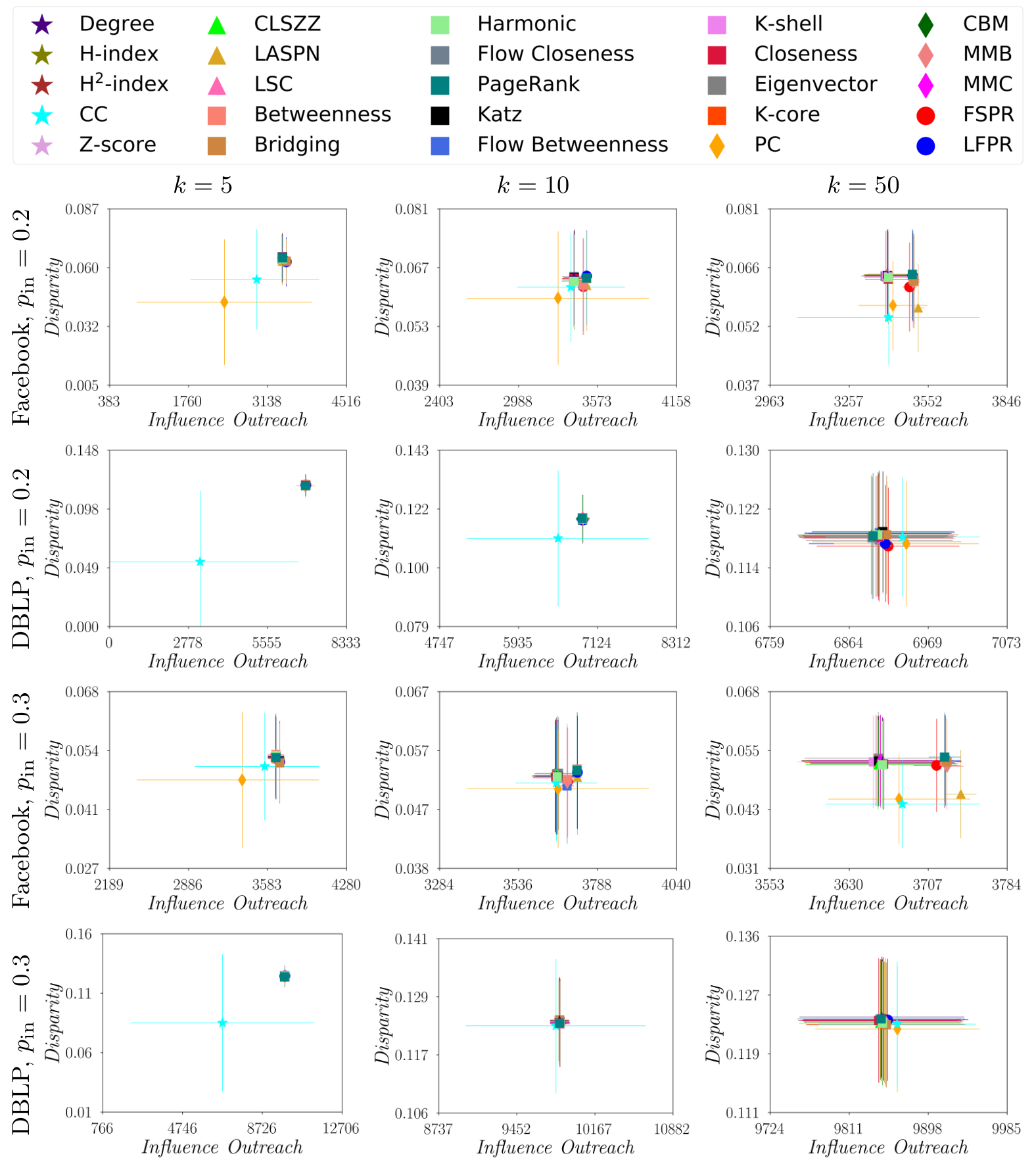}
\caption{{\textbf{Sensitivity to the activation probability, real-world networks:} disparity versus influence outreach under $\ICM{2}$ on Facebook and DBLP with $\pin=0.2$ (top two rows) and $\pin=0.3$ (bottom two rows).}}
\label{fig:sens_real_icm2}
\end{figure}

\begin{figure}
    \centering
    \includegraphics[width=\linewidth]{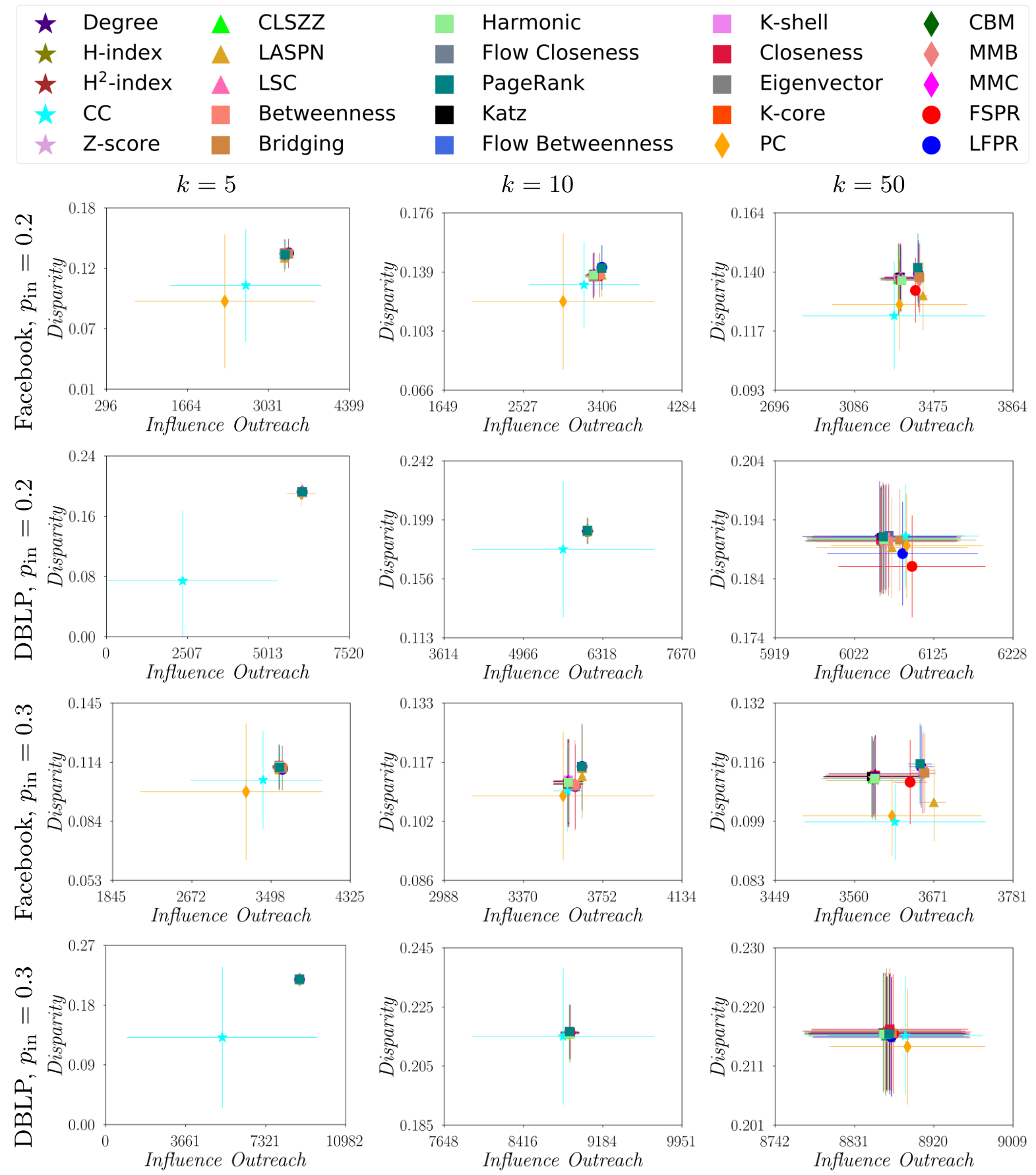}
\caption{{\textbf{Sensitivity to the activation probability, real-world networks:} disparity versus influence outreach under $\ICM{3}$ on Facebook and DBLP with $\pin=0.2$ (top two rows) and $\pin=0.3$ (bottom two rows).}}
\label{fig:sens_real_icm3}
\end{figure}

Figures~\ref{fig:sens_hba_02}--\ref{fig:sens_real_icm3} show the results of the sensitivity analysis of Section~\ref{sec:sens}, in which the ICM experiments are repeated with $\pin\in\{0.2,0.3\}$ and $\pout\in\{\pin,\pin/2,\pin/4\}$ ($k=100$ on the synthetic networks and $k\in\{5,10,50\}$ on the real networks).

\suppressfloats[t]

\FloatBarrier
\section{Complete robustness results}\label{app:robust}
\setcounter{figure}{0}

\begin{figure}[htbp]
    \centering
    \includegraphics[width=0.98\linewidth]{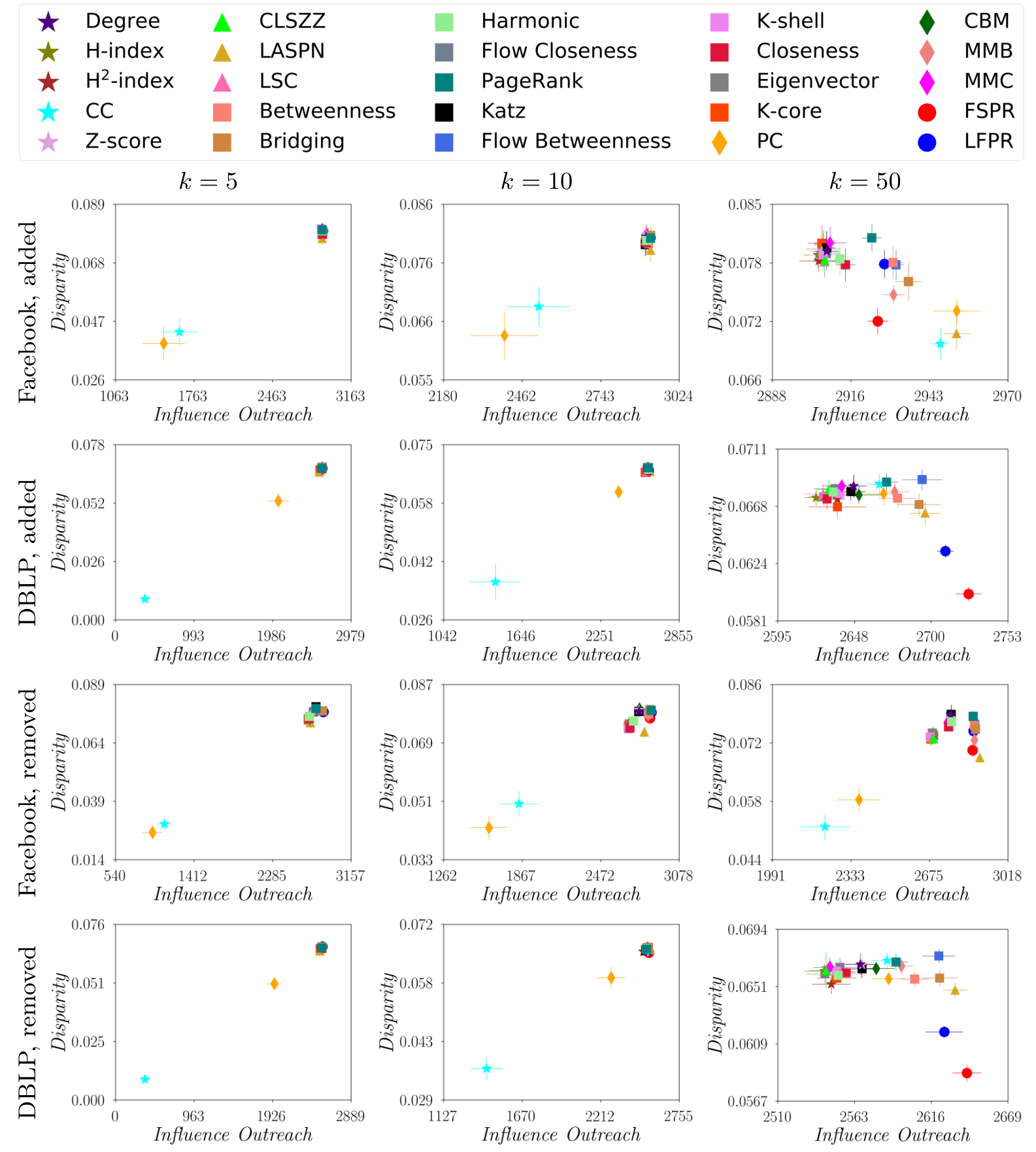}
\caption{{\textbf{Robustness, ICM, $\eta=1\%$:} disparity versus influence outreach under $\ICM{2}$ ($\pin=0.1$) on Facebook and DBLP after randomly adding (top two rows) or removing (bottom two rows) $1\%$ of the edges; seed sets computed on the original networks; averages over $M=20$ perturbed networks.}}
\label{fig:rob_icm_1}
\end{figure}

\begin{figure}
    \centering
    \includegraphics[width=\linewidth]{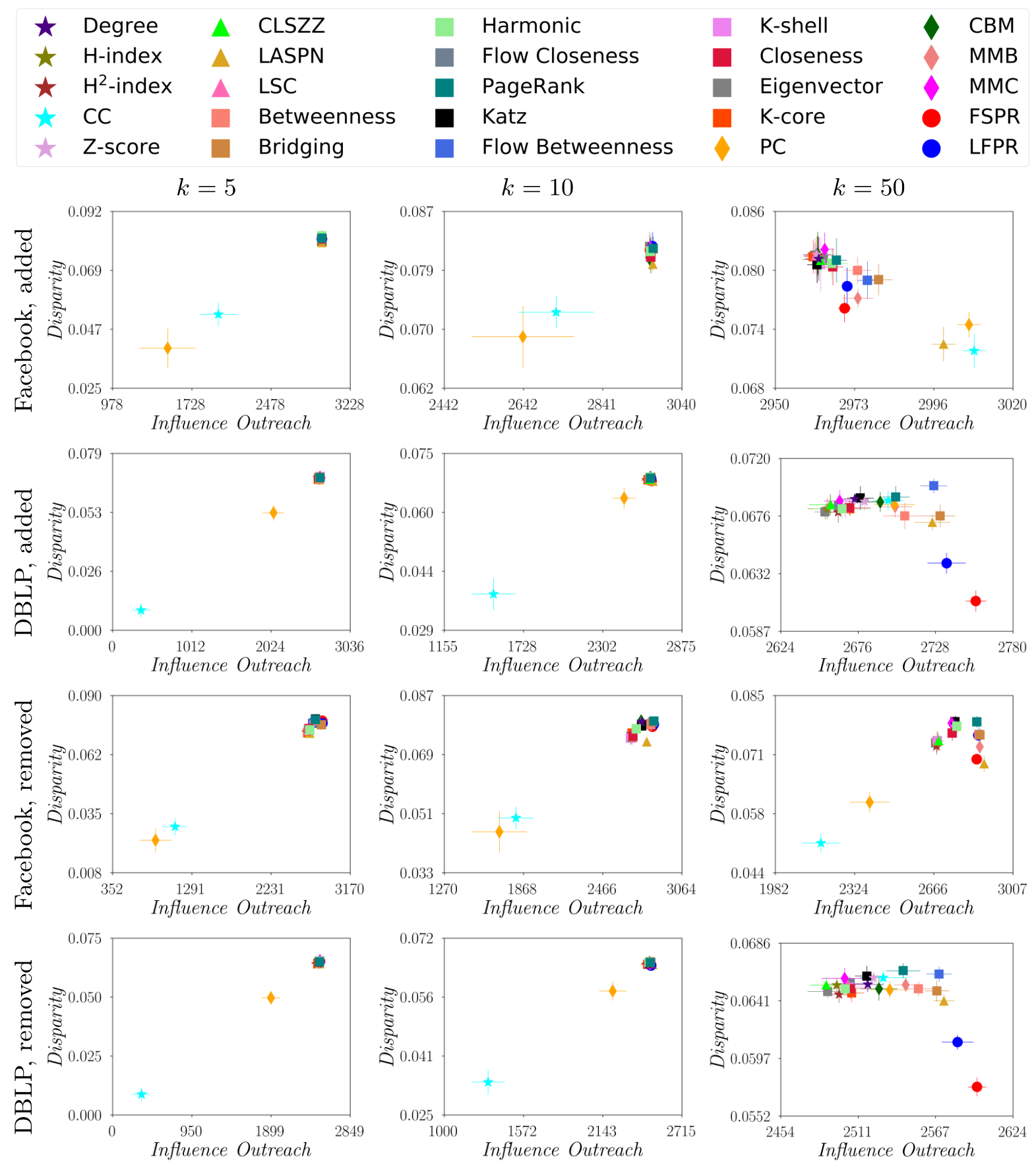}
\caption{{\textbf{Robustness, ICM, $\eta=2\%$:} disparity versus influence outreach under $\ICM{2}$ ($\pin=0.1$) on Facebook and DBLP after randomly adding (top two rows) or removing (bottom two rows) $2\%$ of the edges; seed sets computed on the original networks; averages over $M=20$ perturbed networks.}}
\label{fig:rob_icm_2}
\end{figure}

\begin{figure}
    \centering
    \includegraphics[width=\linewidth]{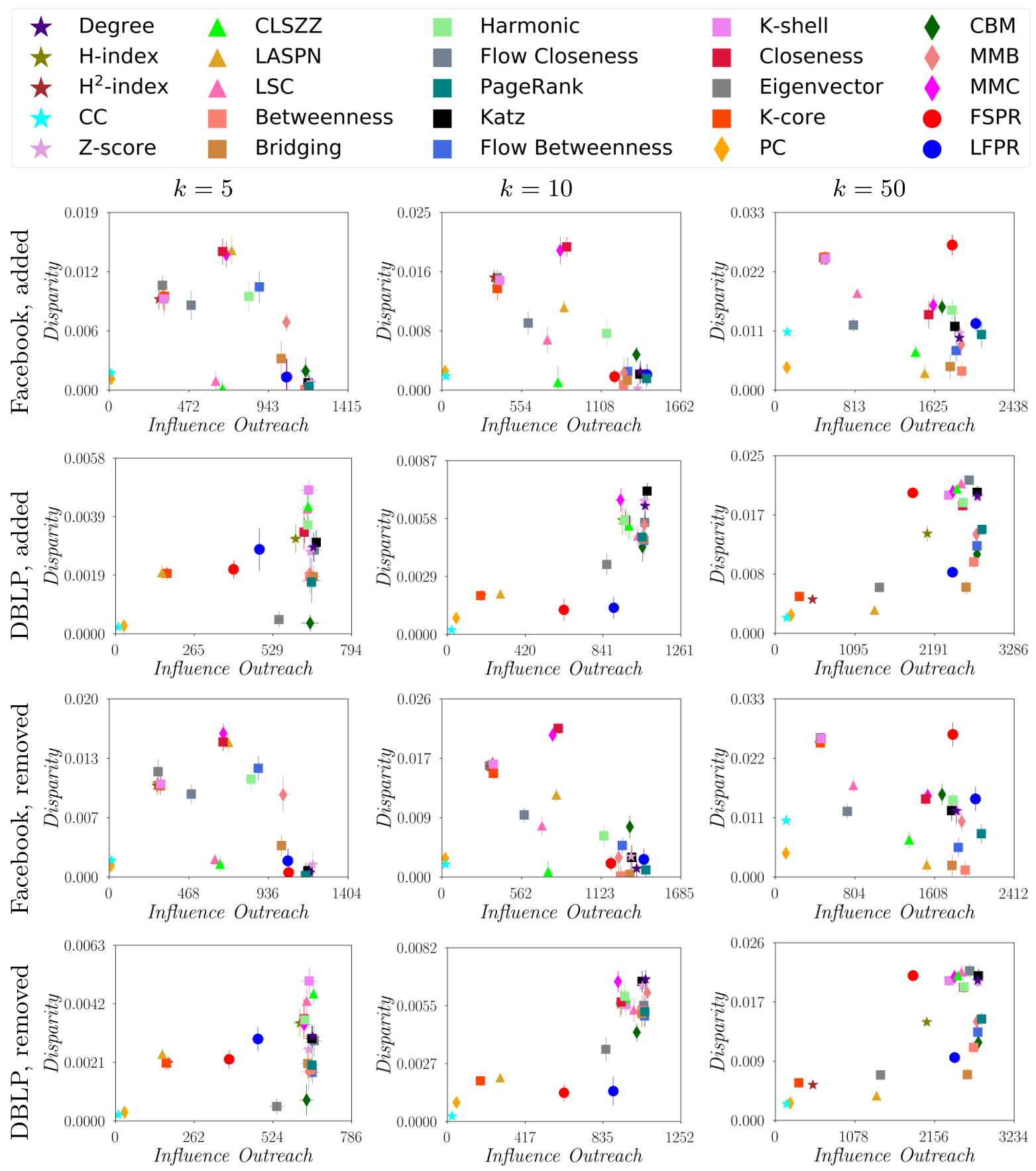}
\caption{{\textbf{Robustness, LTM, $\eta=1\%$:} disparity versus influence outreach under the LTM on Facebook and DBLP after randomly adding (top two rows) or removing (bottom two rows) $1\%$ of the edges; seed sets computed on the original networks; averages over $M=20$ perturbed networks.}}
\label{fig:rob_ltm_1}
\end{figure}

\begin{figure}
    \centering
    \includegraphics[width=\linewidth]{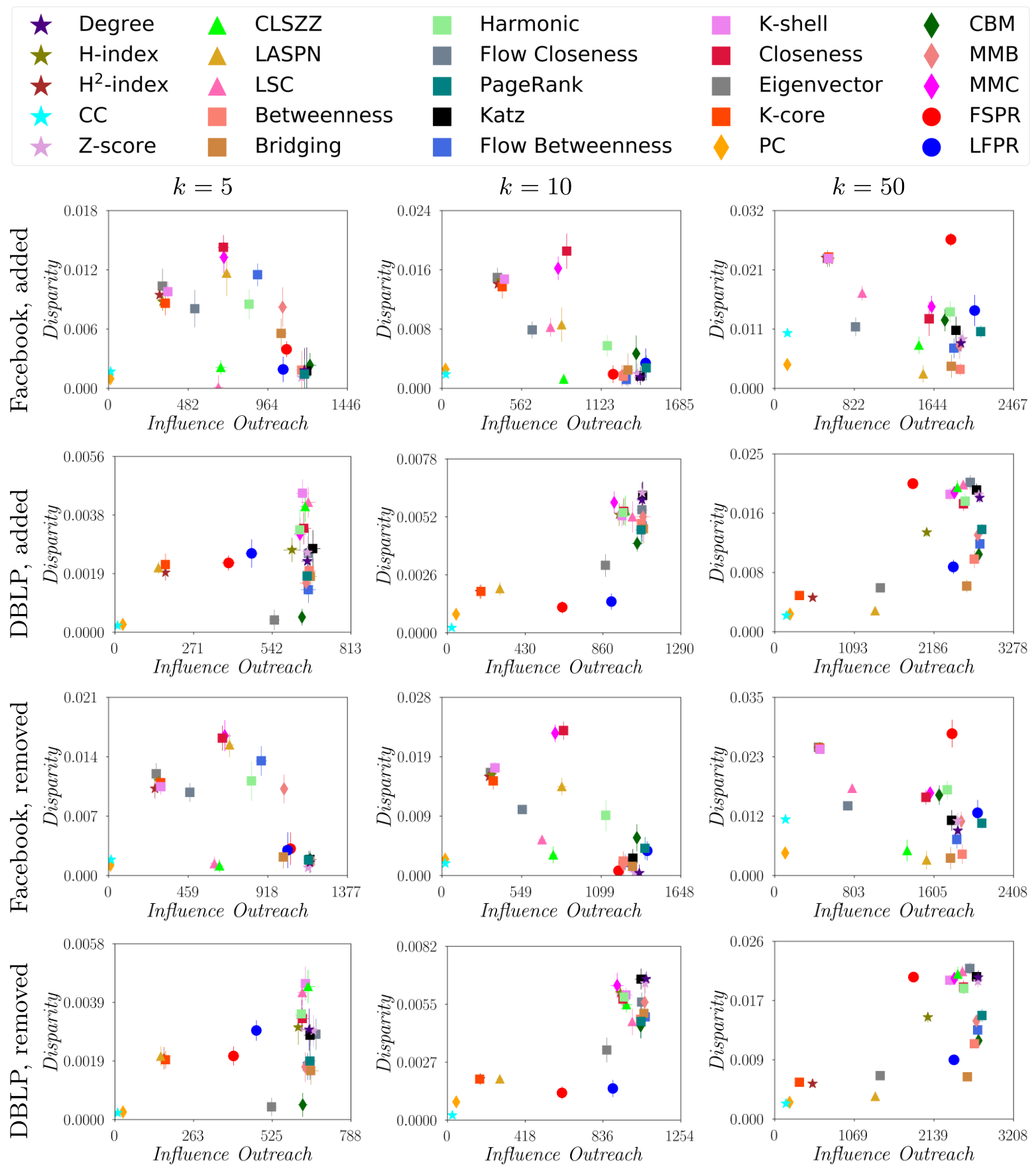}
\caption{{\textbf{Robustness, LTM, $\eta=2\%$:} disparity versus influence outreach under the LTM on Facebook and DBLP after randomly adding (top two rows) or removing (bottom two rows) $2\%$ of the edges; seed sets computed on the original networks; averages over $M=20$ perturbed networks.}}
\label{fig:rob_ltm_2}
\end{figure}

Figures~\ref{fig:rob_icm_1}, \ref{fig:rob_icm_2}, \ref{fig:rob_ltm_1} and~\ref{fig:rob_ltm_2} report the perturbation levels $\eta=1\%$ and $\eta=2\%$ that complement Figs.~\ref{fig:rob_icm_3} and~\ref{fig:rob_ltm_3}. Under $\ICM{2}$, the changes relative to the unperturbed networks grow with $\eta$, and under the LTM they remain small for most measures at every level; both lead to the same conclusions as in Section~\ref{sec:robust}.

\FloatBarrier

\section{Class-wise views}\label{app:classwise}
\setcounter{figure}{0}

\begin{figure}
    \centering
    \includegraphics[width=0.97\linewidth]{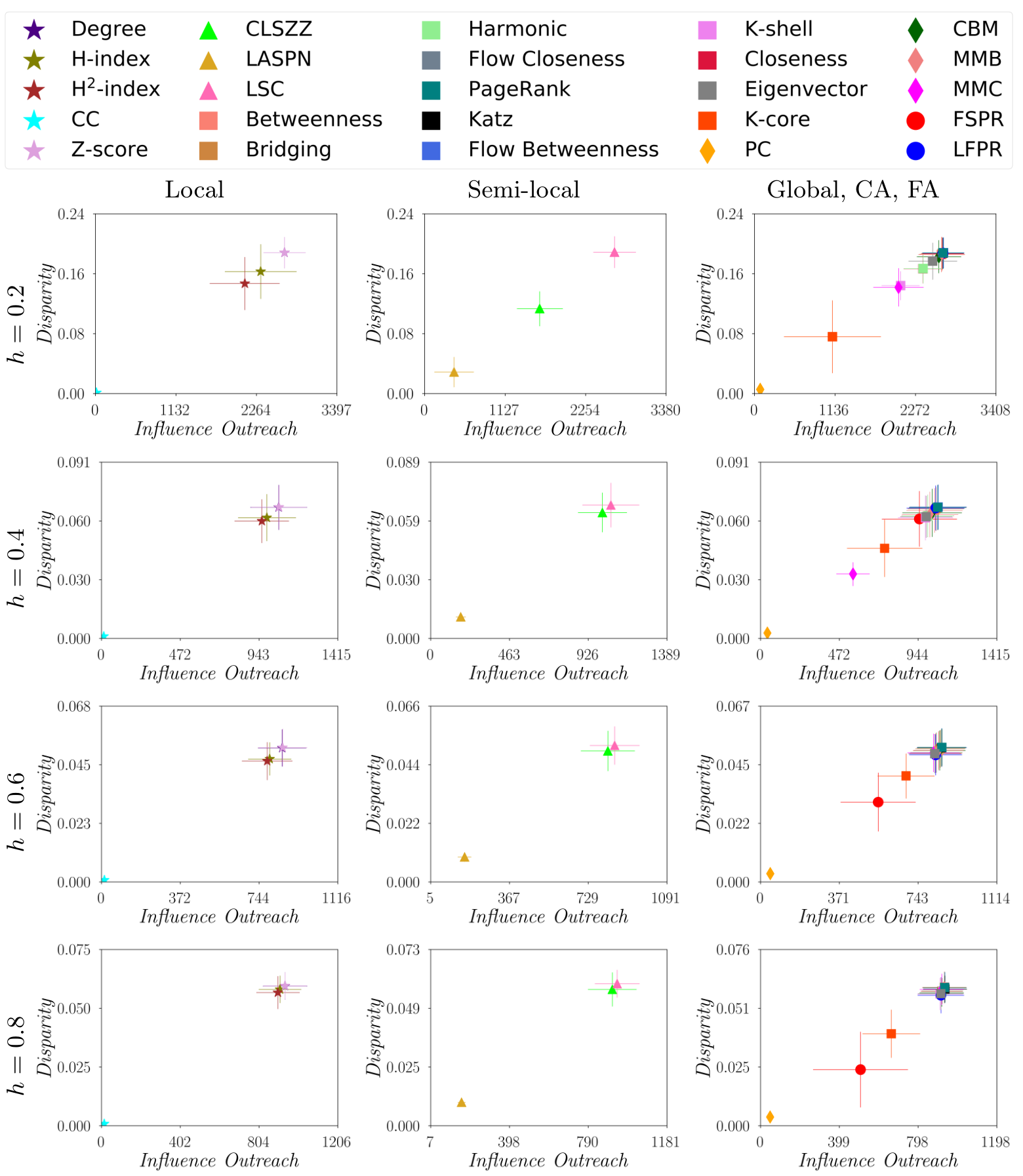}
\caption{Class-wise view of the LTM results on the HBA network ($k=5$).}
\label{fig:cls_hba_ltm}
\end{figure}

\begin{figure}
    \centering
    \includegraphics[width=\linewidth]{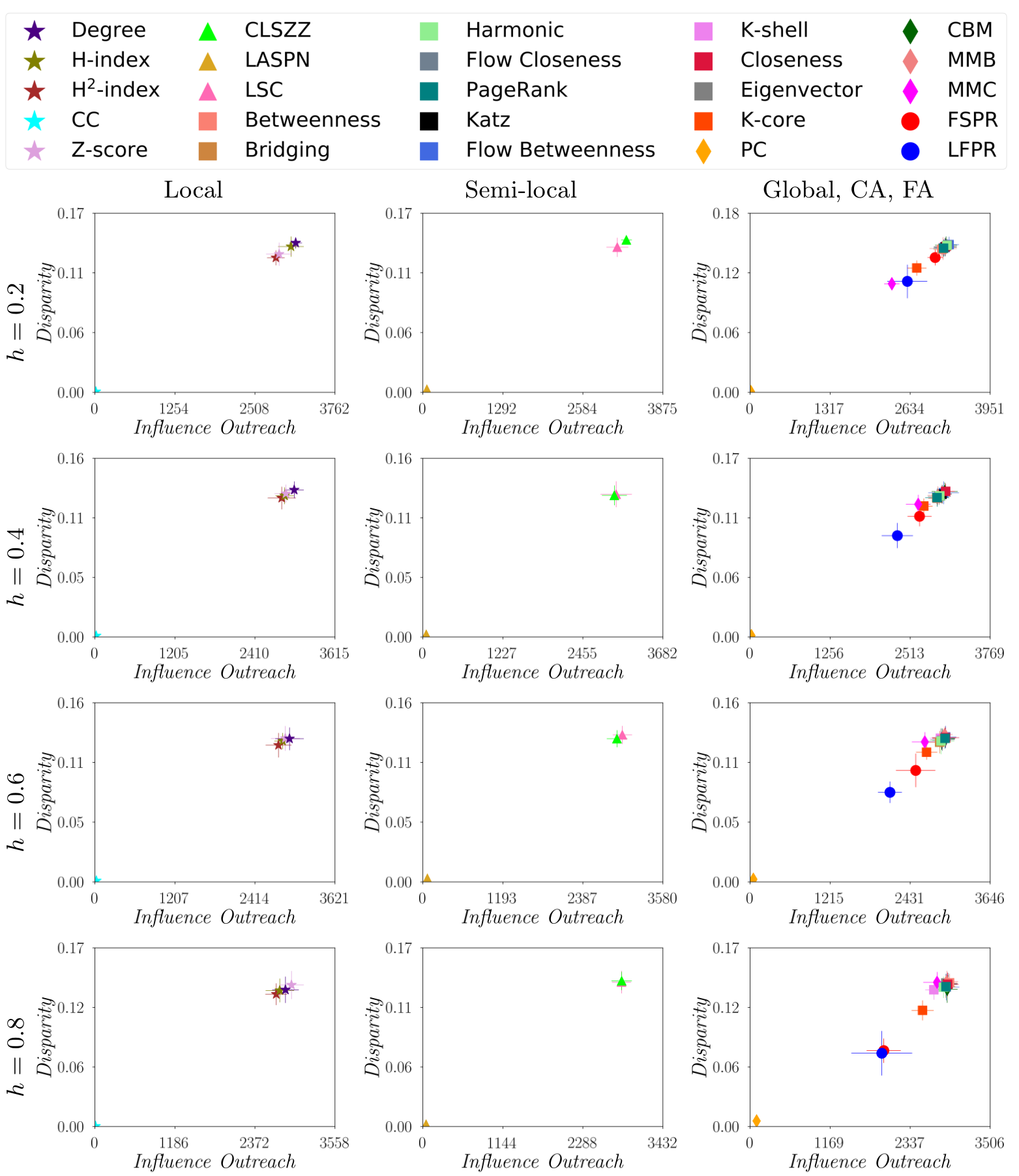}
\caption{Class-wise view of the LTM results on the HICH-BA network ($k=5$).}
\label{fig:cls_hich_ltm}
\end{figure}

Figures~\ref{fig:cls_hba_ltm} and~\ref{fig:cls_hich_ltm} show the LTM results on HBA and HICH-BA separately for local, semi-local, and global measures, in the same way as Figs.~\ref{fig:hba_class} and~\ref{fig:cls_hich_icm2}. They confirm that the spread within each class is comparable to the spread between classes, and they show that, under the LTM, disparity increases with outreach within each class.

\end{appendices}

\end{document}